\documentclass[11pt]{article}
\pdfoutput=1
\usepackage{jcappub,natbib}
\input{colordvi.tex}

\def\la{\mathrel{\raise.3ex\hbox{$<$\kern-.75em\lower1ex\hbox{$\sim$}}}}

\title{Warm dark energy}

\author[a,b]{Gianguido Dall’Agata,}
\author[b]{Sergio Gonz\'alez-Mart\'in,}
\author[c]{Alexandros Papageorgiou,}
\author[a,b]{Marco Peloso}

\affiliation[a]{Dipartimento di Fisica e Astronomia “Galileo Galilei” Universit\`a di Padova, 35131 Padova, Italy}
\affiliation[b]{INFN, Sezione di Padova, 35131 Padova, Italy}
\affiliation[c]{School of Physics and Astronomy, University of Minnesota, Minneapolis, 55455, USA}

\abstract{Motivated by some of the recent swampland conjectures, we study the implementation for the late time acceleration of the Universe of a mechanism developed by Anber and Sorbo in the context of primordial inflation, in which an axion field can slowly roll in a steep potential due to additional friction provided by its coupling to some U(1) gauge field. We first study the realization of this mechanism in N = 2 supergravity models resulting from string compactifications on Calabi--Yau manifolds. We then study the transition between matter domination and the axion domination, and show that indeed the backreaction of the produced gauge field can sufficiently slow the motion of the axion, so to produce the present accelerated era. We finally study the transition from a pre-inflationary matter or radiation domination to primordial inflation. In the regime that we could explore numerically, the evolution is characterized by stages of faster axion roll (and consequent bursts of gauge field amplification) intermitted by stages of slower roll, with a pattern that ``oscillates'' about the steady state Anber and Sorbo solution, but that does not appear to relax to it.}

\begin{document}

\maketitle
\flushbottom

\section{Introduction}
\label{sec:intro} 

Both the inflationary phase in the early epoch and the dark energy component of the current phase require that in the background describing our Universe there is a component acting as (nearly) a positive cosmological constant for some time.
Since we believe that the gravitational interaction should be eventually described by a quantum theory, we need to address the question about the consistency of (meta-)stable positive energy vacua in quantum gravity, or at least in string theory, which gives us a consistent framework where we can pose the question.
The difficulty of constructing consistent string theory models with a de Sitter phase, together with other hints coming from possible quantum gravity constraints on low-energy effective theories, has led first to the de Sitter conjecture \cite{Obied:2018sgi,Andriot:2018wzk,Dvali:2018fqu,Andriot:2018mav,Garg:2018reu,Ooguri:2018wrx,Rudelius:2019cfh}, in the context of the swampland programme \cite{Palti:2019pca}, and then to the Trans-Planckian-Censorship conjecture \cite{Bedroya:2019snp,Bedroya:2019tba}.
The main constraint that arises from these conjectures on effective theories of inflation or dark energy is that either there is a tachyon in the spectrum of the vacuum, with a mass of the order of the cosmological constant, or there is no vacuum and the slope of the inflaton / dark energy (quintessence) potential $V \left( \phi \right)$ satisfies
\begin{equation}\label{dsconj}
	|V'| \geq c \, V/M_p,
\end{equation}
for some constant $c$ of order one.
In this relation prime denotes derivative with respect to the quintessence or inflaton field $\phi$, while $M_p$ is the reduced Planck mass.

While this is a constraint that is generically violated in both inflationary and dark energy models, it has revived the quintessence scenario and called for a renewed and more careful analysis of its realizations.

Phenomenological constraints on quintessence work as upper bounds rather than lower bounds on $|V'/V|$ \cite{Akrami:2018ylq}. This creates a (small) tension between observation and theory. In this work we explore the possibility that a large slope of the quintessence potential can be compatible with the data in presence of a coupling of the quintessence with light fields.

We choose a coupling so that the motion of $\phi$ leads to an amplification of these fields.
These fields are therefore produced at the expense of the kinetic energy of $\phi$, resulting in an additional source of friction for its motion, beside the standard one due to the background expansion.

The idea of exploiting this extra friction to obtain an accelerated expansion in otherwise too steep potentials has been well explored in the context of primordial inflation. The original mechanism has been dubbed warm inflation \cite{Berera:1995ie} and for this reason we call our implementation ``warm dark energy''.
Warm inflation assumes that the particles produced by the inflaton are in a thermal state. This assumption is dropped in the more recent implementations of this idea, as for instance trapped inflation \cite{Green:2009ds}, or the mechanism of gauge field amplification by an axion inflaton \cite{Anber:2009ua}.

In this work, we explore this latter mechanism, developed by Anber and Sorbo in the context of primordial inflation  \cite{Anber:2009ua}, to the case of late time quintessence.
The gauge fields are coupled to an axion-like particle, which couples to the gauge fields by means of the topological term\footnote{Our study is restricted to abelian gauge fields, as in the Anber and Sorbo model \cite{Anber:2009ua}.
Refs.~\cite{Alexander:2016mrw,Alexander:2016nrg} studied the effects of coupling an axionic quintessence field to a massive non-Abelian SU(2) gauge field, that develops a spatial vev (analogously to \cite{Adshead:2012kp}, this allows for a slow roll evolution of the pseudo-scalar under certain conditions).
Different models of dissipative quintessence have been recently considered in refs.~\cite{Dimopoulos:2019gpz,Rosa:2019jci,Lima:2019yyv}.}:
\begin{equation}
S = \int d^4 x \sqrt{-g} \left[ - \frac{1}{2} \left( \partial \phi \right)^2 - V \left( \phi \right) - \frac{1}{4} F_{\mu \nu} F^{\mu \nu} - \frac{\phi}{4 f} F_{\mu \nu} {\tilde F}^{\mu \nu}\right] \;.
\label{action-phi-F-Ft}
\end{equation} 
The extra friction resulting from this interaction has also the effect of reducing the range of field values spanned by the inflaton, and can result in a sub-Planckian inflaton evolution\footnote{The topological interaction between an axion inflaton and gauge fields can have several other phenomenological consequences also in the regime of negligible backreaction, see for instance \cite{Pajer:2013fsa} for a review.}.

For instance, for a typical axion potential $V \propto \cos \left( \phi / f \right)$ (as in the so called model of natural inflation \cite{Freese:1990rb}), this coupling can result in an inflationary expansion also for an axion scale $f$ that is a few orders of magnitude smaller than $M_p$.
On the other hand, a trans-Planckian axion scale of natural inflation, beside being in tension with the latest CMB results \cite{Akrami:2018odb}, can hardly be reconciled with quantum gravity \cite{Banks:2003sx,Kim:2004rp,Silverstein:2008sg,Bachlechner:2014gfa,Rudelius:2015xta,Montero:2015ofa,Brown:2015iha,Bachlechner:2015qja,Heidenreich:2015wga,Kaplan:2015fuy,Hebecker:2015zss}.

Given the renewed interest in quintessence models because of the swampland programme, we revisit here axion models coupled to vector fields as a possible solution of the tension between swampland criteria and observations of the current phase of expansion of the Universe.
We therefore investigate potentials satisfying (\ref{dsconj}), with $c$ of order one, trying to make them compatible with observation thanks to the vector field couplings (\ref{action-phi-F-Ft}).

As in the inflationary application \cite{Anber:2009ua}, the last term in (\ref{action-phi-F-Ft}) results in a significant gauge field amplification.
This amplification occurs at the expense of the kinetic energy of the scalar field, and therefore it backreacts on the scalar field motion.
Anber and Sorbo provided an analytic approximation for the backreaction term, that is valid only after a sufficiently prolonged period of accelerated expansion, with nearly constant $\dot{\phi}$.
This is not the case at the onset of the present accelerated expansion.
For this reason, we resort to a numerical simulation of this effect.
Previous numerical studies either performed full lattice simulations \cite{Cuissa:2018oiw,Adshead:2015pva,Adshead:2019lbr,Agrawal:2018vin}, in which both the gauge and the scalar field are inhomogeneous, or adopted a discretization scheme \cite{Cheng:2015oqa,Cheng:2018yyr,Notari:2016npn,Ferreira:2017lnd}, in which the scalar field is taken to be homogeneous, while a discretized set of modes is evolved for the gauge field.
All these works focus on inflation or (p)reheating after inflation, with the exception of ref.~\cite{Agrawal:2018vin}, which studied the production of dark matter gauge bosons through this mechanism.
We also follow a discretization scheme as this second class of works.

In this work we are also interested in the string theory uplift of warm dark energy models.
The first step to embed them in string theory is to find an appropriate supergravity model that reproduces them at sufficiently low energies.
We therefore identified a class of N=2 theories that produce the action (\ref{action-phi-F-Ft}).
We focus on N=2 supergravity theories because they are a natural framework to describe the effective theories of Calabi--Yau compactifications.
In fact their truncations capture the closed string sector also in the presence of supersymmetric configurations of branes and orientifolds.
We will see that in a fairly general setup one can generalize the N=1 construction of \cite{DallAgata:2018ybl} to N=2 in a way that also clarifies the higher-dimensional origin of some of the ingredients.
Although we do not offer here a full string theoretic construction, we see that extended supersymmetry gives us a better control on the origin of crucial couplings in (\ref{action-phi-F-Ft}), like the inflaton-vector coupling.

The paper is structured as follows.
In Section \ref{sec:supergravity_realizations} we construct N=2 supergravities that contain the model (\ref{action-phi-F-Ft}).
In Section \ref{sec:dynamics} we study the background evolution in this model, and show that the gauge field production can sufficiently slow down the dark energy field, so to lead to a late time acceleration compatible with data.
While this part is mostly focused on the transition period between matter domination and dark energy domination, in the following Section \ref{sec:AS} we study the later dynamics inside the dark energy dominated phase.
We also study the onset of primordial inflation, choosing parameters as in the Anber--Sorbo paper, and considering whether their solution can be dynamically achieved from an earlier period of matter or of radiation domination.
In Section \ref{sec:conclusions} we then summarize and discuss our results.
The paper is concluded by two appendices, in which, respectively, we provide an analytic approximation for the amplified gauge field and we present the rescaled evolution equations used in our numerical integrations.

\section{Supergravity realizations} 
\label{sec:supergravity_realizations}

A generic N=1 supergravity model is defined in terms of a K\"ahler potential $K$, defining the scalar $\sigma$-model, a superpotential $W$, giving the scalar-self interactions and generating masses for the fields in the chiral multiplets, the gauge-kinetic functions $f_{\Lambda \Sigma}$, describing the non-minimal couplings in the vector kinetic terms, and the gauge group $G$.
We are interested in embedding our supergravity models of inflation/quintessence where the inflaton/quintessence field couples to a vector field.
We therefore need at least one chiral multiplet containing our scalar field and a vector multiplet, containing our gauge field.
The details of this minimal construction have been worked out in \cite{DallAgata:2018ybl} for a rather general choice of scalar potentials and we therefore use it as a starting point for our discussion.

In this section we want to go beyond the minimal construction, asking ourselves what are the N=2 supergravity models that can give rise to the required couplings between the inflaton/quintessence field and vector fields.
There are several reasons to look for such extension.

The first one is to see if and how extended supersymmetry constrains such scenarios.
More interestingly, we would like to better understand the string theory origin of models of this type.
While 4-dimensional supergravity models with minimal supersymmetry like those in \cite{DallAgata:2018ybl} are fairly easy to construct, their stringy origin is usually very difficult to assess.
In particular, quite often, fields that have an axionic coupling to the vector fields do not have canonical kinetic terms and therefore most of the studies on inflation/quintessence scenarios with vector field couplings should be re-evaluated.
On the other hand, generic string compactifications scenarios use Calabi--Yau manifolds as internal spaces, supplemented by various types of branes, fluxes and orientifolds.
This means that the closed string sector is described by an N=2 supergravity theory in 4 dimensions, which is then truncated to N=1 by the presence of orientifolds.
Analyzing N=2 supergravity models we can therefore address if the inflaton/quintessence field can come from the closed-string sector and in case of positive answer, identify the relevant compactification scenario.

To recover the N=1 models of \cite{DallAgata:2018ybl} as truncations of some N=2 supergravity, we have to imagine that our N=1 supermultiplets derive from either N=2 vector or hypermultiplets.
As a first step, in the following we focus on scenarios that involve only N=2 vector multiplets.
We therefore analyze N=2 supergravities that contain the gravity multiplet $(g_{\mu\nu},\psi_\mu^A,A_\mu)$, coupled to $n_V$ vector multiplets $(A_{\mu}^i, \lambda^i, z^i)$, $i=1,\ldots,n_V$.
Note that in N=2 supergravity the gravity multiplet, in addition to the graviton $g_{\mu\nu}$ and to the two gravitini $\psi_\mu^A$, $A=1,2$, contains also a vector field, the graviphoton, which generically mixes with the other vector fields.
This imposes several restrictions to the scalar self-couplings and to the couplings of the scalars to the vector fields.

The $\sigma$-model described by the scalars in the vector multiplets is a K\"ahler manifold, but of a restricted type.
In detail, its K\"ahler potential depends on the scalar fields only via projective coordinates $X^\Lambda(z)$.
The reason of this restriction is the fact that complex scalar fields $z^i$ appear in vector multiplets, but there are $n_V +1$ vectors because of the graviphoton and they all mix under duality transformations.
This means that the correct way to describe vectors is by means of a duality-covariant set of fields $A_\mu^\Lambda$, where $\Lambda = 0,1,\ldots, n_V$, so as to include the graviphoton.
At the same time, since duality transformations mix up all the vector fields, the scalar fields associated to them should transform accordingly, hence the need of new projective coordinates $X^\Lambda$, holomorphic functions of $n_V$ complex coordinates $z^i$.
Actually, duality transformations can mix vector fields with their duals $A_{\mu \Lambda}$ and therefore one has to introduce also ``dual'' holomorphic coordinates $F_{\Lambda}(z^i)$, which, in specific frames, can be deduced from a prepotential function $F(X)$, by taking its derivatives with respect to the projective coordinates $F_{\Lambda} = \frac{\partial F(X)}{\partial X^{\Lambda}}$.
The prepotential function and its derivatives (denoted by additional pedices involving capital greek indices) determine all the couplings of the vector multiplet fields.
The outcome is that the K\"ahler potential must have the following restricted form\footnote{We use natural units conventions such that $M_p=1$ for the whole discussion on the N=2 models.
We will introduce back the Planck mass when we revert to N=1 notation.} \cite{Ceresole:1995ca,Andrianopoli:1996cm}
\begin{equation}\label{N2kahler}
	K = - \log \left[i\left(\overline{X}^{\Lambda} F_{\Lambda} - X^{\Lambda} \overline{F}_{\Lambda}\right)\right].
\end{equation}
Once the number of vector multiplets has been fixed, any arbitrary holomorphic function $F(X)$, which scales quadratically $F \to \lambda^2 F$ when $X \to \lambda X$, is a legitimate prepotential from which one can derive the various couplings in the action.
However, for simplicity and because it comes as the natural outcome of Calabi--Yau compactifications, we will restrict the prepotential function to the case of very special K\"ahler manifolds, i.e.
those with prepotentials of the form
\begin{equation}\label{cubicprep}
	F(X) = d_{ijk} \frac{X^i X^j X^k}{X^0},
\end{equation}
where $d_{ijk}$ is a constant totally symmetric tensor that describes the various intersection numbers of the cycles on the Calabi--Yau manifold.
The K\"ahler potential in terms of the physical fields can be recovered by gauge-fixing the projective coordinates (for instance $X^0 = 1$, $X^i = z^i$):
\begin{equation}\label{Kdzzz}
	K=-\log\left[-i \, d_{ijk}(z^i-\bar{z}^{\bar \imath})(z^j-\bar{z}^{\bar \jmath})(z^k-\bar{z}^{\bar k})\right].
\end{equation}

As discussed previously, the prepotential fixes all the couplings of the vector multiplets and hence it also fixes the non-minimal gauge couplings to the scalar fields 
\begin{equation}
	e^{-1}{\cal L}_{kin} = {\cal I}_{\Lambda \Sigma}(z,\bar{z}) F_{\mu\nu}^\Lambda F^{\mu\nu \Sigma} + {\cal R}_{\Lambda \Sigma}(z,\bar z) \, F_{\mu\nu}^\Lambda \tilde{F}^{\mu\nu\Sigma},
\end{equation}
with $\tilde{F}^{\mu\nu} = \frac12\, \epsilon^{\mu\nu\rho\sigma} F_{\rho \sigma}$.
These couplings are dictated by the N=2 gauge-kinetic matrix ${\cal N} = {\cal R} + i {\cal I}$, with ${\cal I}$ negative-definite, whose explicit form can be deduced from the second derivatives of the prepotential $F_{\Lambda \Sigma} = \frac{\partial^2 F(X)}{\partial X^\Lambda \partial X^\Sigma}$.
The explicit expression is the following \cite{Andrianopoli:1996cm}:
\begin{equation}\label{gaugekin2}
	{\cal N}_{\Lambda\Sigma} = \bar{F}_{\Lambda\Sigma}+ \frac{(F_{\Lambda\Gamma}-\bar{F}_{\Lambda\Gamma})(F_{\Sigma\Delta} - \bar{F}_{\Sigma\Delta})X^\Gamma X^\Delta}{\left(F_{\Gamma\Delta}-\bar{F}_{\Gamma\Delta}\right)X^\Gamma X^\Delta},
\end{equation}
where repeated indices are summed over.

In order to identify the relevant N=2 models, we need to consider which fields survive the truncation process to N=1.
For instance, the truncation to N=1 will get rid of the graviphoton $A^0_\mu$ (because the N=1 gravity multiplet does not contain any vector field) and therefore one needs at least two additional vector multiplets in the model, one from which we recover a chiral multiplet containing the inflaton/quintessence field and one from which we recover the surviving gauge field.
Luckily, the rules for N=2 supergravity truncations to N=1 have been worked out in detail in \cite{Andrianopoli:2001gm}.
We therefore refer to that paper for the derivation of such rules, while we simply apply them in the following.

As mentioned above, we focus our analysis to the N=2 models with a cubic prepotential of the form (\ref{cubicprep}).
These depend on a totally symmetric tensor $d_{ijk}$, whose possible form has been studied and classified in a series of papers by Van Proeyen, de Wit and collaborators \cite{deWit:1990na,deWit:1991nm,deWit:1992wf}.
In particular one can check the form of the gauge-kinetic function (\ref{gaugekin2}) and of the K\"ahler potential (\ref{Kdzzz}) and see which models have fields with canonical kinetic terms coupling linearly to the vector fields through the matrix ${\cal R}$.
After a straightfoward analysis one can check that the simplest model one can construct that leads to the wanted truncation requires $n_V = 3$, which implies that one has an extra chiral multiplet in the resulting N=1 truncation, whose scalar is going to play the role of a stabilizer field.
In detail, one can use a homogeneous manifold with 
\begin{equation}\label{ddd}
	d_{122} = - d_{133} = 1/3 \;, 
\end{equation}
and keep in the truncation the scalar fields $z^1$ and $z^3$, while truncating $z^2 = 0 = A_\mu^0 = A_\mu^1 = A_\mu^3$.
We verified that the rules for a consistent truncation given in \cite{Andrianopoli:2001gm} are satisfied.
In fact, for $X^2(z) = z^2 = 0$, the K\"ahler covariant derivatives of the holomorphic sections with respect to the truncated scalar identically vanish
\begin{equation}
	 \quad [D_2 X^0]_{z_2 = 0} = [D_2 X^1]_{z_2 = 0} = [D_2 X^3]_{z_2 = 0} = 0.
\end{equation}
The scalar $\sigma$-model metric factorizes and the terms involving $z^2$ also identically vanish
\begin{equation}
	g_{1\bar{2}}|_{z_2 = 0} = g_{1\bar{3}}|_{z_2 = 0} = 0.
\end{equation}
Finally, the gauge kinetic functions that mix the vector fields we truncated with the surviving one also vanish
\begin{equation}
	{\cal I}_{20}|_{z_2 = 0} = {\cal I}_{21}|_{z_2 = 0} = {\cal I}_{23}|_{z_2 = 0} = 0.
\end{equation}
These conditions fully decouple the residual vector multiplet from the chiral multiplets in the N=1 supersymmetry rules.

The result of the truncation gives a N=1 model with a gauge-kinetic coupling that is linear in the surviving chiral multiplet scalar field as in \cite{DallAgata:2018ybl}
\begin{equation}\label{gaugecoupling}
	f_{22} = 2\,\overline{{\cal N}}_{22} = 4\,z^1.
\end{equation}
As mentioned above, we actually want more, because we need it to be linear in the quintessence scalar field, which should be canonically normalized.
In order to see this, we split the $z^1$ field into its real and imaginary part, while we also introduce for later convenience a reparameterization of the other scalar field in terms of the stabilizer $S$:
\begin{equation}\label{scalars}
	z^1 = \Phi + i = \phi + i\, (1+ \alpha), \qquad z^3 = i \frac{1+S}{1-S}.
\end{equation}
We will shortly see that the field $\phi$ has a canonical kinetic term if $\alpha = 0$ and it couples linearly to the topological term of the surviving vector field once (\ref{gaugecoupling}) is taken into account.

The peculiar parameterization of the N=2 scalars $z^i$ in terms of the N=1 fields as in (\ref{scalars}) has been chosen in a way that allows to obtain a model such that $S$ acts as a stabilizer field to set both $\alpha = S = 0$ consistently in the equations of motion.
This can be achieved as follows.
The K\"ahler potential deriving from truncating (\ref{N2kahler}) is
\begin{equation}
	\begin{split}
	K = &- M_p^2\, \log \left[\frac12\left(1-i \frac{\Phi - \bar{\Phi}}{2 M_p}\right)\right] - 2M_p^2\,\log\left[2\left(1-\frac{|S|^2}{M_p^2}\right)\right] \\[2mm]
	&+ 2 \log\left[\left(1-\frac{S}{M_p}\right)\left(1-\frac{\bar{S}}{M_p}\right)\right].
	\end{split}
\end{equation}
The last term can be removed by a K\"ahler transformation $K \to K + h + \bar{h}$, where $h = -2 \log\left(1-\frac{S}{M_p}\right)$.
In the new frame we can then fix the K\"ahler potential to 
\begin{equation}\label{Kahler1}
	K = - M_p^2\, \log \left[\frac12\left(1-i \frac{\Phi - \bar{\Phi}}{2 M_p}\right)\right] - 2M_p^2\,\log\left[2\left(1-\frac{|S|^2}{M_p^2}\right)\right].
\end{equation}
The scalar potential follows from the superpotential in the usual way
\begin{equation}\label{V1}
	V = e^{K/M_p^2} \left(|DW|^2 - 3 \frac{|W|^2}{M_p^2}\right).
\end{equation}
In order to achieve a consistent truncation to $\alpha = S = 0$, we can follow a procedure similar to the one outlined in \cite{Kallosh:2010xz} and propose a superpotential that has a linear dependence on $S$, so that 
\begin{equation}\label{supo}
	W = \Lambda^2 S \;{\cal F}\left(\frac{\Phi}{M_p}\right),
\end{equation}
where ${\cal F}$ is an arbitrary holomorphic function of $\Phi$, which we subject to conditions coming from the equations of motion and $\Lambda$ is a mass-dimension 1 parameter.
The scalar field sector of our theory is described by the $\sigma$-model metric 
\begin{equation}
	e^{-1} {\cal L}_{scal} = -\left(2 - i\, \frac{\Phi - \bar{\Phi}}{2 M_p}\right)^{-2} \partial_\mu \Phi \partial^\mu \bar{\Phi} - 2 \left(1 - \frac{|S|^2}{M_p^{2}}\right)^{-2} \partial_\mu S \partial_\mu \bar{S},
\end{equation}
with scalar potential $V$ following from inserting (\ref{Kahler1}) and (\ref{supo}) in (\ref{V1}).
The special choice of superpotential guarantees that $S=0$ is a critical point and that
\begin{equation}
	\partial_{\alpha} V|_{S=0=\alpha} = 0 \quad \Leftrightarrow \quad -|{\cal F}|^2 + i\, {\cal F}' \overline{\cal F} -i\, {\cal F} \overline{\cal F}' = 0,
\end{equation}
where prime means derivative of the argument.
This can be further simplified positing
\begin{equation}
	{\cal F} = e^{-\frac{i}{2}\frac{\Phi}{M_p}} \; u\left(\Phi/f\right),
\end{equation}
where $u$ is a holomorphic function of $\Phi$ and $f$ is a mass-dimension 1 parameter.
The critical point at $\alpha=0$ is assured if
\begin{equation}
	u' \bar{u} - u \bar{u}' = 0
\end{equation}
and the truncated potential at $S=\alpha=0$ is
\begin{equation}
			V|_{S=0=\alpha} = \Lambda^4 \, \left|u\left(\frac{\phi}{f}\right)\right|^2,
\end{equation}
which can be used to embed almost any potential in our construction.
For instance, the potential 
\begin{equation}
	V = \Lambda^4 \left(1 - \cos \frac{\phi}{f}\right)
\end{equation}
follows from the choice $u = \sin \frac{\Phi}{2f}$.

While having a consistent truncation, it is not obvious that setting $S=\alpha=0$ gives a good effective theory.
Although we do not have a general behaviour of the masses as functions of ${\cal F}$ that guarantees us that our consistent truncation is also a good effective theory, we can work out the specifics of the models of interest to see that it becomes true with some tuning.
For the cosine potential, for $f = 10^{-4} M_p$, at $\phi = \pi$ we have that $\phi$ is tachyonic, $\alpha$ is clearly stabilized, but $S$ is tachyonic, too.
\begin{figure}[ht]\label{fig:masses}
 \centering
 \includegraphics[width=0.5\textwidth]{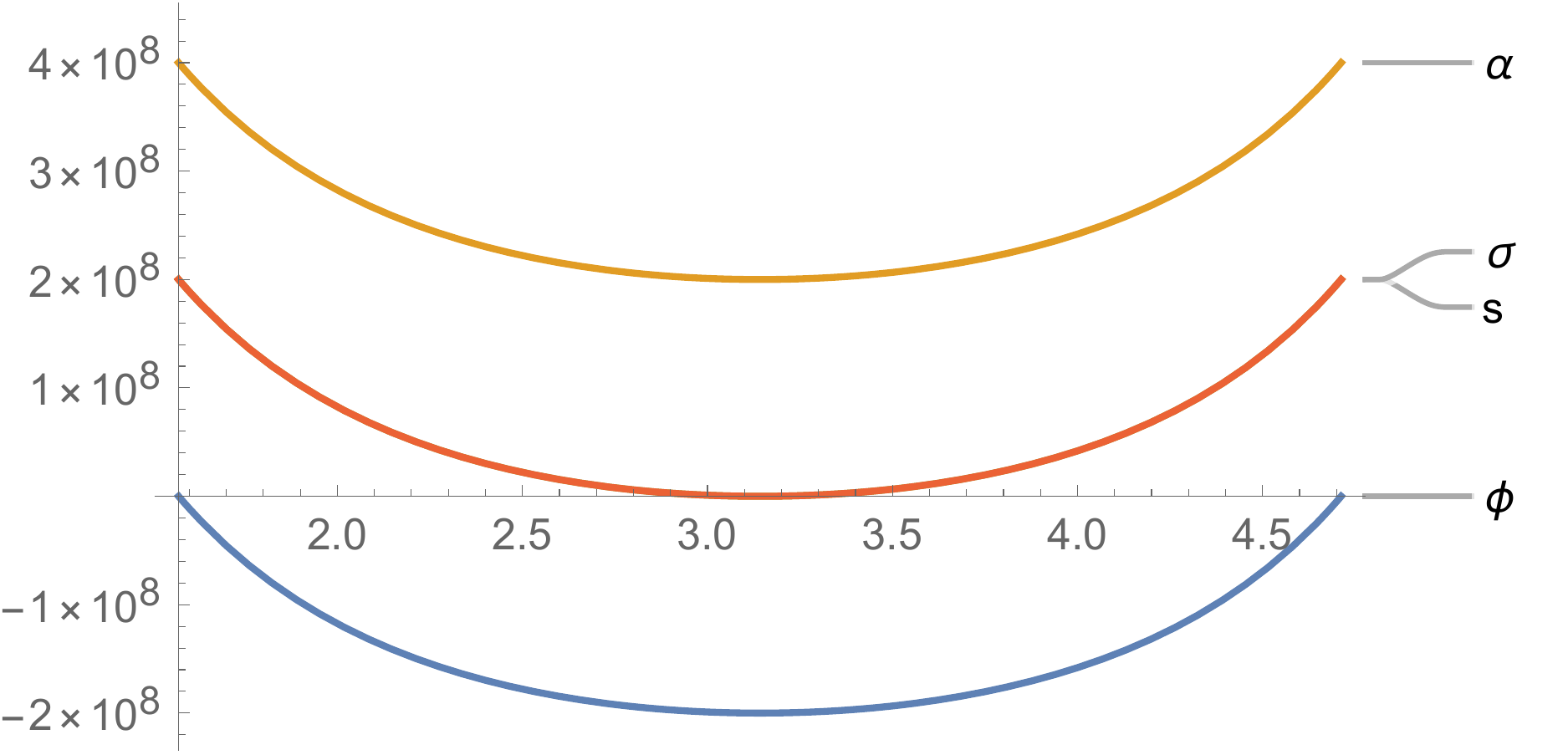}
 \caption{Normalized masses of the fields for the cosine potential}
 \end{figure}
This is usually corrected by higher-order corrections in the K\"ahler potential that generate an additional term of order $|S|^4/M^4$ inside the second log, giving $S$ a mass of order $M$ (see for instance \cite{Lee:2010hj,Ferrara:2010yw}).
However, as one can see from Figure~\ref{fig:masses}, the ratio of normalized masses is such that the scalar fields will immediately roll in the $\phi$ direction, while $S$ develops a huge mass.
In fact all the plots in Figure~\ref{fig:masses} are in terms of the dimensionless parameter 
\begin{equation}\label{massratio}
	M_p^2 \frac{V''}{V}
\end{equation} 
and the scale of the plot shows that very rapidly $S$ acquires a huge mass.
When the field $\phi$ reaches $\pi/2$ it becomes massless.
At that point the other fields are much heavier.

The situation is a bit more complicated for the exponential potential 
\begin{equation}
	V = \Lambda^4 e^{-\frac{\phi}{f}},
\end{equation}
but it can also be solved with some additional tuning.
In this case it is indeed difficult to achieve a consistent truncation and at the same time preserving a hierarchy in the masses, because of the nature of the exponential potential.
If one naively sets ${\cal F} = e^{-\frac{i \, w}{2}\frac{\Phi}{M_p}}$, the fields $\alpha$ and $\phi$ acquire masses of the same order.
However, if one posits
\begin{equation}
	{\cal F} = e^{-\frac{i \, w}{2}\frac{\Phi}{g}}
\end{equation}
with $w$ developing a small imaginary part $w = 1 - i \frac{g}{f}$, with $g \lesssim f/10$, then $S$ and $\alpha$ have the mass ratio (\ref{massratio}) of order $10^{10}$, while $\phi$ is almost massless.
In this case we do not have a consistent truncation at $\alpha = 0$, but the fields are stabilized by the huge masses they acquired.

Since we discussed K\"ahler potentials of restricted form because of their origin from string theory compactifications on Calabi--Yau manifolds, we can ask ourselves if the models we just presented can actually be obtained in such string theory reductions.
Effective theories for Calabi--Yau compactifications with branes and orientifolds have been discussed in detail in \cite{Grimm:2004uq,Jockers:2004yj,Grimm:2004ua}.
While it is not difficult to check that there are Calabi--Yau manifolds with the right geometric structure to give rise to the prepotential (\ref{cubicprep}) with (\ref{ddd}), one should be careful about the natural units of the coupling between the inflaton/quintessence scalar $\phi$ and the vector fields.
Although in a dual setup, the analysis of \cite{Grimm:2007hs} shows that such couplings are always of order $1/(2\pi)^2$ for all gauge fields in the closed string sector and can be enhanced by a factor N when considering gauge fields on $N$ branes, provided we identify them.
One therefore would need a rather large number of branes to make the couplings work, but this in turn would require to take into account the backreaction due to their presence, thus questioning the supergravity approximation.
We therefore conclude that the full uplift of these models is still an open question, though we provided here some of the necessary ingredients, at least in the low-energy effective theory.


\section{Dynamical evolution and accelerated expansion from dissipation} 
\label{sec:dynamics} 

Let us now discuss the cosmological consequences of the models studied in the previous Section.
We consider only the bosonic part of the action, including only the dynamically relevant fields.
In this view, the study presented in this and in the next section can be applied also beyond the supergravity constructions that we have studied so far, and it relies on the starting action (\ref{action-phi-F-Ft}).
We want to study whether this model can account for the present accelerated expansion of the Universe, for the cases in which the potential of a scalar field $\phi$ in this model respects the condition in (\ref{dsconj}), and it is therefore too steep to lead to accelerated expansion, in absence of particle production.
We show that a typical axionic coupling between the scalar field in this model $\phi$ and a U(1) gauge field can allow for sufficient dissipation, that slows down the scalar field to a sufficiently level so to behave as an effective (slowly varying) cosmological constant.

This section is divided in five parts.
In Subsection \ref{subsec:gauge} we present the Anber--Sorbo mechanism for gauge field amplification and backreaction.
In Subsection \ref{subsec:example} we present and discuss the concrete potential that we study in this work.
The following three subsections are devoted to study the dynamics of this model, from the early matter dominated phase to the onset of the late time acceleration.

\subsection{Gauge field amplification from a rolling axion, and backreaction} 
\label{subsec:gauge} 

We consider the action 
\begin{equation}
S = \int d^4 x \sqrt{-g} \left[ - \frac{1}{2} \left( \partial \phi \right)^2 - V \left( \phi \right) - \frac{1}{4} F_{\mu \nu} F^{\mu \nu} - \frac{\phi}{4 f} F_{\mu \nu} {\tilde F}^{\mu \nu} \right] \;, 
\end{equation} 
where $\phi$ is an axionic field that enjoys a shift symmetry $\phi \rightarrow \phi + C$, which is respected by its interaction with a U(1) gauge field, and which is broken by the potential term $V$.

In this action, $F_{\mu \nu}$ is the gauge field strength, and ${\tilde F}^{\mu \nu} \equiv \frac{\epsilon^{\mu \nu \alpha \beta}}{2\sqrt{-g}} \, F_{\alpha \beta}$ is its dual (where the tensor $\epsilon^{\mu \nu \alpha \beta}$ is totally anti-symmetric, and it is normalized to $\epsilon^{0123} = 1$).
We assume a flat isotropic and homogeneous (FLRW) geometry, with line element $d s^2 = a \left( \tau \right) \left( - d \tau + \delta_{ij} \, d x^i d x^j \right)$, where $a$ is the so called scale factor of the Universe, and $\tau$ is conformal time.
We assume that $\phi$ is a cosmologically relevant field.
 Therefore, to respect the FLRW geometry, $\phi$ only depends on time, $\phi = \phi \left( \tau \right)$ up to small fluctuations, that we disregard in this work.
The motion of $\phi$ then modifies the dispersion relations for the gauge field: the gauge field mode function obeys the equation 
\begin{equation}
\left( \partial_\tau^2 + k^2 \mp 2 a H k \xi \right) A_\pm \left( \tau ,\, k \right) \;\;,\;\; \xi \equiv \frac{\partial_\tau \phi}{2 a f H} \;\;, 
\label{A-eom}
\end{equation} 
where $A_+$ and $A_-$ are, respectively, the right-handed and the left-handed gauge field polarizations, and $k$ is the comoving momentum of the gauge field mode.
Finally, $H \equiv \frac{\partial_\tau a}{a^2}$ denotes the Hubble rate.

As seen from eq.~(\ref{A-eom}), due to the motion $\phi \left( t \right)$ one gauge field polarization becomes tachyonic for momenta smaller than the threshold momentum 
\begin{equation}
k < 2 \, a \left( \tau \right) H \left( \tau \right) \xi \left( \tau \right) = 
\frac{\partial_\tau \phi}{f} \equiv k_{\rm thr} \left( \tau \right) \;, 
\end{equation} 
where we have assumed that $\partial_\tau \phi> 0$, so that the tachyonic mode is\footnote{For $\partial_\tau \phi < 0$ the two polarizations interchange their role.}~$A_+$. 
For notational convenience, from now on we disregard the $A_-$ mode, and we relabel $A_+$ as $A$.
For sufficiently large coupling (sufficiently small $f$) this leads to a strong gauge field amplification, which in turn can backreact on the background dynamics.
The produced gauge field modifies the evolution equation for $\phi$ (following from the extremization of (\ref{action-phi-F-Ft}) with respect to $\phi$) through the typical axionic coupling \cite{Anber:2009ua}
\begin{equation} 
\frac{\partial^2 \phi}{\partial \tau^2} + \frac{2}{a} \, \frac{\partial a}{\partial \tau} \, \frac{\partial \phi}{\partial \tau} + a^2 
\, \frac{\partial V}{\partial \phi} = \frac{a^2}{f} \, \vec{E} \cdot \vec{B} \;\;.
\label{eom-phi}
\end{equation} 
It also modifies the Friedman equation (namely, the $00$ component of the Einstein equation) for the evolution of the scale factor through its energy density \cite{Anber:2009ua}
\begin{equation} 
\left( \frac{1}{a} \, \frac{\partial a}{\partial \tau} \right)^2 = \frac{a^2}{3 M_p^2} \left[ \frac{1}{2 a^2} \left( \frac{\partial \phi}{\partial \tau} \right)^2 + V + \frac{\rho_{m,{\rm in}} \, a_{\rm in}^3}{a^3} + \frac{E^2+B^2}{2} \right] \, , 
\label{eom-00}
\end{equation} 
where we have also included the contribution of the energy density of matter, which scales as $\rho \propto a^{-3}$, and the suffix `in' refers to a moment in the matter dominated era.
In these expressions, we have used the standard electromagnetic notation, with 
\begin{equation} 
\vec{E} \cdot \vec{B} = -\frac{1}{4 \pi^2 a^4} \int d k \, k^3 \, \frac{\partial}{\partial \tau} \left\vert A \right\vert^2\,, \qquad
 \frac{E^2+B^2}{2} = \frac{1}{4 \pi^2 a^4} \int d k \, k^2 \, \left[ \left\vert \frac{\partial A}{\partial \tau} \right\vert^2 
 + k^2 \, \left\vert A \right\vert^2 \right] \;, 
 \end{equation} 
simply for notational convenience, without necessarily requiring that the U(1) field is the Standard Model photon (or the hypercharge gauge boson).
As we show in the next subsection below in a concrete example, the main backreaction effect is on the dynamics of $\phi$, via eq.~(\ref{eom-phi}).
The backreaction term is controlled by the parameter $\xi$, which is in turn related to the speed of $\phi$ (for a constant $\phi$, the last term in (\ref{action-phi-F-Ft}) is a topological term, which does not contribute to the equations of motion).
This generates a friction term analogous to the second term in eq.~(\ref{eom-phi}) (the so called ``Hubble friction'').
As we show below, this additional friction can slow down the motion of $\phi$ and allow for accelerated expansion, in a potential that would be too steep to lead to acceleration in absence of this effect.

\subsection{A concrete example} 
\label{subsec:example} 

Let us now study how the mechanism presented above works for a specific example.
For definiteness, we consider an exponential potential 
\begin{equation}
V = V_0 \, {\rm e}^{-\frac{\lambda \, \phi}{M_p}} \;.
\label{potential}
\end{equation} 
We consider a sufficiently early time, when the energy density is dominated by that of matter, and the scalar $\phi$ is nearly frozen, due to Hubble friction, with a negligible energy.
Eventually, the matter energy density is diluted by the expansion of the Universe, and the scalar field becomes dynamically relevant.
For $\lambda < \sqrt{3}$, the scalar field eventually comes to dominate over matter, and acquires an equation of state (pressure over energy density) \cite{Ferreira:1997hj} 
\begin{equation}
w_\phi \equiv \frac{p_\phi}{\rho_{\phi}} = \frac{\lambda^2}{3} - 1 \;.
\label{w-late}
\end{equation} 
This value is obtained only after the scalar field comes to dominate.
Moreover, we are for the moment disregarding the coupling between $\phi$ and the gauge field.

Accelerated expansion requires $w < -\frac{1}{3}$.
This is however not enough to agree with the data.
For this potential, comparison with data provides an upper bound on $\lambda$.
Marginalizing over the scale of the potential, one obtains $\lambda \la 0.49,\, 0.80 ,\, {\rm and} \, 1.02$ at $68\%$, $95\%$, and $99.7\%$, respectively \cite{Akrami:2018ylq}.
This is incompatible with the generic expectation coming from string theory, where $\lambda \gtrsim \sqrt6$ \cite{Obied:2018sgi,Bedroya:2019snp}.
In our numerical study, we fix $\lambda =1$, leading to $w_\phi = -2/3$ when the scalar field dominates.
This is compatible with the data only at $3 \sigma$, and, we therefore study whether the additional friction from this production can allow for a more phenomenologically acceptable expansion law.

As can be verified a-posteriori, due to the friction from the Hubble expansion and the gauge field amplification, the evolution of the scalar field only spans a small region of the potential, and we could perform the present study for a linearization $V = \alpha + \beta \phi$ of a generic potential.
However, the current presentation is focused on the potential (\ref{potential}) for definiteness.
We solve the background evolution equations for the scalar field, for the scale factor, and a set of $N$ equations (\ref{A-eom}) for a grid of $N$ different modes of the gauge field (each one characterized by a distinct comoving momentum $k$).
We solve these equations in three ways, progressively including the various physical contributions.
Firstly, we disregard the backreaction of the produced gauge fields, and we include only the energy density of matter in the evolution for the scale factor.
This allows to obtain an analytic solution for the scalar field and scale factor evolution, that is accurate only at very early times, when the $\left\{ \phi ,\, \vec{A} \right\}$ sector is completely subdominant\footnote{We do not include any radiation term, which would be dominant at even earlier times, since our interest is on the onset of the current accelerated stage of the Universe.}. 
This allows us to obtain the initial conditions for the numerical evolutions that we consider next.
Secondly, we include the energy density of the scalar field in the evolution equations, but we still disregard the backreaction of the gauge fields.
We now solve also the equations for the gauge fields in this approximation, and we evaluate the backreaction terms as a diagnostic (we compute it, but we do not include it in the evolution equation).
This allows us to find the moment at which the approximation breaks down and, more importantly, the range of momenta of gauge field modes that dominate the backreaction term at this moment.
Thirdly, we evolve the full set of equations, making sure that the sampled range of momenta for the gauge fields covers the dynamically relevant regions, and that the set of probed modes is sufficiently dense.

We discuss these three solutions in the next three subsections.

\subsection{Analytic solutions for $\rho_m \gg V$ and negligible backreaction} 
\label{subsec:solutions1}

We consider an initial time $\tau_{\rm in}$ deep inside the matter dominated regime.
With no loss of generality we can always set 
\begin{equation}
a_{\rm in} = 1\;,\qquad \phi_{\rm in} = 0 \;, 
\label{initial-a-phi}
\end{equation}
at this initial time.
The first condition is allowed by the fact that the normalization of the scale factor is arbitrary (and unphysical) in a flat Universe.
The second condition is allowed by the fact that any constant initial value for the scalar field can be reabsorbed in the value of $V_0$ in the exponential potential (\ref{potential}).
At these early times, the motion of the scalar field is extremely small due to Hubble friction, and $\phi$ essentially contributes to the expansion of the Universe as a cosmological constant term $V_0$.
Therefore, the ratio between the energy density of matter and $V_0$, 
\begin{equation}
{\bar \rho}_m \equiv \frac{\rho_{m,{\rm in}}}{V_0} \;, 
\end{equation}
is the most immediate physical quantity that can be used to parametrize the initial time.
Keeping only the energy density of matter at the r.h.s.
of eq.~(\ref{eom-00}), one obtains the solution 
\begin{equation}
a = \left( \frac{\tau}{\tau_{\rm in}} \right)^2 \;, \qquad \tau_{\rm in} = \frac{2 \sqrt{3}M_p}{V_0^{1/2}} \, \frac{1}{{\bar \rho_m}^{1/2}} \;, 
\label{a-early}
\end{equation} 
where an integration constant has been fixed to set the first of (\ref{initial-a-phi}).
To obtain the early time solution for the scalar field, we 
take equation (\ref{eom-phi}), we insert in it the solution (\ref{a-early}), we disregard the backreaction term, and we approximate the potential derivative as $\frac{\partial V}{\partial \phi} = - \frac{\lambda \, V_0}{M_p} \, {\rm e}^{-\frac{\lambda \phi}{M_p}} \simeq - \frac{\lambda \, V_0}{M_p}$ as it is appropriate at the early times, when $\phi \ll M_p$.
The resulting equation is then solved by 
\begin{equation} 
\phi = \frac{2}{9} \, \frac{\lambda}{\bar \rho_m} \, M_p \, \left[ \left( \frac{\tau}{\tau_{\rm in}} \right)^6 - 1 \right] \;, 
\label{phi-early}
\end{equation} 
where one integration constant has been used to remove a decreasing mode (which, if included at times $\ll \tau_{\rm in}$, would have become negligible at $\tau_{\rm in}$), and another one to set the second of (\ref{initial-a-phi}).

\subsection{Solutions with negligible backreaction} 
\label{subsec:solutions2}

We now improve over the solutions presented in the previous subsection by consistently solving eqs.
(\ref{eom-phi}) and (\ref{eom-00}) in absence of the backreaction terms (these are therefore exact solutions in the $f \rightarrow \infty$ limit).
 The parameters $V_0$, $\rho_{m,{\rm in}}$ and $M_p$ can be then removed from the evolved system if we rescale $\phi = M_p \, \varphi$ and if we use the dimensionless combination 
\begin{equation}
{\tilde \tau} \equiv \frac{{\bar \rho}_m^{1/3} \, V_0^{1/2}}{\sqrt{12} \, M_p} \, \tau \;, 
\label{tilde-tau}
\end{equation} 
as time variable, the evolution equations become: 
\begin{equation}
\left\{ \begin{array}{l} \displaystyle
\frac{d^2 \varphi}{d {\tilde \tau}^2} + \frac{2}{a} \, \frac{\partial a}{\partial {\tilde \tau}} \, \frac{\partial \varphi}{\partial {\tilde \tau}} - \frac{12}{{\bar \rho}_m^{\,2/3}} \, a^2 \, \lambda \, {\rm e}^{-\lambda \varphi} = 0 \\ \\ \displaystyle
\left( \frac{1}{a} \, \frac{\partial a}{\partial {\tilde \tau}} \right)^2 = \frac{1}{6} \, \left(  \frac{\partial \varphi}{\partial {\tilde \tau}} \right)^2 + \frac{4}{{\bar \rho}_m^{\,2/3}} \, a^2 \, {\rm e}^{-\lambda \varphi} + 4 \frac{\bar \rho_m^{\,1/3}}{a} 
\end{array} \right.
\;.
\label{system2}
\end{equation}

The initial time (\ref{a-early}) corresponds to the rescaled initial time ${\tilde \tau}_{\rm in} = {\bar \rho}_m^{-1/6} $.
To estimate when the transition between matter domination and scalar field domination occurs, we equate the last two terms in the second of (\ref{system2}), corresponding, respectively, to the potential energy of the scalar field and to the energy density of matter.
As we are interested in a parametric estimate, we set $\varphi =0$ in this comparison, and obtain 
\begin{equation}
a_{\rm transition} \simeq {\bar \rho}_m^{1/3} \;\;\; \Rightarrow \;\;\; {\tilde \tau}_{\rm transition} \simeq 1 \;, 
\label{tau-transition}
\end{equation} 
where the early time analytic solution (\ref{a-early}) has been used in the second estimate.
This final estimate justifies the rescaling done in (\ref{tilde-tau}).
Namely, we chose a dimensionless time variable that evaluates to $\simeq 1$ at the equality between matter and the scalar field.

\begin{figure}[ht!]
\centerline{
\includegraphics[width=0.32\textwidth,angle=0]{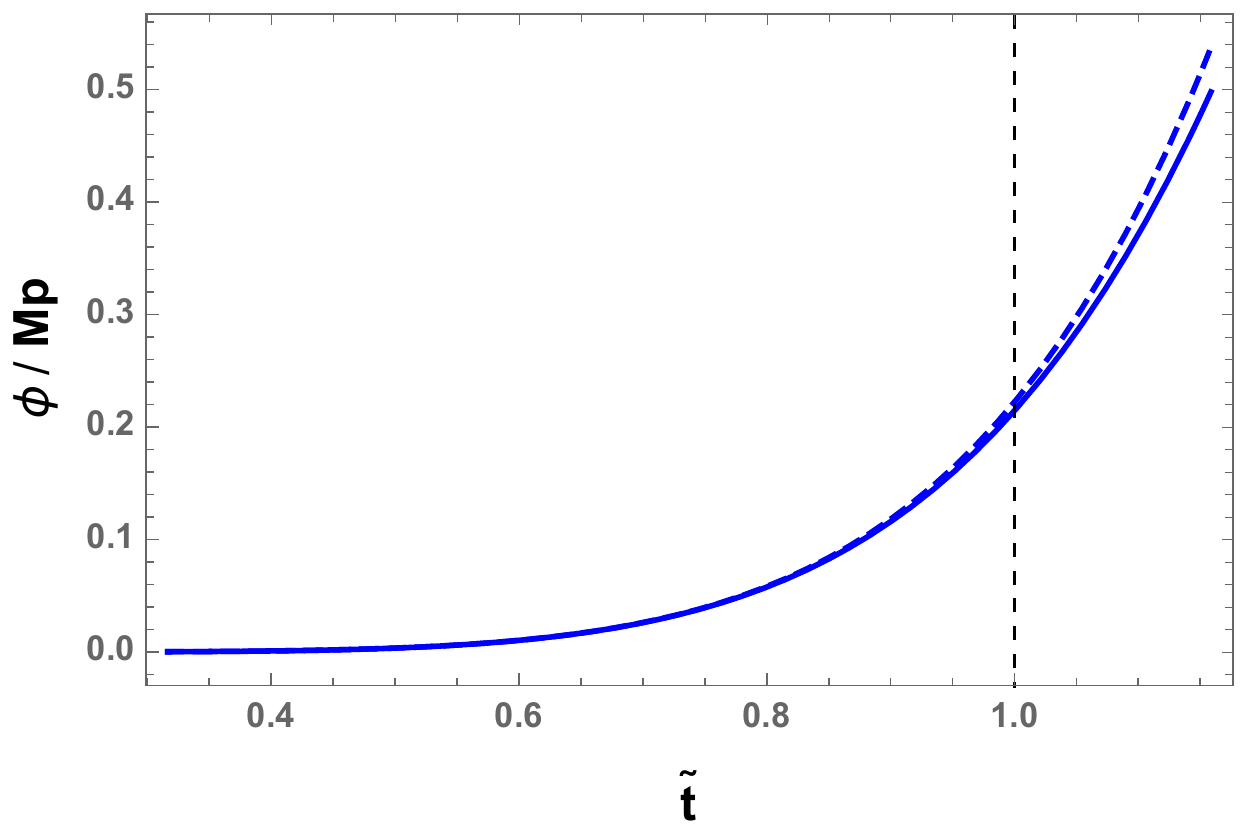}
\includegraphics[width=0.32\textwidth,angle=0]{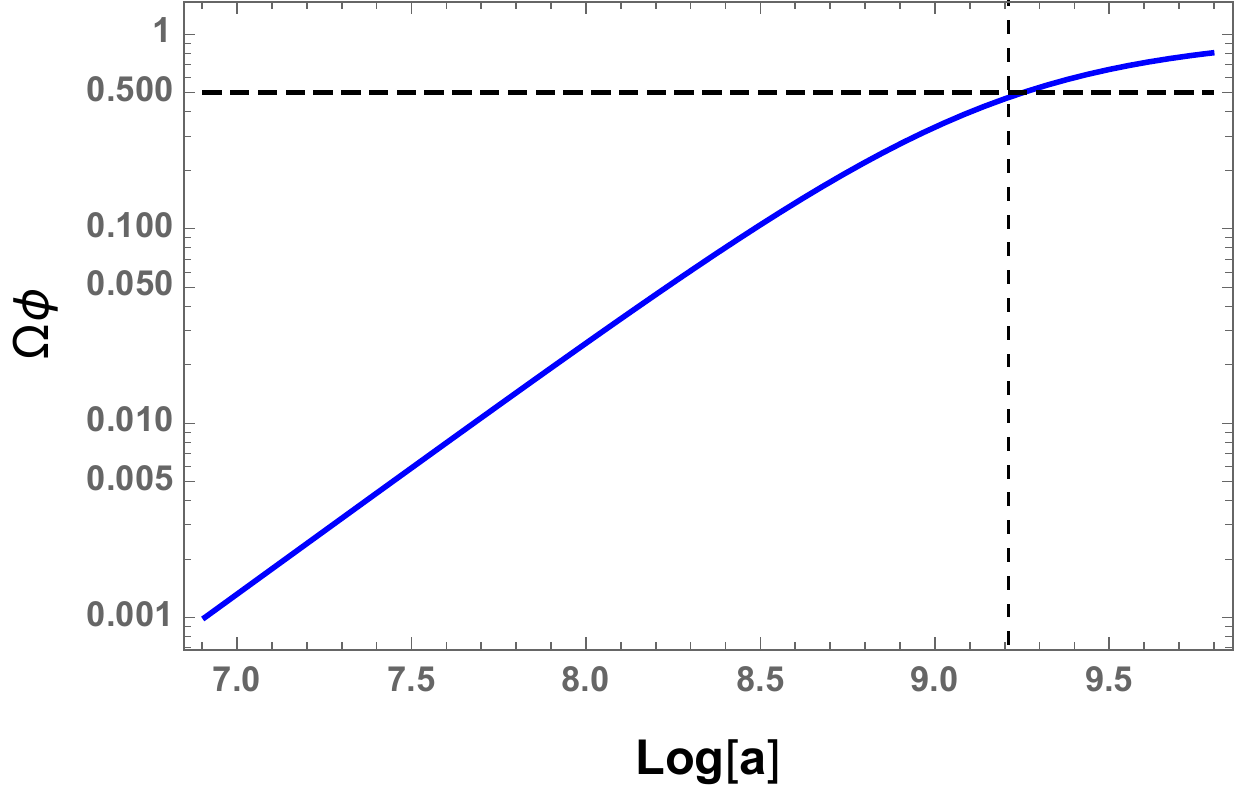}
\includegraphics[width=0.32\textwidth,angle=0]{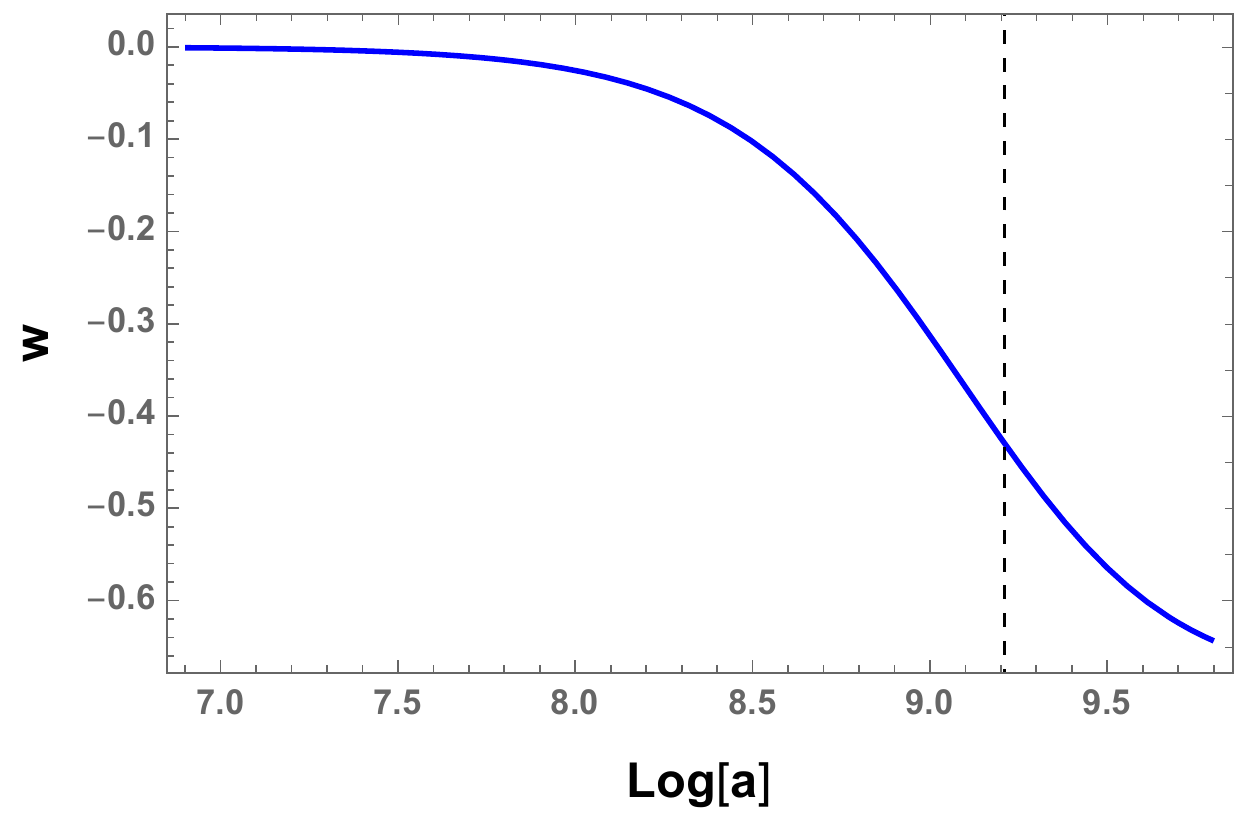}
}
\caption{\textit{Left panel}: evolution of the scalar field from the system (\ref{system2}) (solid line) compared with the early time analytic solution (\ref{phi-early}) (dashed line).
\textit{Central panel}: fractional density of the scalar field.
\textit{Right panel}: total equation of state (\ref{eos}).
All panels show a portion of the numerical evolution, from the moment in which $\rho_\phi = 10^{-3} \, \rho_m$ to the moment in which $\rho_\phi = 4 \rho_m$ (as a comparison, the ratio between the present energy density of the dark energy and of matter is about 2.2 \cite{Aghanim:2018eyx}).
In all panels, the dashed vertical line denotes our estimate (discussed in the text) at which $\rho_\phi = \rho_m$ (as seen from the central panel, this estimate is very accurate).
}
\label{fig:system2}
\end{figure}

We solved the system of equations (\ref{system2}) numerically, fixing ${\bar \rho}_m = 10^{12}$ and, as we wrote above, $\lambda = 1$.
The conditions (\ref{initial-a-phi}) provide the initial conditions for this evolution.
In the left panel of Figure \ref{fig:system2} we show the evolution of the scalar field.
We see that $\phi \ll M_p$ in the dynamical range of interest, justifying the statement that we could have obtained analogous results by simply considering a linearization $V = \alpha + \beta \phi$ rather than the exponential potential (the backreaction due to the gauge field production that we include in the next subsection further decreases the range spanned by $\phi$).
In the central panel of the figure we show the evolution of the fractional energy of the scalar field 
\begin{equation}
\Omega_\phi = \frac{\rho_\phi}{\rho_\phi + \rho_m} \;.
\end{equation} 
Finally, in the right panel of the figure we show the evolution of the total equation of state 
\begin{equation}
w_{\rm tot} = \frac{p_{\rm tot}}{\rho_{\rm tot}} = \frac{1}{3} - \frac{2 \, a}{3} \, \frac{\partial^2 a}{\partial \tau^2} \, \Big/ \left( \frac{\partial a}{\partial \tau} \right)^2 \;.
\label{eos}
\end{equation} 
We recall that $w < - \frac{1}{3}$ leads to accelerated expansion, and we note that the equation of state indeed starts from the early matter dominated value $w=0$ and it then evolves to the value $-2/3$ at late times, in agreement with eq.~(\ref{w-late}).
In the central and in the right panel, the evolutions as shown are a function of the number of e-folds of expansion, defined as 
\begin{equation}
N \equiv \ln a \;, \qquad N_{\rm in} = 0 \;.
\label{efolds}
\end{equation} 

\subsubsection{Growth of the backreaction term}
\label{subsubsec:backgrowth}

We study here the backreaction effects of the gauge field production, encoded by the last two terms in eqs.
(\ref{eom-phi}) and (\ref{eom-00}).
We evaluate them using the numerical solutions for $\phi$ and $a$ derived in absence of this backreaction.
Therefore, the solution for the backreaction terms that we obtain is valid only as long as it is negligible.
This computation allows us to study the growth of the backreaction terms, and to find the value of the axion decay constant $f$ for which the backreaction becomes important close to the transition point between the matter field and the scalar field domination.
Moreover, it also allows us to understand what is the relevant range for the gauge field modes that we need to cover in the full numerical evolutions that we study in the next section (namely, what are the modes that dominate the backreaction terms when they become dynamically relevant).

We start from eq.~(\ref{A-eom}) for the gauge field modes, that we rewrite as 
\begin{equation}
\frac{d^2 A}{d \tau^2} + \left( k^2 - \frac{k}{f} \, \frac{d \phi}{d \tau} \right) A = 0 \;.
\label{eq-A}
\end{equation}
We can insert in this expression the time derivative $\frac{d \phi}{d \tau}$ from the numerical system evolved in the previous subsection, or the early time analytic solution (\ref{phi-early}), valid during the matter dominated regime.
In the latter case, rescaling the time as in eq.~(\ref{tilde-tau}), we obtain 
\begin{equation}
\frac{d^2 A}{d {\tilde \tau}^2} + \frac{12 M_p^2}{{\bar \rho}_m^{2/3} \, V_0} 
\left( k^2 - \frac{2 \, \lambda}{3 \sqrt{3}} \frac{k}{f} \, {\bar \rho}_m^{1/3} \, V_0^{1/2} \, {\tilde \tau}^5 \right) A = 0 \;.
\end{equation} 
We see that at sufficiently early times the first term dominates the parenthesis, and the mode is stable.
The second term becomes dominant at sufficiently large ${\tilde \tau}$, leading to the amplification of the gauge mode.
At any fixed time, the maximum momentum amplified is 
\begin{equation}
k_{\rm max} \left( \tau \right) = \frac{1}{f} \, \frac{d \phi}{d \tau} \simeq \frac{2 \, \lambda}{3 \sqrt{3}} \frac{1}{f} \, {\bar \rho}_m^{1/3} \, V_0^{1/2} \, {\tilde \tau}^5 \,.
\label{kmax}
\end{equation} 

We normalize the comoving momentum to the value of $k_{\rm max}$ at the estimated transition time ${\tilde \tau} = 1$, 
\begin{equation}
{\tilde k} \equiv \frac{k}{ \frac{2 \, \lambda}{3 \sqrt{3}} \frac{1}{f} \, {\bar \rho}_m^{1/3} \, V_0^{1/2} } \;.
\end{equation} 
In these variables, the equation for the gauge field modes becomes 
\begin{equation}
\frac{d^2 A}{d {\tilde \tau}^2} + \left( \frac{4 \, \lambda \, {\tilde k}}{3 \, {\tilde f}} \right)^2 \left[ 1 - \frac{{\tilde \tau}^5}{\tilde k} \right] A = 0 \;, 
\label{A-eq-early}
\end{equation} 
where the axion decay constant has also been rescaled according to ${\tilde f} \equiv f / M_p$.

In Appendix \ref{app:Asol-early} we present an analytic estimate for the solution of this equation, and for the corresponding backreaction on the evolution of the scalar field.
 As a measure of the backreaction, we take the ratio between the fourth and third term in eq.~(\ref{eom-phi}), 
\begin{equation}
\left( \frac{\partial V}{\partial \phi} \right)^{-1} \frac{\vec{E} \cdot \vec{B}}{f}  = - \left( \frac{\partial V}{\partial \phi} \right)^{-1} \frac{1}{4 \pi^2 a^4 \, f} \int d k \, k^3 \, \frac{\partial}{\partial \tau} \left\vert A \right\vert^2 \equiv \int d {\tilde k} \; {\cal B}_{EB} \;\;, 
\label{back}
\end{equation} 
which is smaller than one in the regime of negligible backreaction, and of ${\rm O } \left( 1 \right)$ when backreaction is important.
Thanks to the above rescalings, we know that we need to study the backreaction at ${\tilde \tau} = {\rm O } \left( 1 \right)$ (the moment of approximate equality between the matter and the scalar field) and for momenta ${\tilde k}$ up to $ {\rm O } \left( 1 \right)$ (the highest momentum excited at this time).

\begin{figure}[ht!]
\centerline{
\includegraphics[width=0.32\textwidth,angle=0]{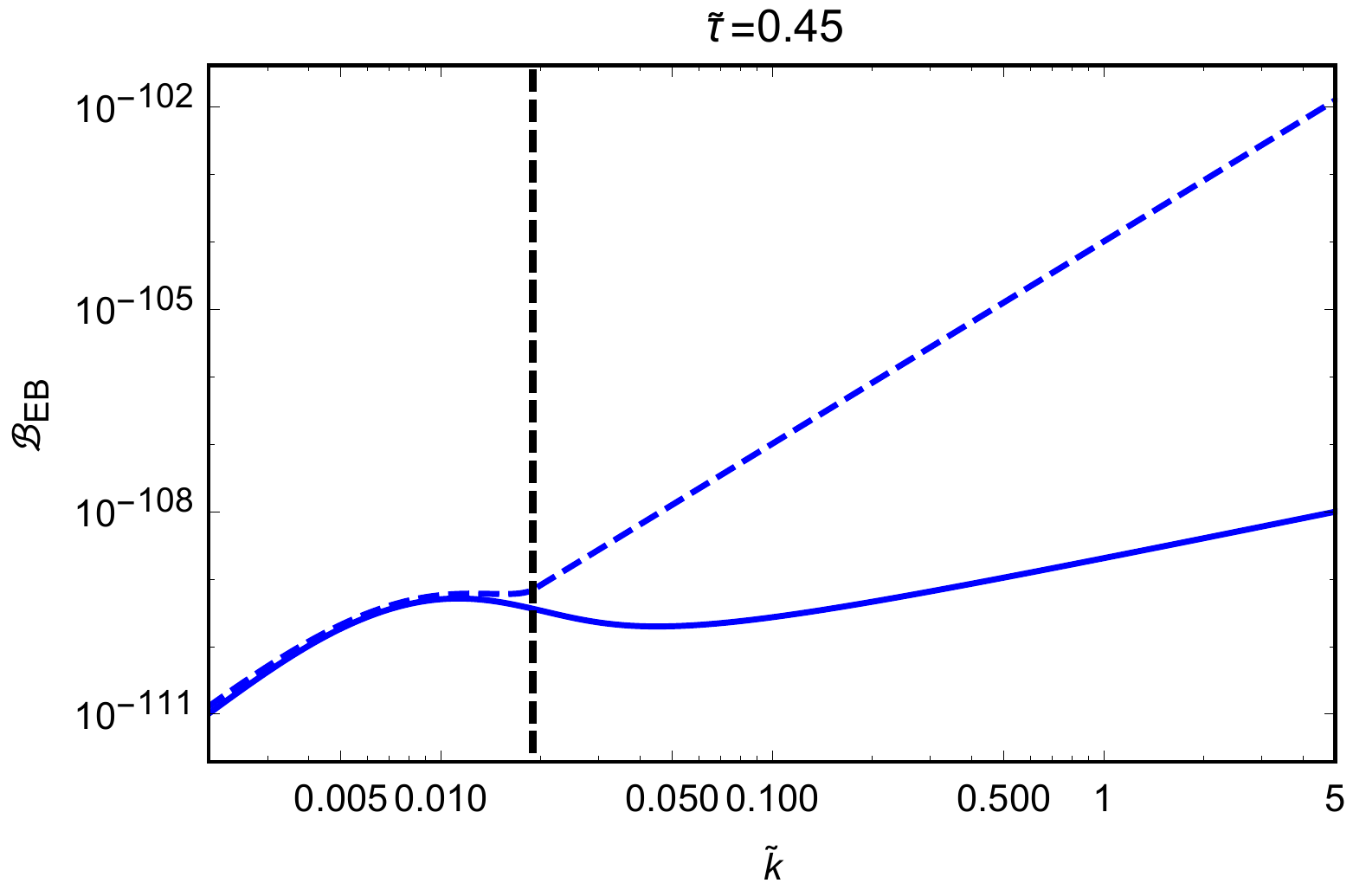}
\includegraphics[width=0.32\textwidth,angle=0]{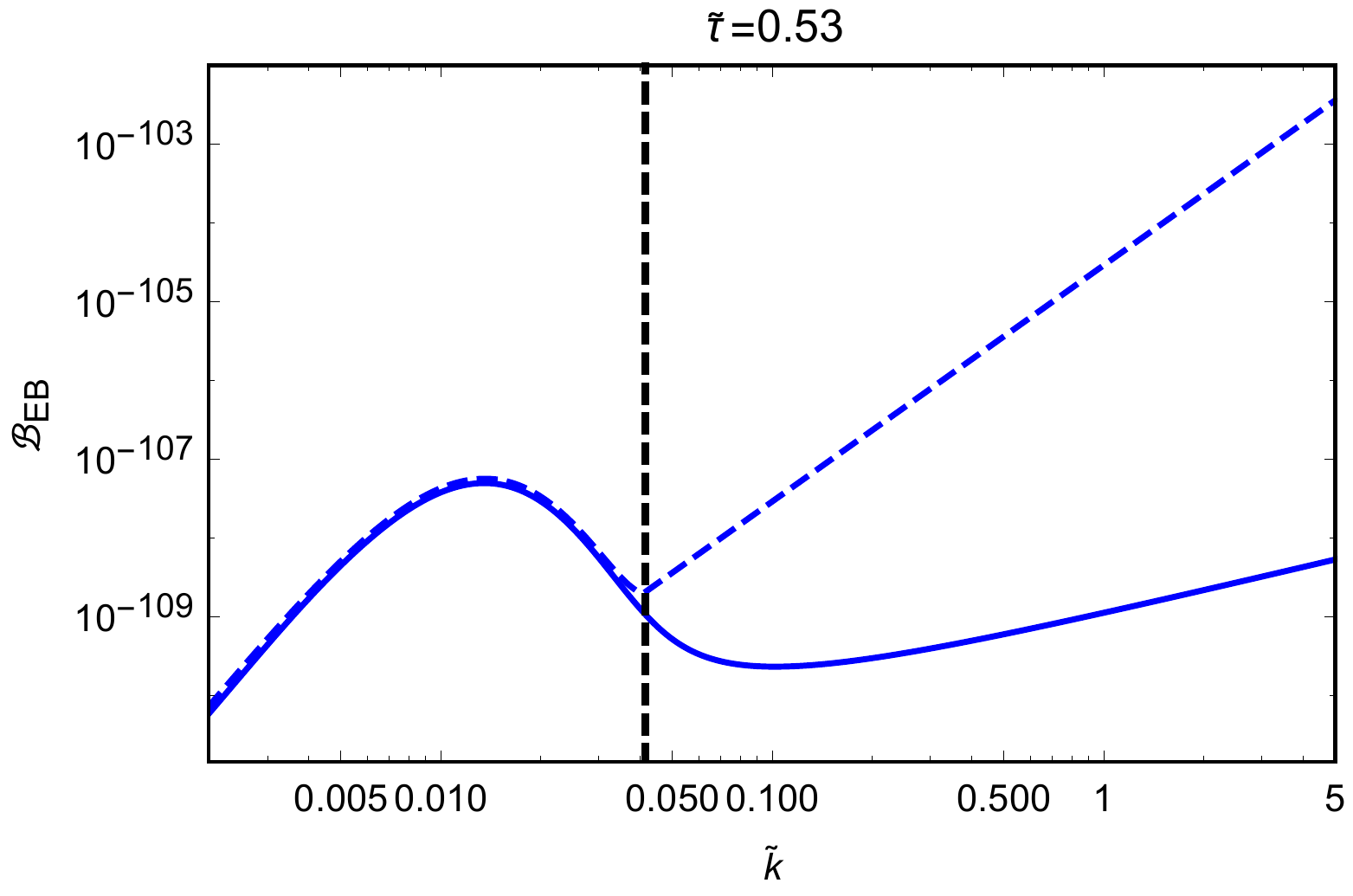}
\includegraphics[width=0.32\textwidth,angle=0]{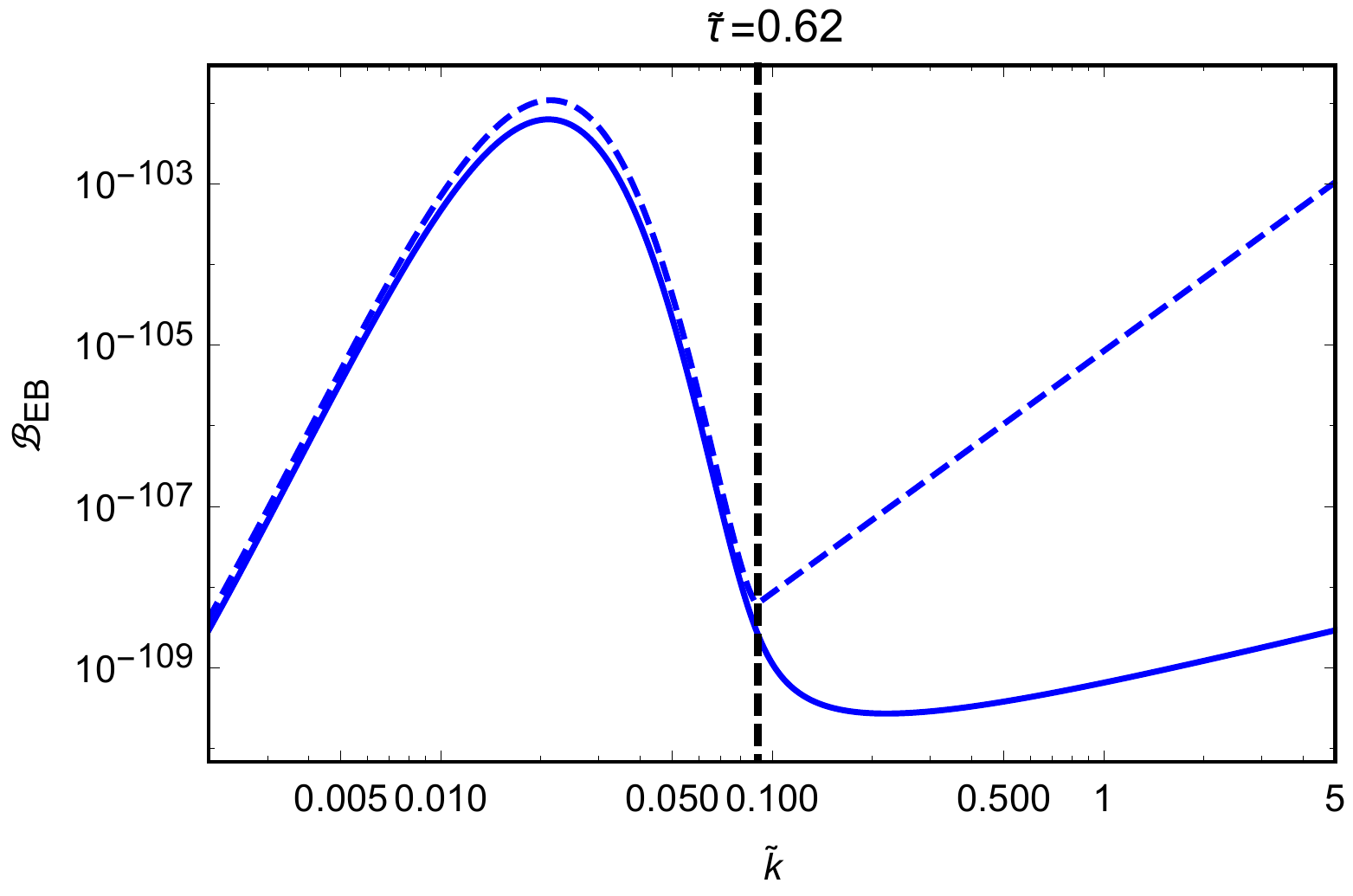}
}

\centerline{
\includegraphics[width=0.32\textwidth,angle=0]{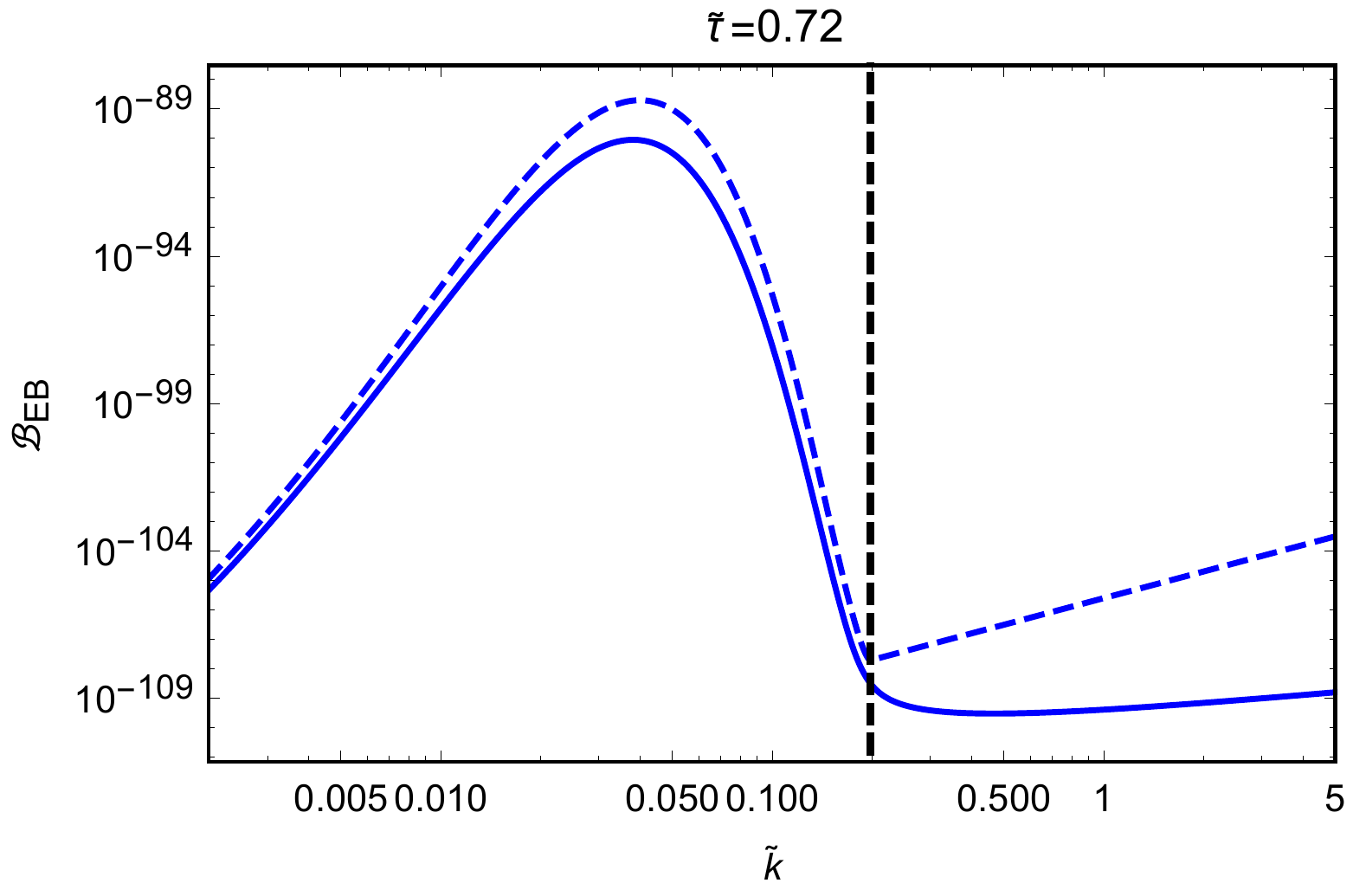}
\includegraphics[width=0.32\textwidth,angle=0]{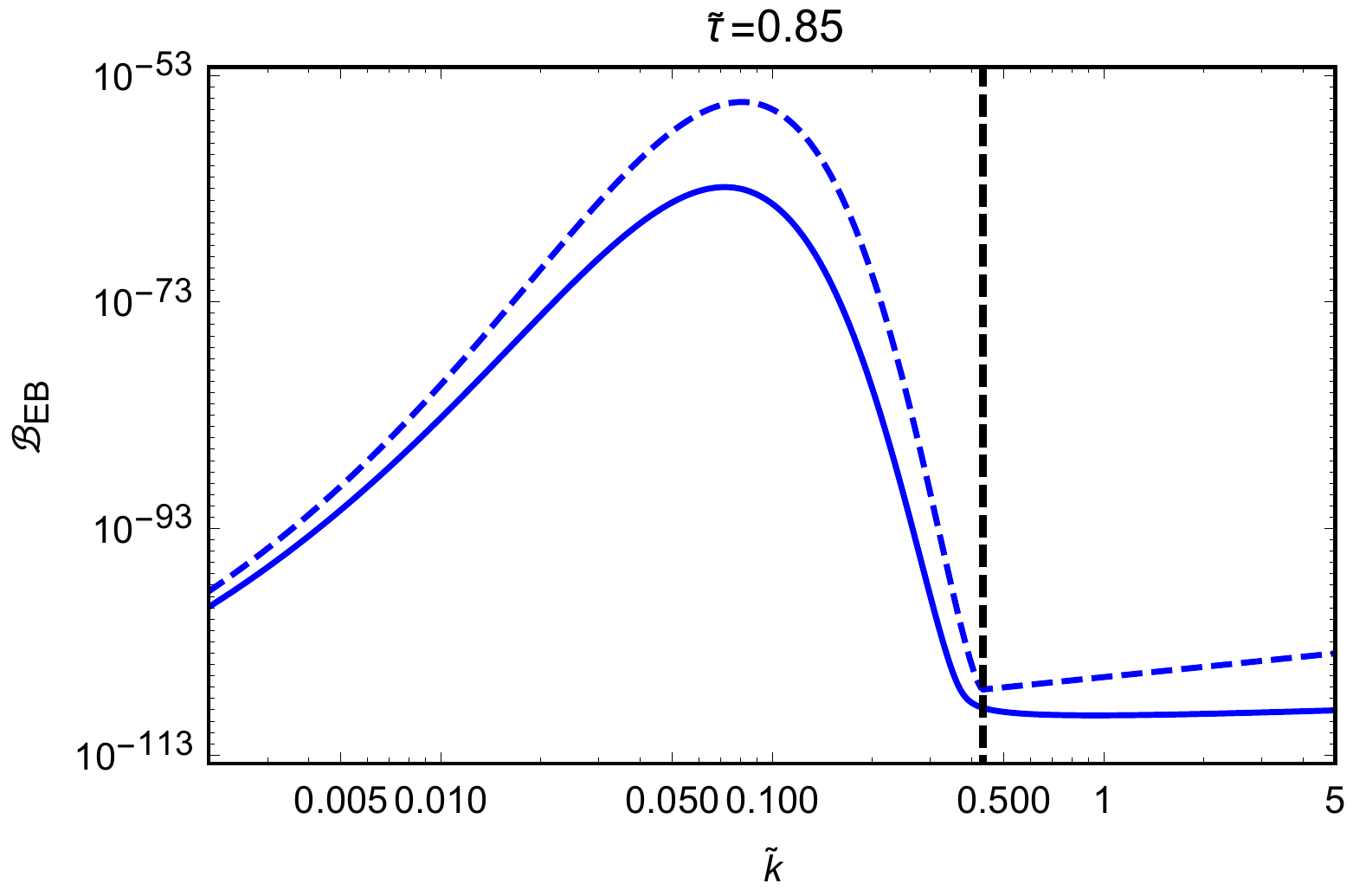}
\includegraphics[width=0.32\textwidth,angle=0]{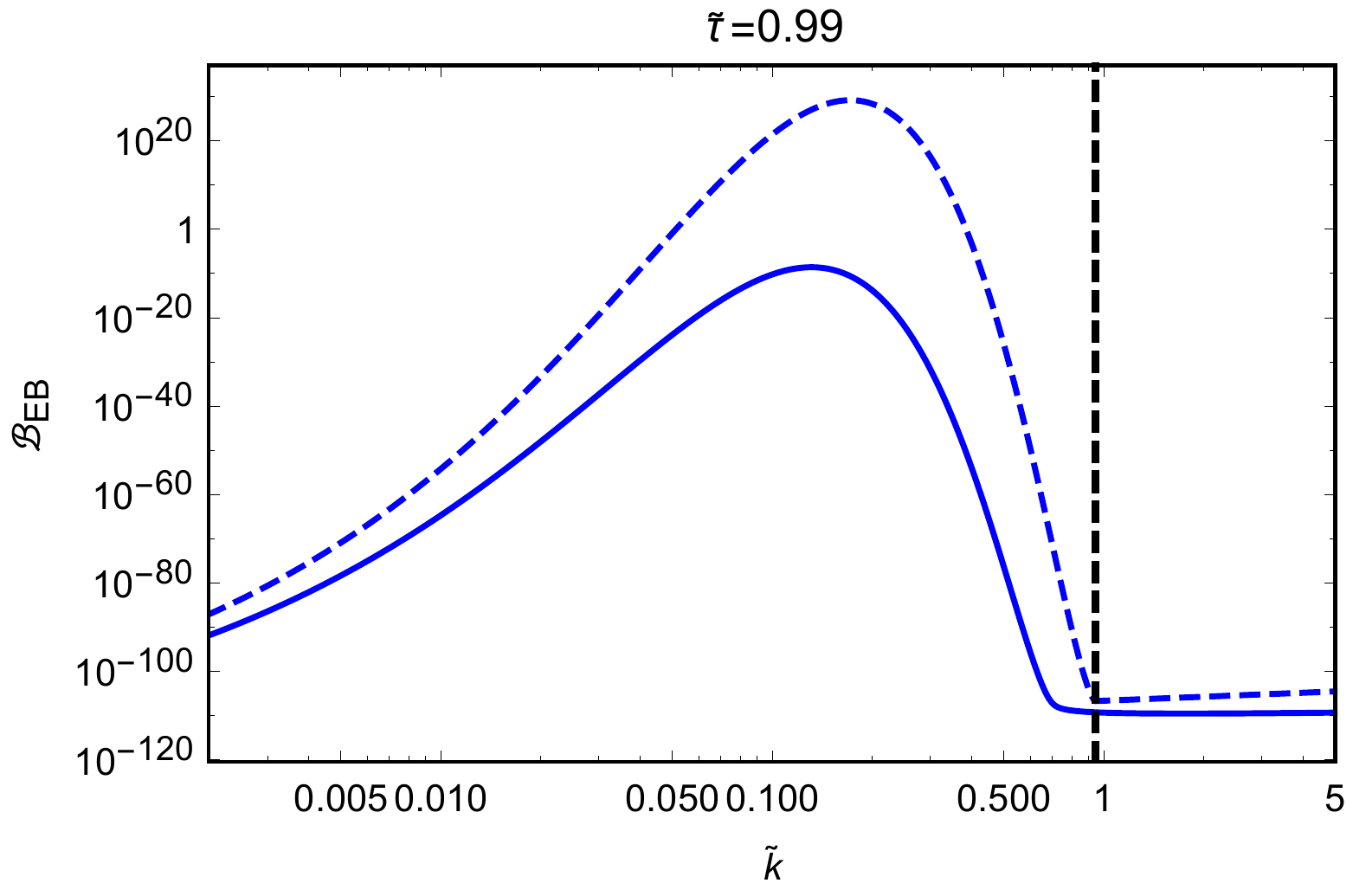}
}
\caption{Comparison for the integrand defined in (\ref{back}) using the analytic approximation given in eq (\ref{BEB}) (dashed line) vs.
using the numerical solution of the system of equations (\ref{system2}) (solid line), for increasing values of the (rescaled) time.
We choose $f=5 \cdot 10^{-4} M_p$ for the coupling constant and $V_0=10^{-120} M_p^4$ for the amplitude of the scalar field potential.
The vertical line separates the unstable (on the left) from the stable (on the right) modes, according to the analytic approximation (\ref{kmax}).
}
\label{fig:integrand}
\end{figure}

In Appendix \ref{app:Asol-early} we provide an analytic estimate for the backreaction term (see eqs.
(\ref{A2p-approx}) and (\ref{BEB})).
In figure \ref{fig:integrand}, we plot the integrand ${\cal B}_{EB}$ obtained from that estimate, against the one obtained from the numerical evolution of the system (\ref{system2}).
As discussed above, the integrand is normalized so that the backreaction becomes relevant when the area underneath the curve approaches one.
In all cases shown the integral under the solid line is much smaller than one.
The vertical line shown in the plots separates the unstable (on the left) from the stable (on the right) modes, according to the analytic approximation (\ref{kmax}).
 As explained in the appendix, our analytic estimate holds for the unstable regime.
 We note that, in this region, the approximated line matches the exact one well at early times, while the two lines start to diverge as ${\tilde \tau}$ approaches one.
This is due to the fact that the analytic approximation assumes a matter dominated background, which stops to be true at ${\tilde \tau} \simeq 1$.
In the stable region to the right of the line, the modes are still close to the vacuum state, and we should also add the contribution of the $A_-$ mode, that would cancel against the one shown here in the high momentum vacuum regime.
We do not implement this cancellation (namely, we do not include the backreaction of the modes $A_-$ in our computations), because this region always contributes a negligible amount of backreaction.

We see from the figure that, even if the analytic approximation does not reproduce the correct amplitude of the backreaction at ${\tilde \tau} \simeq 1$, it still continues to correctly provide the momentum range of the unstable modes.
Therefore, we can use these runs to understand the dynamically relevant range of momenta that must be included in the full numerical simulations that we discuss in the next subsection.

\subsection{Solutions with backreaction included} 
\label{subsec:solutions3}

The final step in our analysis is to solve the full system of equations including the backreaction term in the equation of motion of the scalar field.
We want to investigate what values of the coupling constant $f$ produce a sufficiently large backreaction that affects the motion of the scalar field before the present moment.
The additional friction changes the equation of state parameter of the scalar field $w_\phi\equiv\frac{p_\phi}{\rho_\phi}$ with respect to what is obtained in the absence of backreaction (formally, in the $f \rightarrow \infty$ limit).

The system of equations to be evolved as well as all the necessary re-scaling of the variables are presented in Appendix \ref{app:eqs}.
In order to compute the backreaction term in the equation of motion of the scalar field, we discretize the momenta in equally spaced intervals in $\log k$ space.
The contributions of the various momenta are then added up using the trapezoid rule.
We varied the number of discreet $k$ modes between ${\cal O} \left(10\right)$ and ${\cal O} \left(1000\right)$ and determined that there is negligible difference in the evolution for anything larger than approximately $500$ modes.

\begin{figure}[ht!]
\centerline{
\includegraphics[width=0.5\textwidth,angle=0]{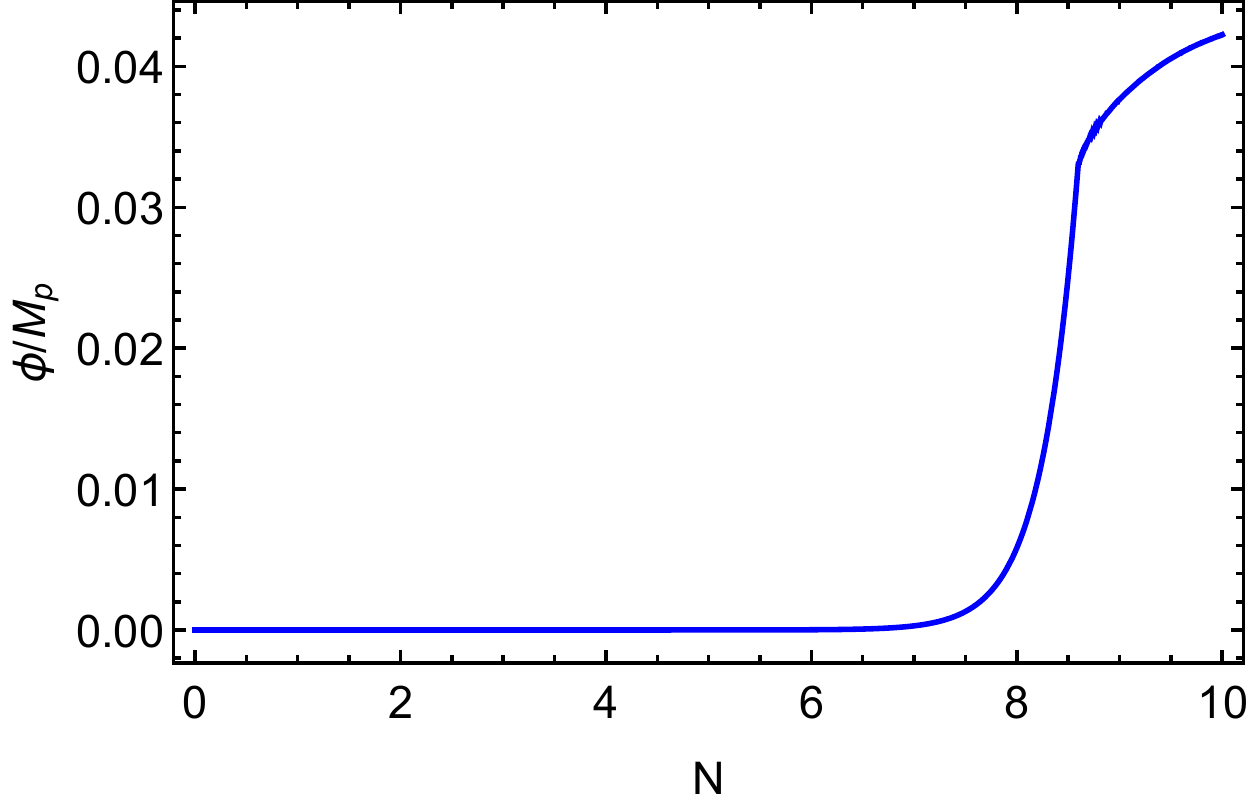}
\includegraphics[width=0.5\textwidth,angle=0]{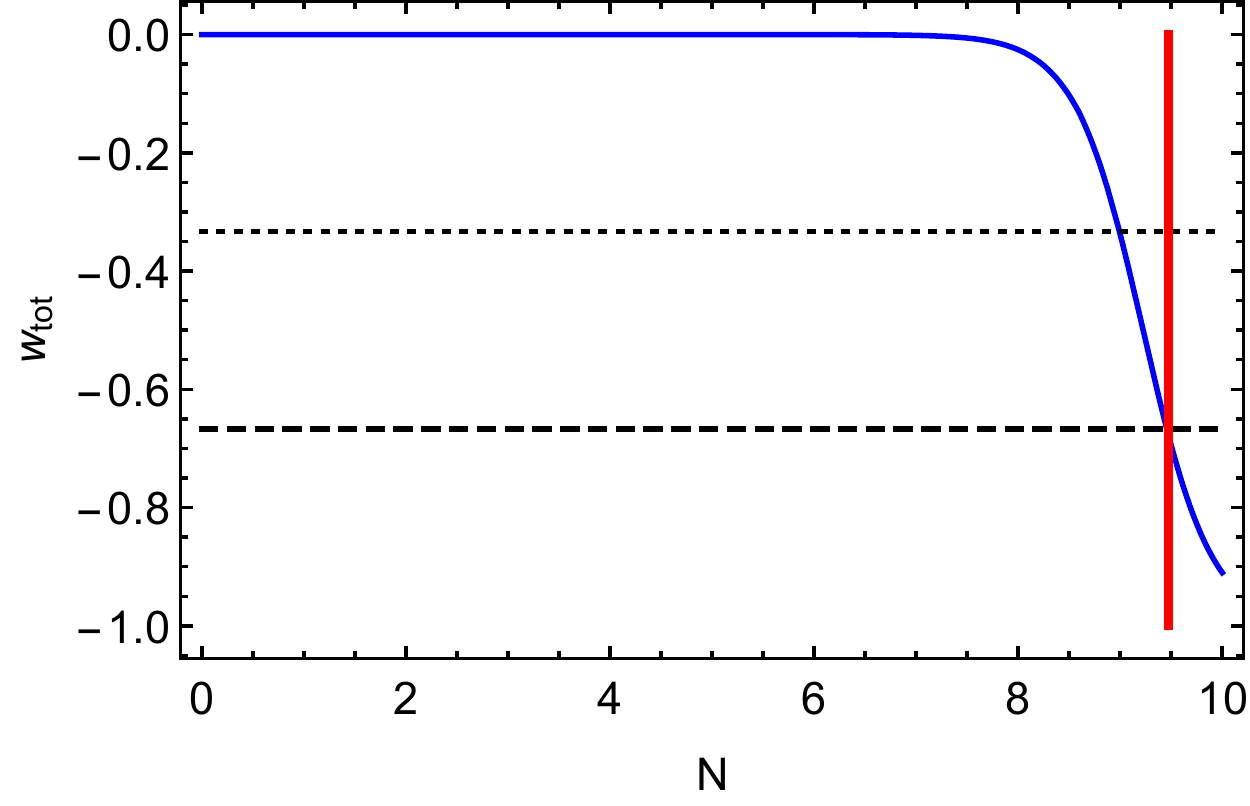}
}

\centerline{
\includegraphics[width=0.5\textwidth,angle=0]{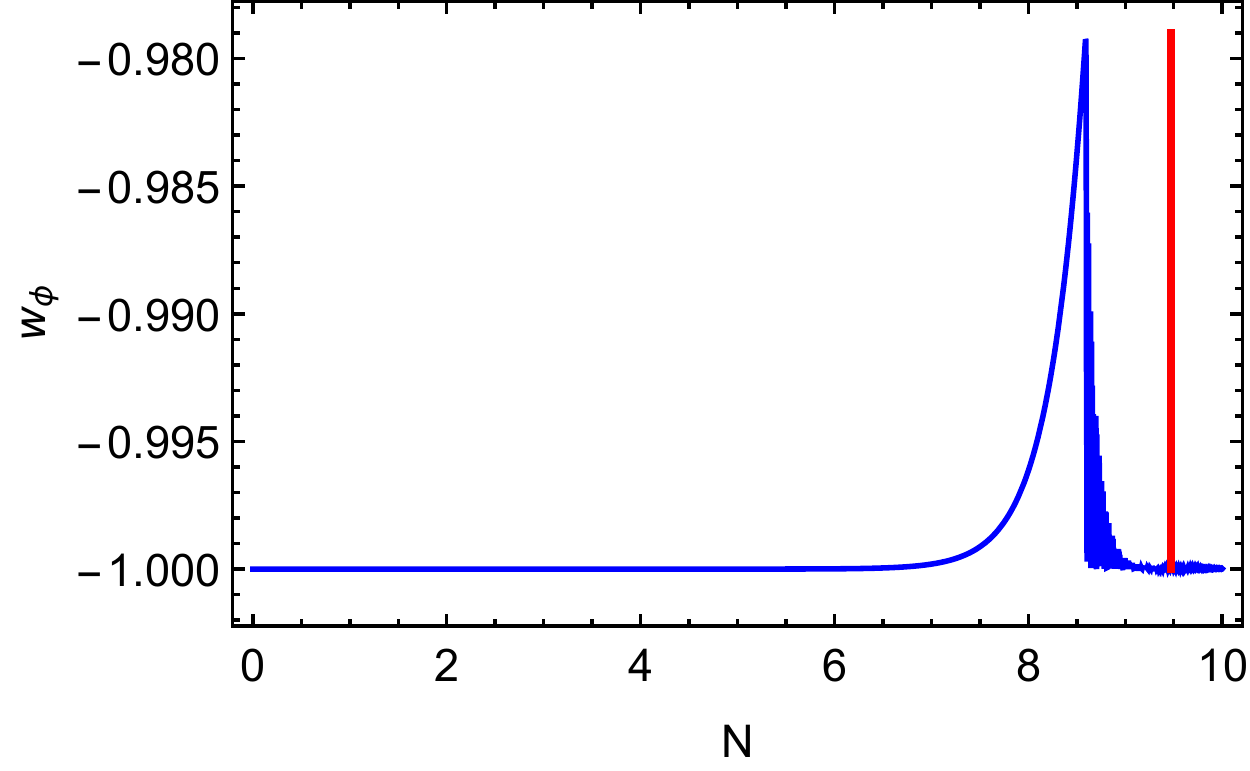}
\includegraphics[width=0.5\textwidth,angle=0]{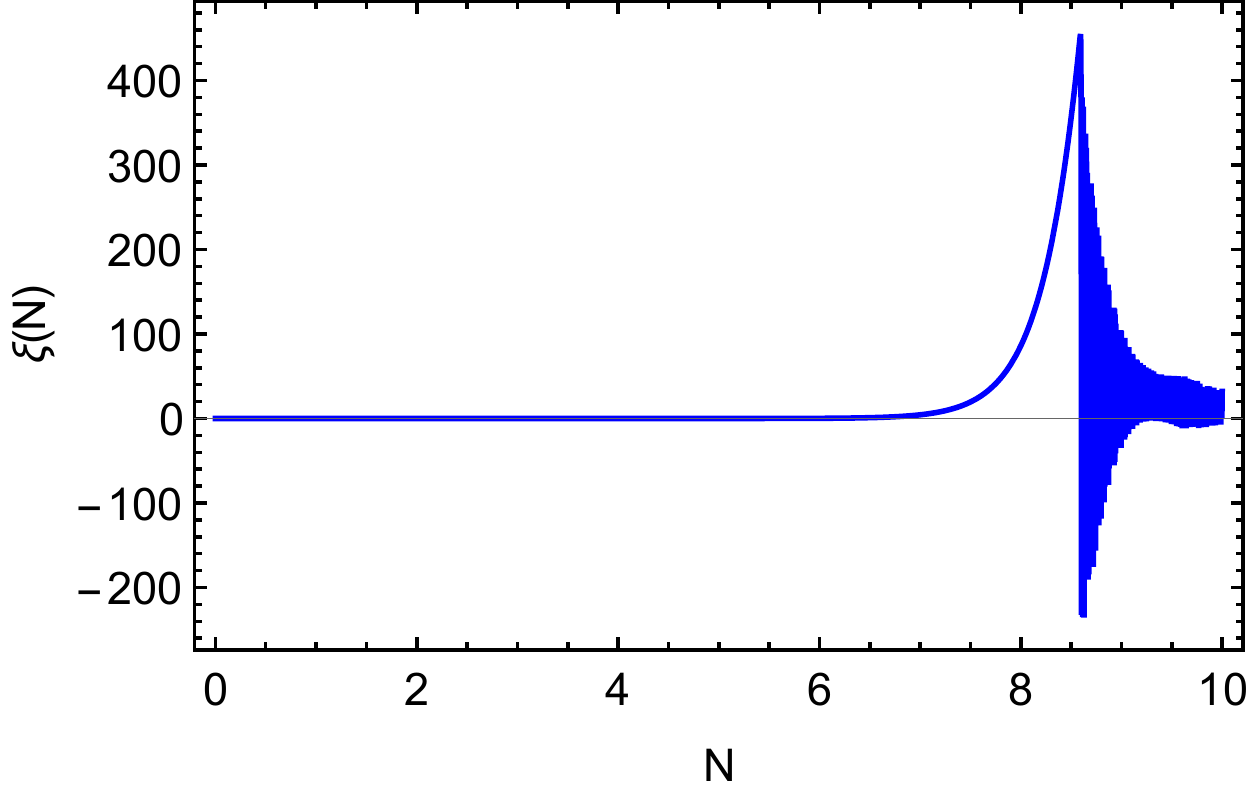}
}
\caption{The parameters used in this evolution are $\tilde{f}=10^{-4}$, $V_0=10^{-120} M_p^4$, $\lambda=1$ and $\bar{\rho}_m=10^{12}$.
\textit{Top left panel}: This is the evolution of the scalar field as a function of the number of e-folds.
The backreaction is negligible until the kink that appears at around $N\sim 8.6$ e-folds.
At that point the backreaction slows down the field appreciably.
\textit{Top right panel}: Evolution of the total equation of state as defined in (\ref{eos}).
As expected the total equation of state is initially that of a matter dominated Universe, and it eventually transitions into accelerated expansion.
The dotted line in the figure is the threshold that has to be crossed for accelerated expansion and the dashed line is the asymptotic value of the equation of state in the absence of backreaction as given by (\ref{w-late}).
The red line is the present moment.
\textit{Bottom left panel}: Evolution of the equation of state of the scalar field.
Again the moment the backreaction becomes dominant is visible by the sudden change in value that occurs at $N\sim 8.6$.
The red line again denotes the present moment.
\textit{Bottom right panel}: The particle production parameter defined in (\ref{A-eom})
}
\label{fig:samplerun}
\end{figure}

In Figure \ref{fig:samplerun} a number of quantities of interest are displayed for a sample run of our numerical code.
The showed run terminates about half an e-fold after the ``present moment'', which is defined as the moment in which the ratio of dark energy to matter is the observed one.
The ``present moment'' is denoted in the figures with a vertical red line.
We have reserved a more in-depth discussion of the evolution deep into the accelerated regime for the next section \ref{sec:AS}.

\begin{figure}[ht!]
\centerline{
\includegraphics[width=0.5\textwidth,angle=0]{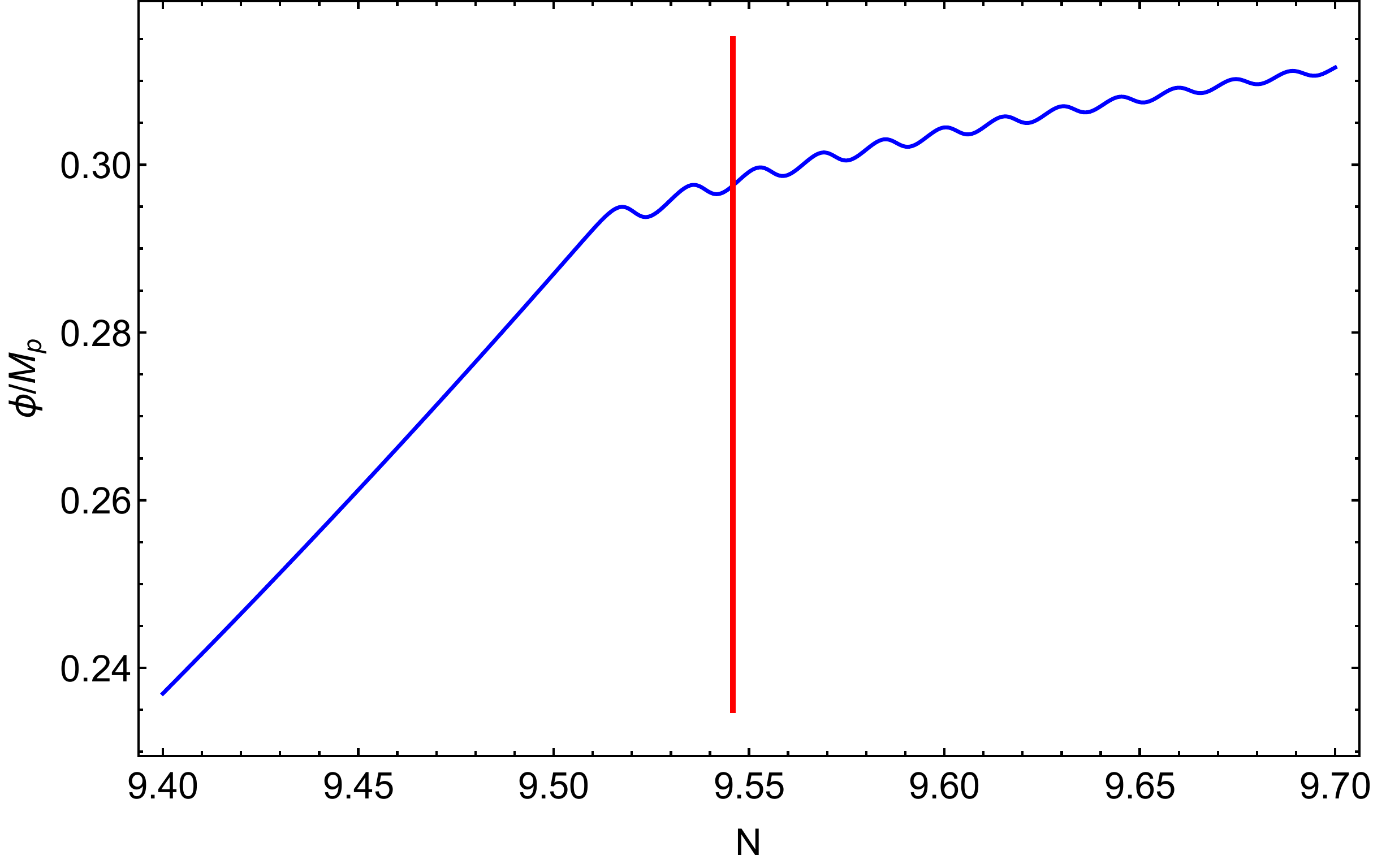}
\includegraphics[width=0.5\textwidth,angle=0]{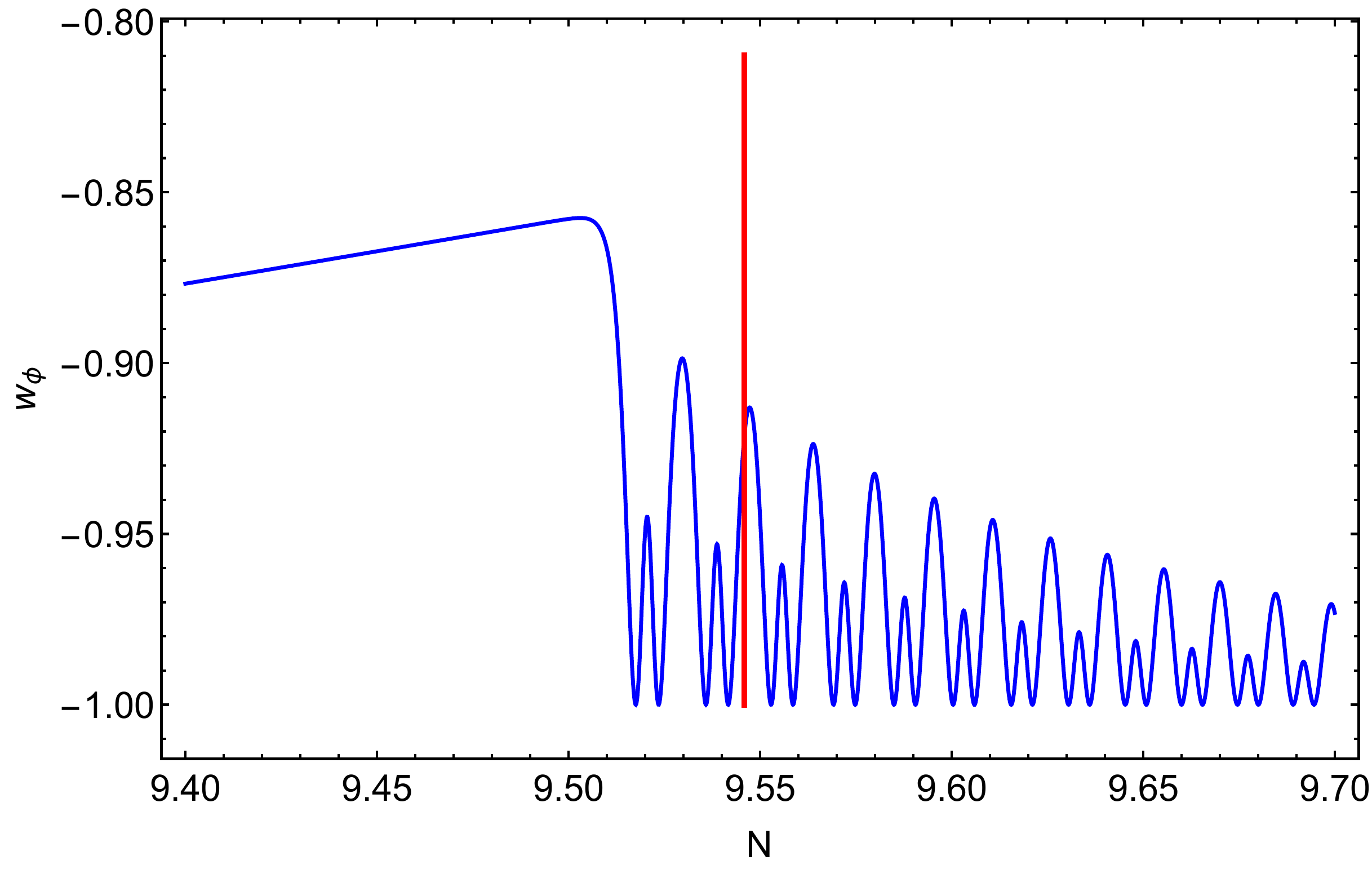}
}
\caption{Evolution of the scalar field and of its equation of state for the value $\tilde{f}=8.7\cdot 10^{-4}$ of the coupling constant.
We are displaying a narrow view of the moment in which backreaction becomes relevant.
The oscillatory behaviour of the equation of state requires a dedicated data comparison study for this choice of $\tilde{f}$.
}
\label{fig:wphi}
\end{figure}

The main important feature that emerges from the numerical results is that an acceptable equation of state for $w_\phi$ can indeed be obtained from this model due to the backreaction.
Specifically, we notice from the bottom left panel that, before the backreaction becomes relevant (namely, for $N \la 8.5$), $w_\phi$ is climbing toward the late time attractor (\ref{w-late}).
Backreaction adds a strong friction to the motion of $\phi$, that brings $w_\phi$ back toward $-1$.
We notice from the plots that backreaction manifests itself very suddenly (we verified that this is not an artifact of a poor sampling of the modes in the numerical evolution), rather than in a smooth way, due to its exponential growth.
The scalar field reacts to this by performing some oscillations about an overall growing solution.
These oscillations are too rapid to be seen clearly in figure \ref{fig:samplerun} but it is possible to see a thickening of the blue line immediately after backreaction has become dominant in the equation of motion for the scalar field.
The oscillations are more clearly visible in Figure \ref{fig:wphi}, which is a zoomed version of the previous figure, centered around the moment in which backreaction becomes important.

\subsubsection{Comparison with the data}

The above results indicate that, even if one starts from a scalar field potential that does not provide a sufficiently fast acceleration (a large value of $\lambda$ in the specific case (\ref{potential}) that we are studying), a sufficiently strong amplification of the vector field can result in an equation of state $w_\phi$ sufficiently close to $-1$, as required by data.This suggests that, for any choice of $\lambda$, there can be a threshold value $f_{\rm thr} \left( \lambda \right)$ below which the mechanism can result in an acceptable phenomenology.
Finding the precise threshold value however requires a dedicated data comparison between this model and the data, which is beyond the scope of this paper.
In particular, an input for the comparison with the supernovae data is the luminosity distance $d_L = \frac{a_0^2}{a} \int_a^{a_0} \frac{ d a'}{a^{'2} \, H \left( a ' \right)}$.
 The distance can be derived from the expansion law $a \left( N \right)$, which in turn is one to one related to the evolution of the scalar field equation of state $w_\phi \left( N \right)$.
In the present case, several choices of $f$ results in significant oscillations of $w_\phi$, so that the simplest parametrizations existing in the literature cannot be used.
This is for instance the case for the example shown in Figure \ref{fig:wphi}, in which ${\tilde f} = 8.7 \cdot 10^{-4}$ was chosen.

Even without a dedicated data comparison, we can obtain an interval in which we expect to find the threshold value.
Specifically, we can obtain a value $f_1$ for which we can be highly confident that the mechanisms agrees with data, and a greater value $f_2$ for which the mechanism is not in agreement with the data (we recall that increasing the value of $f$ results in a smaller backreaction).
This will allow us to indicate that the threshold value is between $f_1$ and $f_2$.
The value of $f_2$ can be obtained with relative ease.
Specifically, by numerically evolving the system as described in the previous subsection, we find that for 
$\tilde{f} = 9.25\cdot 10^{-4}$, or higher, the backreaction becomes noticeable only at the present time (we recall that by `present time' we mean the moment in the evolution in which the ratio between the energies of the dark energy and of matter is the observed one).
This is phenomenologically indistinguishable from the case of no gauge production, which, as we discussed above, is compatible with the data only at $\simeq 3 \, \sigma$ (we recall that we are assuming $\lambda =1$).
Therefore $\tilde{f}_2 = 9.25\cdot 10^{-4}$.

\begin{figure}[ht!]
\centerline{
\includegraphics[width=0.5\textwidth,angle=0]{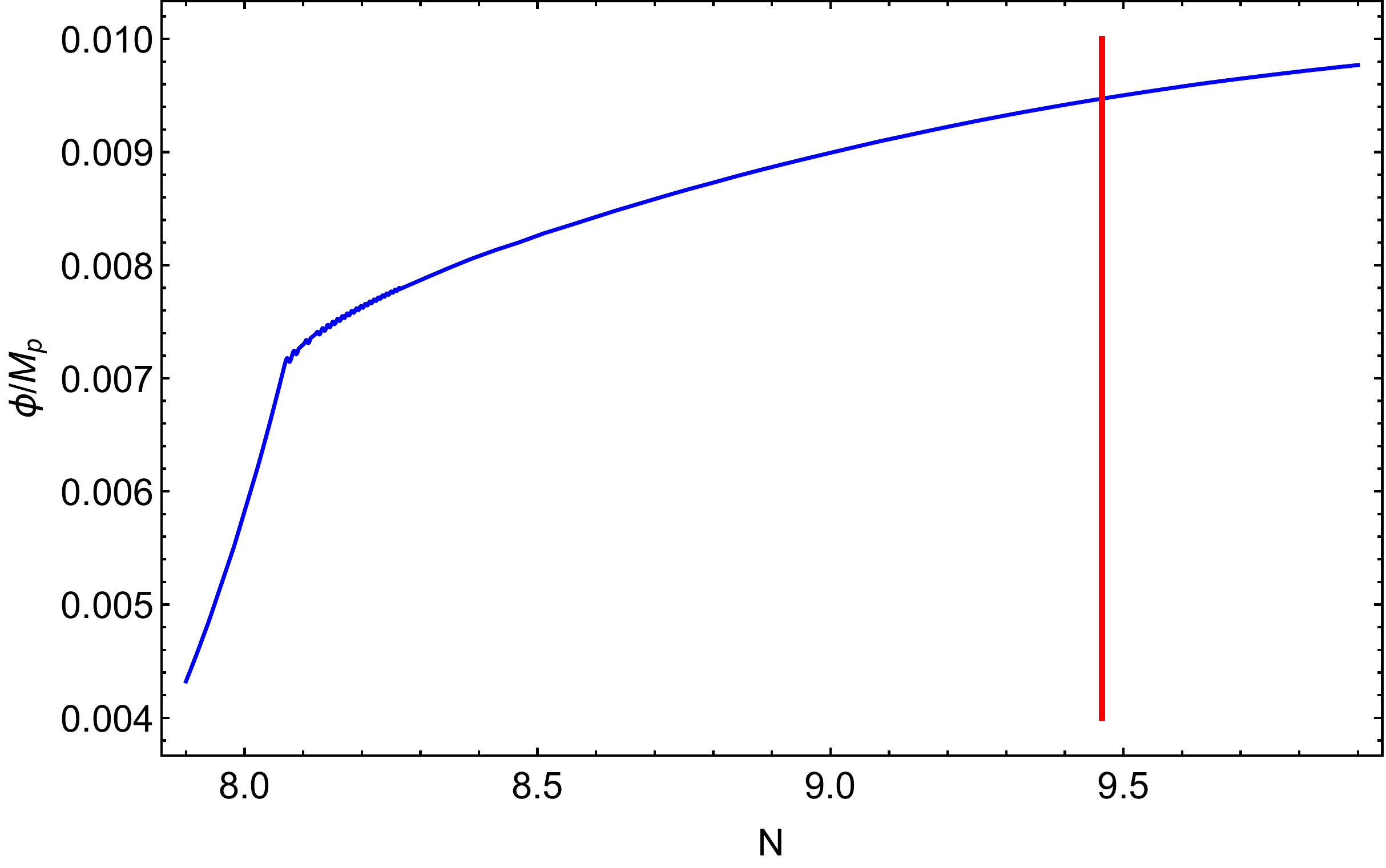}
\includegraphics[width=0.5\textwidth,angle=0]{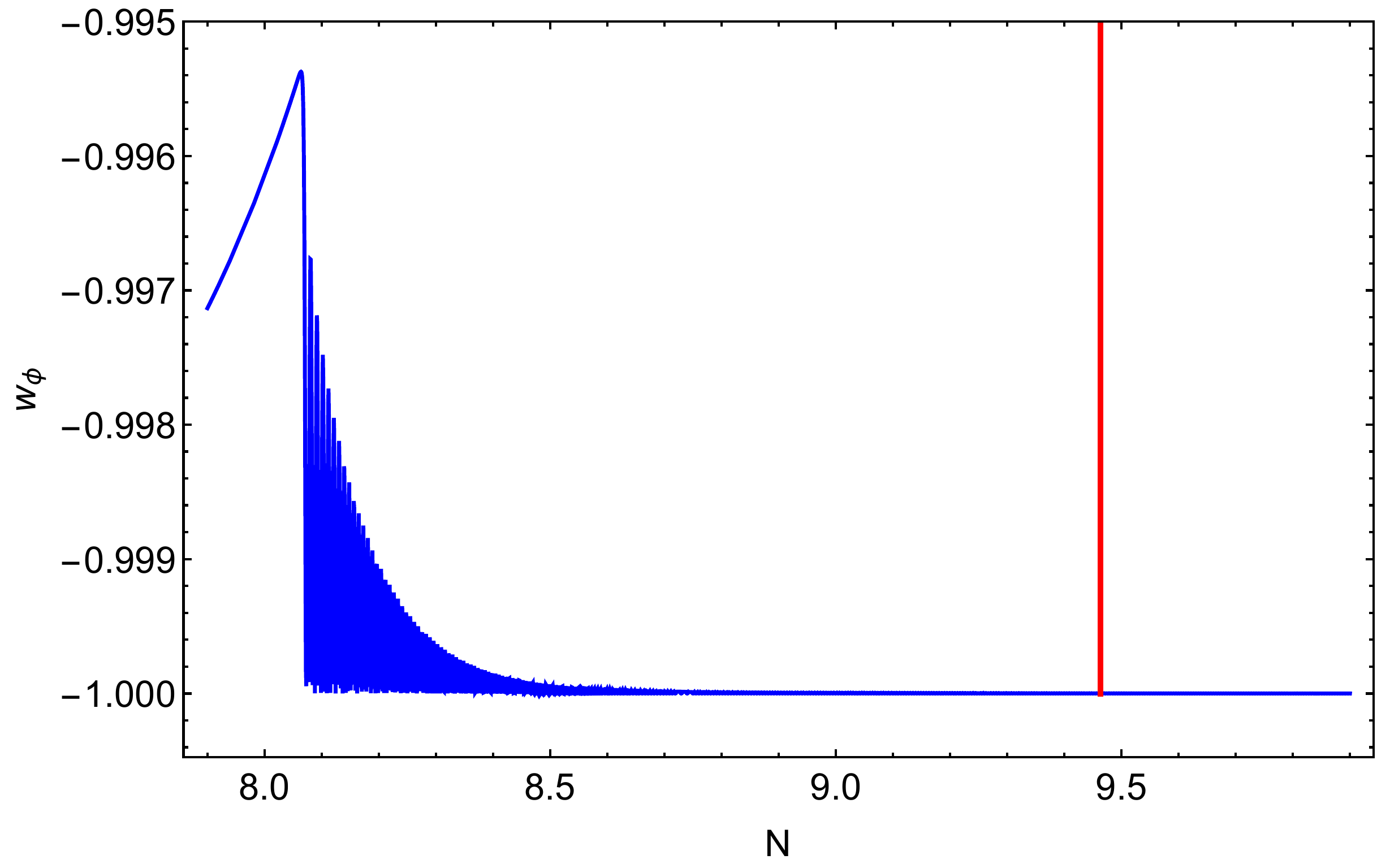}
}
\caption{Analogous to the previous figure, but for ${\tilde f} = 2.2\cdot 10^{-5}$.
Due to the stronger coupling, backreaction becomes important well before the present time, and the late time evolution of $w_\phi$ can be linearized as in eq.~(\ref{parametrization}).}
\label{fig:wphi2}
\end{figure}

To find $f_1$, we consider a sufficiently strong production so that $w_\phi$ reaches a smooth evolution by the present time, with very suppressed oscillations.
In this case, the late time evolution of $w_\phi$ is well described by the parametrization 
\begin{equation}
w_\phi \left( a \right) = w_0 + \left( 1- \frac{a}{a_0} \right) w_a \;, 
\label{parametrization}
\end{equation}
which consists in linearizing the late time evolution of equation of state of the scalar field as a function of cosmological redshift.
We can then use the results of \cite{Aghanim:2018eyx} which, in Figure 30, shows the allowed region in the parameter space $(w_0,w_a)$ for different choice of data.
We choose the dataset that results in the most stringent constraints.

\begin{figure}[ht!]
\centerline{
\includegraphics[width=0.33\textwidth,angle=0]{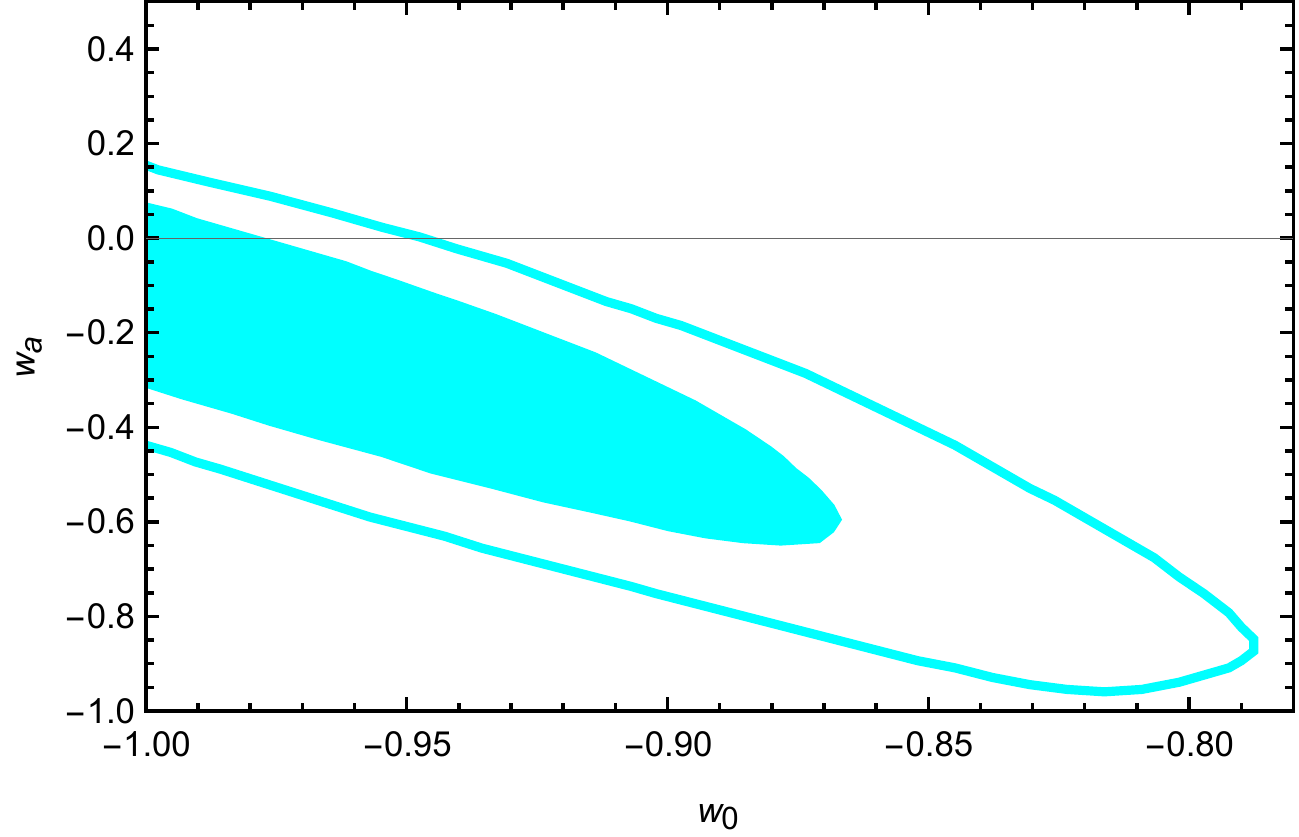}
\includegraphics[width=0.33\textwidth,angle=0]{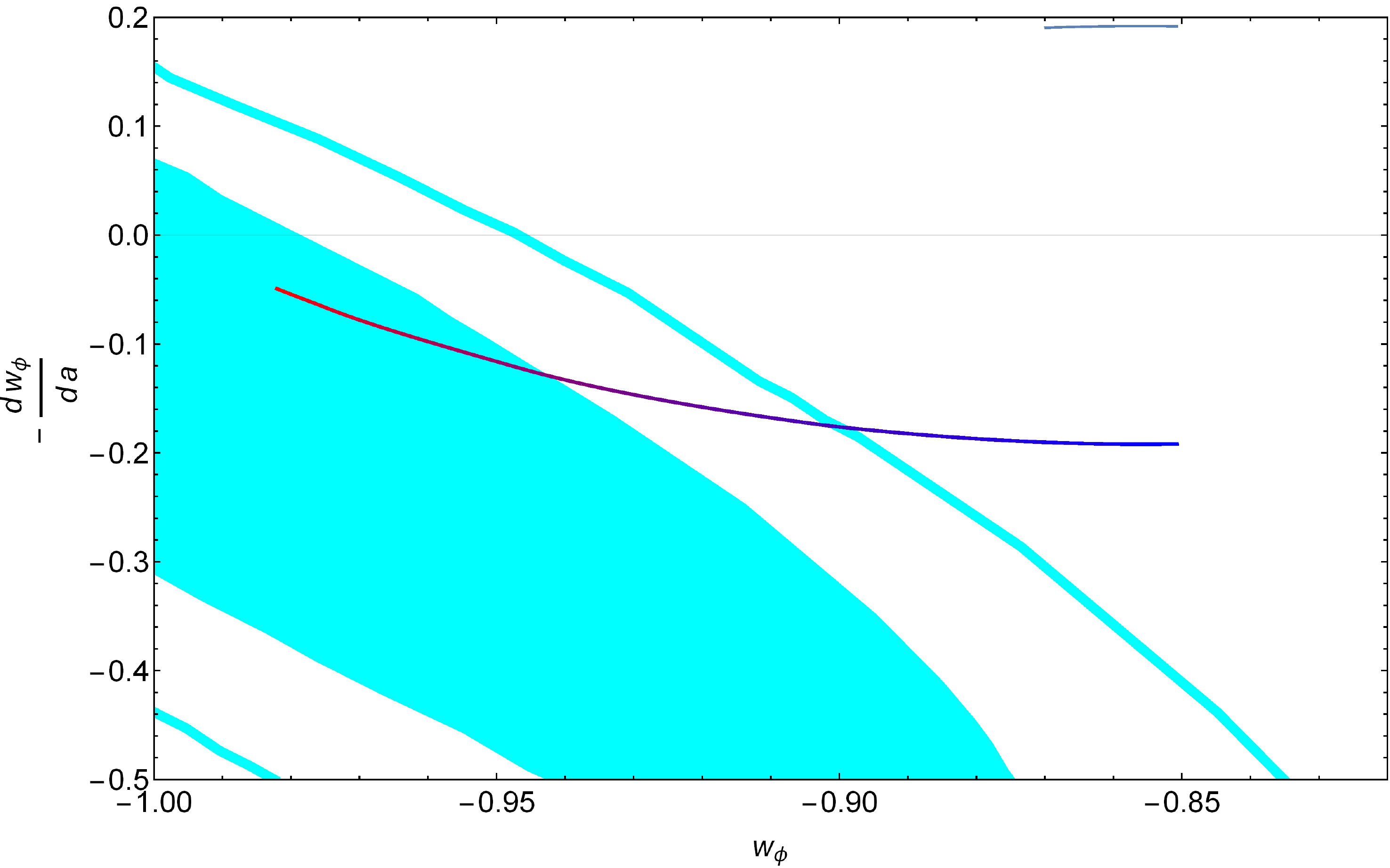}
\includegraphics[width=0.33\textwidth,angle=0]{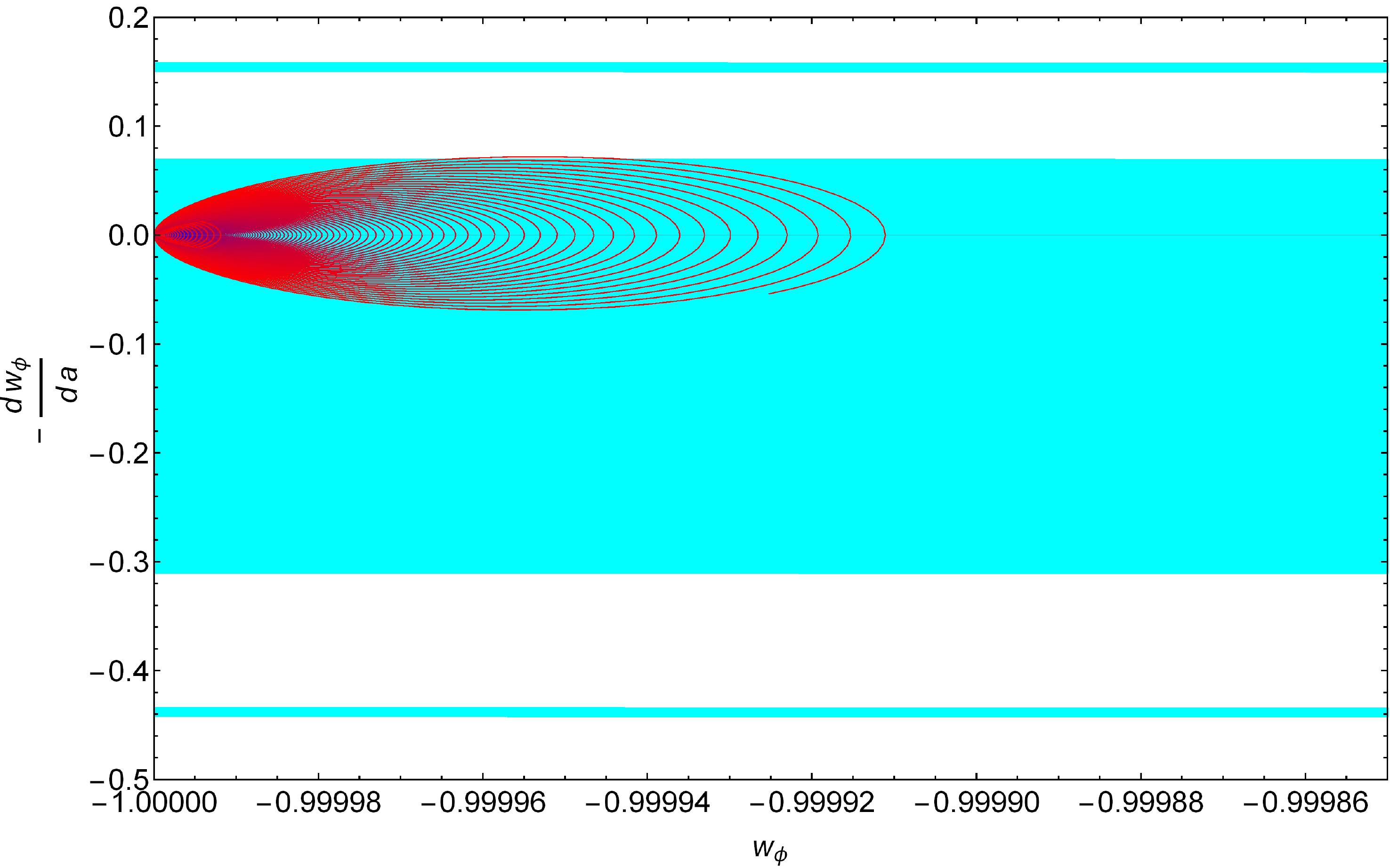}
}
\caption{Left panel: $w_0 \geq -1$ portion of the $1$ and $2 \sigma$ regions from Figure 30 of \cite{Aghanim:2018eyx}, for the data analysis based on the parametrization (\ref{parametrization}) of the scalar field equation of state.
Central and right panel: portion of the same region (for different rescaling of the axis) together with a curve showing the evolution of $w_\phi$ and of its derivative for one e-fold before the present time.
(The curves transition from red to blue where red is the earliest time and blue is the present moment.) The the central panel is for ${\tilde f} = \tilde{f}_1 = 9.25\cdot 10^{-4}$, leading to a significant evolution of $w_\phi$, so that the parametrization (\ref{parametrization}) cannot be employed for this choice of ${\tilde f}$ (we know that this choice is ruled out by data, since it leads to an appreciable backreaction only after the present time).
The right panel is for ${\tilde f} = \tilde{f}_1 = 2.2\cdot 10^{-5}$.
This leads to a much smaller change in $w_\phi$ in the last e-fold (see also Figure \ref{fig:wphi2}), and to an evolution inside the $1\sigma$ allowed region.}
\label{fig:planckandprediction}
\end{figure}

Fig.~\ref{fig:wphi2} shows the evolution of the scalar field and of its equation of state for a coupling for which the oscillations in 
$w_\phi$ took place well before the present time, and for which we argue that the late time evolution can be sufficiently well described by (\ref{parametrization}) for the purpose of data comparison.
We support this claim through Figure \ref{fig:planckandprediction}.
In the left panel of the figure we reproduce the $1$ and $2\sigma$ contours from Figure 30, of \cite{Aghanim:2018eyx}, focusing on the $w_0 \geq -1$ region.
In the central panel we superimpose to these contours a line that encodes the evolution of $w_\phi$ and of its derivative wrt the scale factor, for the choice ${\tilde f} = {\tilde f}_2 = 9.25\cdot 10^{-4}$.
We recall that for this choice, backreaction is important only after the present time, and the model is incompatible with data.
The evolution encoded in the line lasts for one e-fold before the present time, with earlier times on the left of the line, and later times on the right (in the color version of the figure, the color of the line changes from blue to red as time progresses).
Due to the strong evolution of $w_\phi$ and of its derivative the parametrization (\ref{parametrization}) cannot be employed, and the main purpose of this panel is to provide an immediate comparison with the right panel.
In the right panel we show an analogous evolution of $w_\phi$ for the choice $\tilde{f} = 2.2\cdot 10^{-5}$.
We note the very different scale shown now in the figure.
As can also be seen from Figure \ref{fig:wphi2}, the equation of state is extremely close to $-1$ during the last e-folds, and its derivative has fast oscillations, that remain within the allowed $1\sigma$ region.
Although, strictly speaking, the parametrization (\ref{parametrization}) requires an exactly linear evolution of $w_\phi \left( a \right)$ we see from the horizontal scale that the fast and small oscillations of $\frac{d w}{d a}$ around zero nearly average out on the net evolution of $w_\phi \left( a \right)$.
Due to this, and due to the fact that evolution is always within the $1\sigma$ allowed region, we conclude that this choice is compatible with the data. Therefore we set $\tilde{f}_1 = 2.2\cdot 10^{-5}$. 

To summarize, a rigorous data comparison for this mechanism requires a dedicated analysis, due to the nontrivial and oscillatory nature of $w_\phi \left( a \right)$.
We know that the mechanism will be compatible with the data for $f$ smaller than a threshold value (which depends on the curvature of the potential, which in our case is controlled by the parameter $\lambda$).
From our numerical study, performed with $\lambda =1$, that have argued that the threshold value should be between $ 2.2\cdot 10^{-5} \, M_p$ and 
$9.25\cdot 10^{-4} \, M_p$.
A more refined computation of the precise threshold value $f_{\rm thr}$ can be obtained with a dedicated analysis of the data for the present model.

\section{Application to primordial inflation and the road to the Anber--Sorbo solution} 
\label{sec:AS} 

In addition to the case studied in the previous section, we are interested in applying our numerical code to the transition between a pre-inflationary era and inflation.
In the original Anber and Sorbo paper \cite{Anber:2009ua}, the model is studied under the assumption that the backreaction term is always the main source of friction of the motion, and that the inflaton has a nearly constant speed, determined by the interplay between this term and the derivative of the inflaton potential.
This results in a slow roll inflationary solution, with nearly constant inflaton speed.
The backreaction term is computed in \cite{Anber:2009ua} by disregarding the departure from de Sitter expansion, and by providing a convenient approximation for the mode functions of the amplified gauge fields.
One obtains \cite{Anber:2009ua} 
\begin{equation}
\frac{d^2 \phi}{dt^2} +3 H \frac{d \phi}{dt}+V'\left(\phi\right)= -\frac{{\cal I} }{f}\left(\frac{H}{\xi}\right) e^{2 \pi \xi} \;, 
\label{eom-AS}
\end{equation}
where the last term is the one due to backreaction, and the numerical coefficient is ${\cal I}\equiv 7!/(2^{21} \pi)\simeq 2.4\times 10^{-4}$.
Equating the backreaction term with the derivative of the inflaton potential leads to \cite{Anber:2009ua} 
\begin{equation}
\xi \simeq \frac{1}{2 \pi} \, \ln \left[\frac{9}{{\cal I}}\frac{M_p^4\, f\, |V'\left(\phi\right)|}{V^2\left(\phi\right)}\right] \;, 
\label{AS-attractor}
\end{equation}
which is the Anber--Sorbo (AS) solution in the regime of slow roll, sustained by the gauge field amplification.

We want to understand whether the Anber--Sorbo inflationary solution can be dynamically attained if one starts from a pre-inflationary period of matter domination, during which the gauge field amplification, and its back reaction is negligible.
Loosely speaking, this analysis extends to later times the one performed in the previous sections, since so far we have been interested to at most one e-fold of accelerated expansion (corresponding to the present dark energy domination) while now we want to discuss the evolution well inside the inflationary regime.

\subsection{Numerical considerations}

The system of equations to be evolved as well as all the necessary re-scaling of the variables are presented in Appendix \ref{app:eqs}.
Evolving our numerical code deep into the inflationary era poses a numerical challenge.
The reason for this is that the more e-folds of inflation we cover, the greater is the momentum range that corresponds to the unstable modes that have to be included in the momentum integral in the equation of motion of the inflaton.
The momentum threshold between stable and unstable modes is given by 

\begin{equation}
k_{\rm th}=\frac{a \dot{\phi}}{f} \;. 
\end{equation}

If, as a first approximation, we ignore the variation of $\dot{\phi}$ during inflation, we can see that during the course of inflation the threshold is increasing.
Assuming that we evolve $N\simeq15$ e-folds of inflation, the value $k_{\rm th}$ will be $a={\rm e}^{15}\simeq 3\times 10^{6}$ times greater at the end compared to the initial value of the threshold.
Evolving gauge modes of really high momenta slows the code down significantly since we are evolving gauge modes which spend most of the evolution in the stable regime, in which the evolution of the mode is highly oscillatory.
 
An additional problem associated to the high momenta is that the integral giving the backreaction is UV-divergent (as we did not renormalize the modes, since this was not needed for the computations performed in the previous section\footnote{A renormalization scheme was recently provided in \cite{Ballardini:2019rqh}, with results very similar to those obtained by simply cutting off the stable UV modes.}).
Therefore, if we include a too large range of high momentum mode, instead of integrating over the instability associated with particle production, one integrates over a contribution associated with modes which are still in their vacuum configuration.
This problem is also apparent from the first two panels of figure \ref{fig:integrand}.
We see that, at early times the integrand has a UV contribution that is dominant compared to the instability associated with particle production.
For inflationary energy scales, and for very large momenta (which are necessary for consistent evolution deep in the inflationary regime) it is possible that the UV contribution might become dominant at early times.
In principle, one way to overcome this problem would be to put a cut-off at the interface between the stable and unstable modes.
We will see that this is not possible in our case because, due to a non monotonic evolution of $\frac{d \phi}{d t}$, there are gauge modes that become stable after they have been enhanced; therefore, their contribution would be incorrectly neglected if we included in the backreaction only modes that are unstable at any given time.

In order to overcome the problems mentioned above we will perform the evolution by choosing a convenient value of the potential\footnote{This value does not lead to the correct normalization of the power spectrum of the scalar perturbations, so we do not claim that this is a `realistic' situation.
Our present goal is to study the results of the numerical evolution for a convenient choice of parameters.} $V_0= 10^{-20} M_p^4$. 
As can be seen by eq (\ref{back}) in order for the backreaction to become dominant the integrand has to overcome a suppression of the order of $\frac{V_0}{M_p^4}$.
The smallness of the potential allows us to choose a higher value of the highest momentum, without risking the UV spurious divergence being dynamically relevant.
Simultaneously it is beneficial to have a not too small value of the potential because one does not have to evolve too many e-folds for the gauge modes to overcome the suppression and thus all the various effects arise sooner.
Our choice of $V_0$ is a compromise between these two opposite requirements.

Another way to improve the performance of the code is to keep the really high momenta ``frozen'' in their vacuum configuration until late in the evolution when they come close to the instability region.
We realize this by separating the evolution into two parts.
In the first part we evolve the numerical system from deep inside the matter dominated period until the early inflationary era, including only gauge modes that are relevant during this time period.
We use the final values of the fields obtained in this run as the initial condition for a second run, that extends more deeply into the inflationary regime.
In this second run we consider a much greater range of high momentum modes, that we expect to become unstable during this second evolution (these additional modes are initialized from their vacuum state; we verified that this is a good assumption at the start of the second evolution).

\subsection{Numerical results for an exponential potential}

\begin{figure}[ht!]
\centerline{
\includegraphics[width=0.5\textwidth,angle=0]{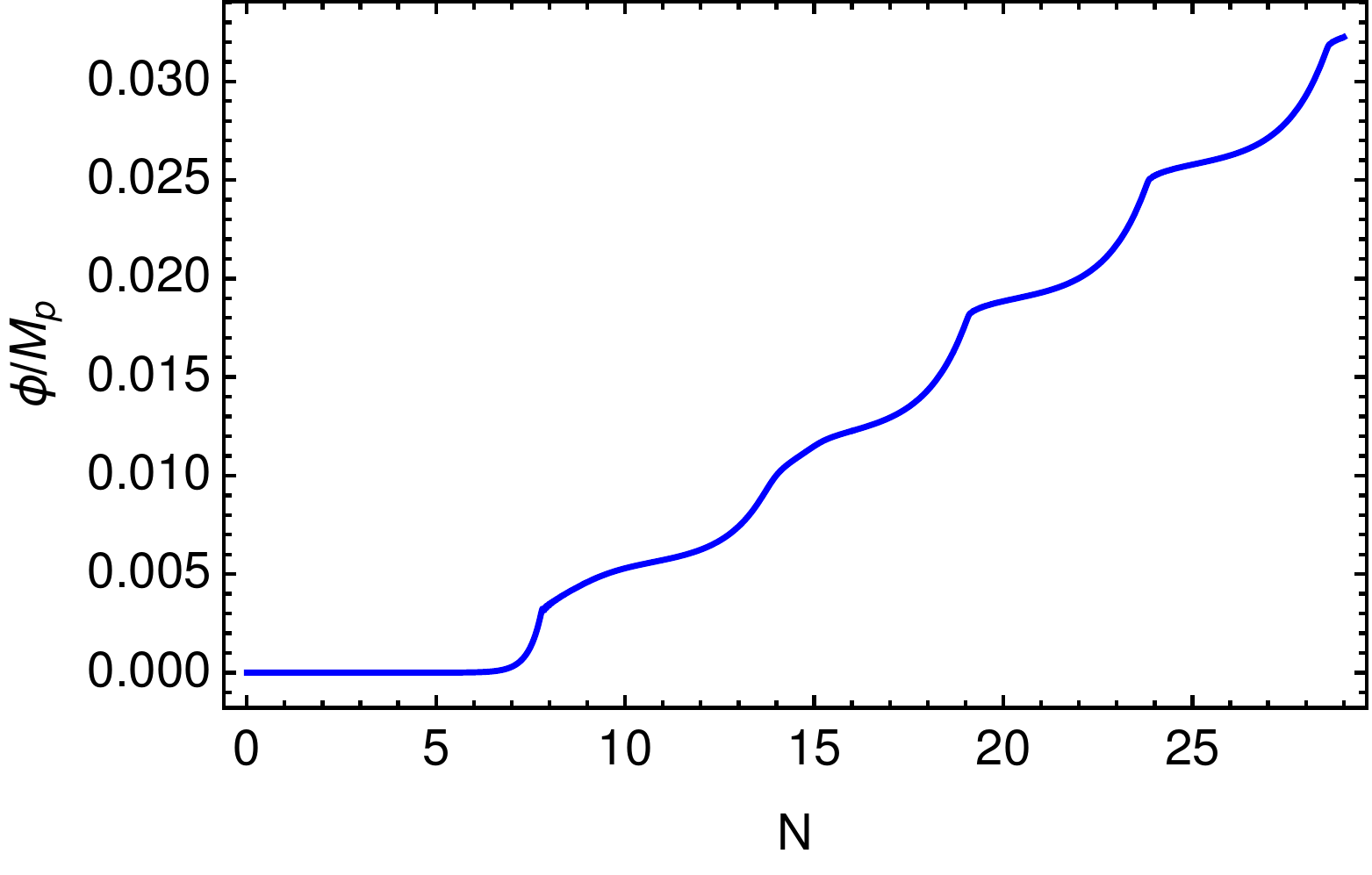}
\includegraphics[width=0.5\textwidth,angle=0]{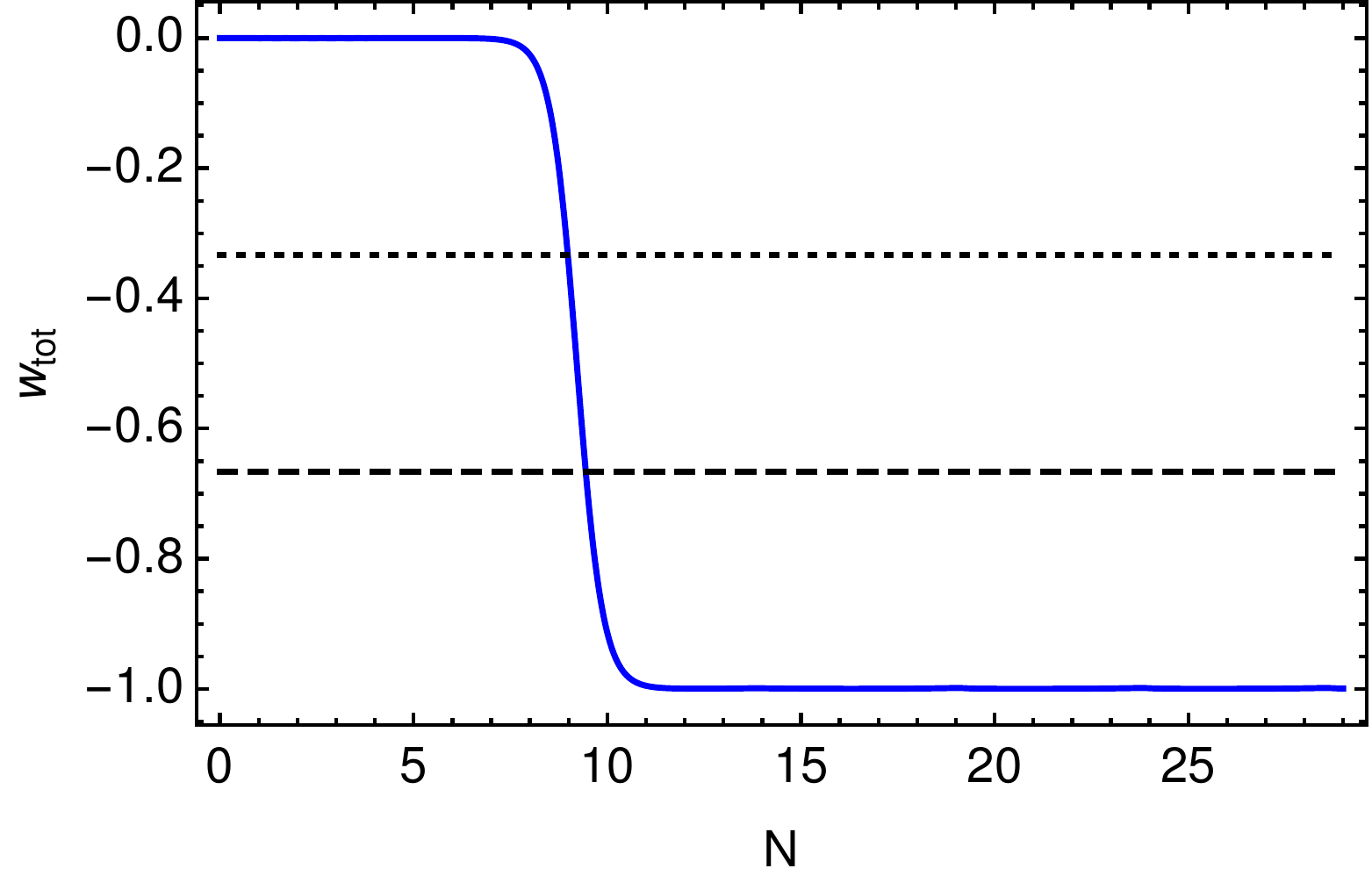}
}
\centerline{
\includegraphics[width=0.5\textwidth,angle=0]{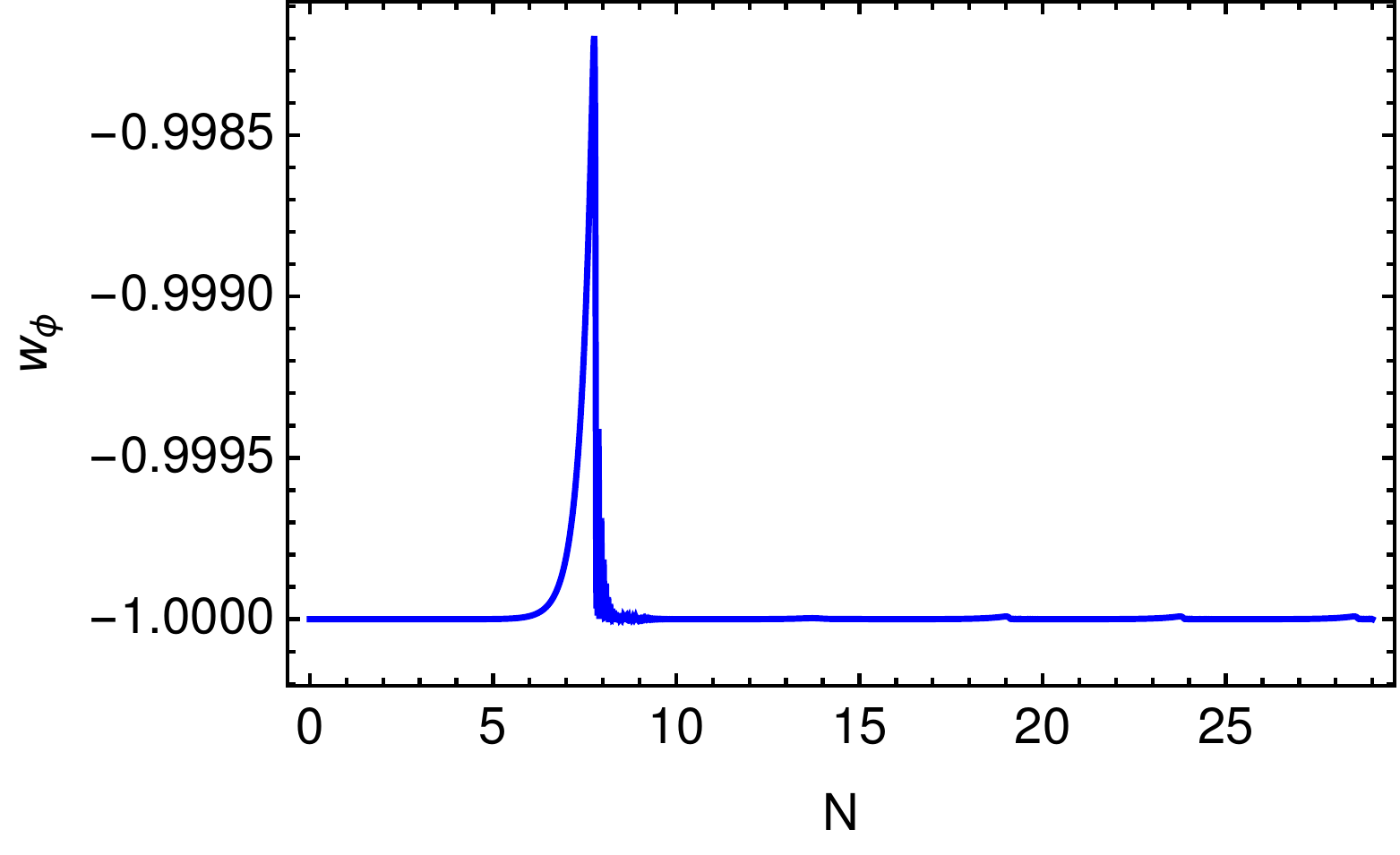}
\includegraphics[width=0.5\textwidth,angle=0]{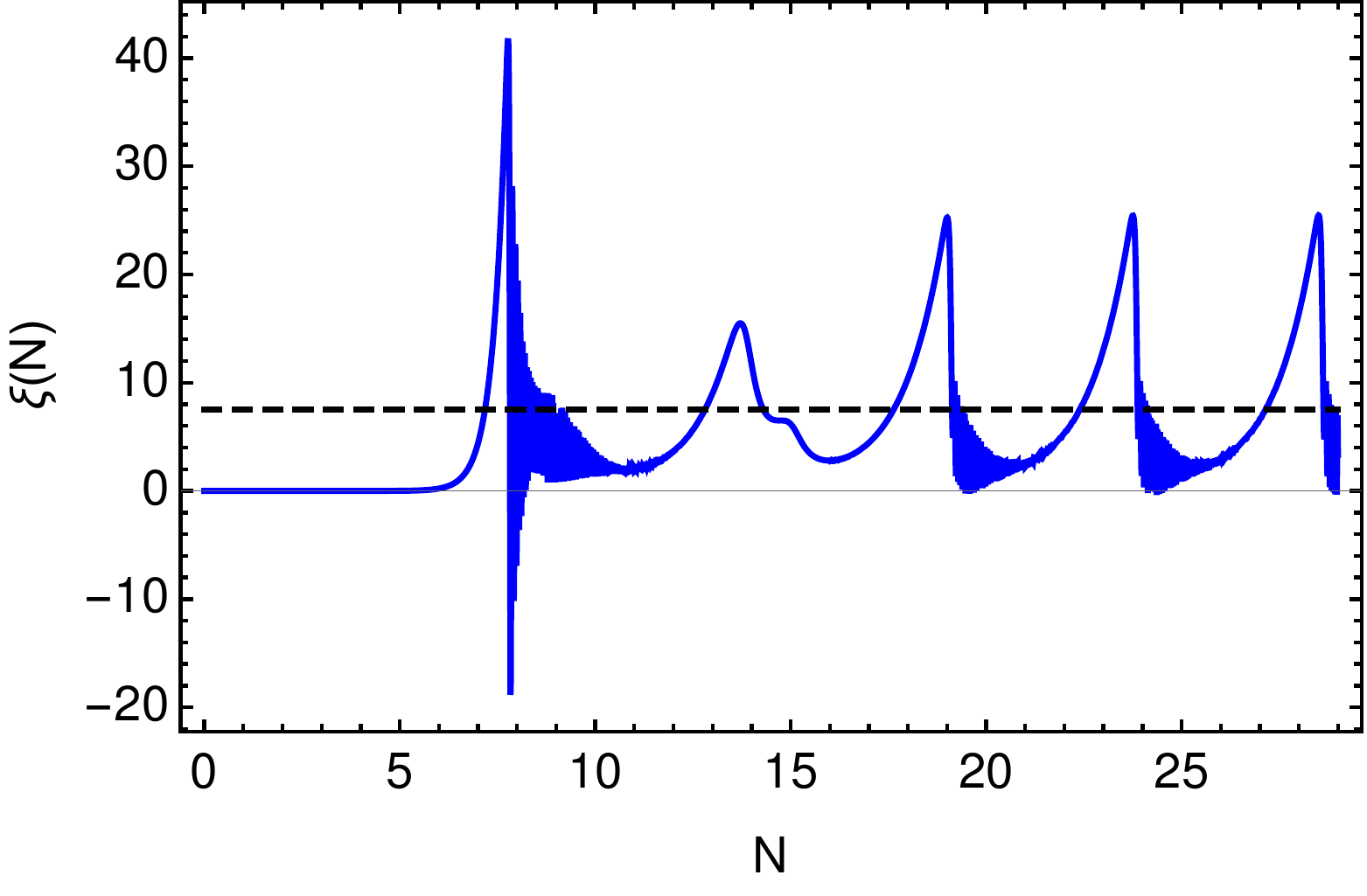}
}
\caption{The parameters used for this run are $V_0= 10^{-20} M_p^4$, $\bar{\rho}_\text{m}=10^{12}$, $\tilde{f}=10^{-4}$ and $w=0$.
\textit{Top left panel}: Evolution of the field $\phi$ in units of $M_p$.
The backreaction is negligible until $N\sim 7.8$.
After that the field undergoes a series of steps in its evolution.
\textit{Top right panel}: The blue line is the total equation of state parameter, the dotted line denotes the threshold for accelerated expansion and finally the dashed line denotes the asymptotic value of the state parameter in the absence of backreaction.
\textit{Bottom left panel}: Evolution of the state parameter of the scalar field.
\textit{Bottom right panel}: Evolution of the particle production parameter defined in (\ref{A-eom}).
The dashed line that is superimposed denotes the value of $\xi$ as predicted by the AS solution (\ref{AS-attractor}).
One can observe a series of spikes after $N \simeq 16$ that appear to be self similar.}
\label{fig:exponential-AS}
\end{figure}

\begin{figure}[ht!]
\centerline{
	\includegraphics[width=0.33\textwidth,angle=0]{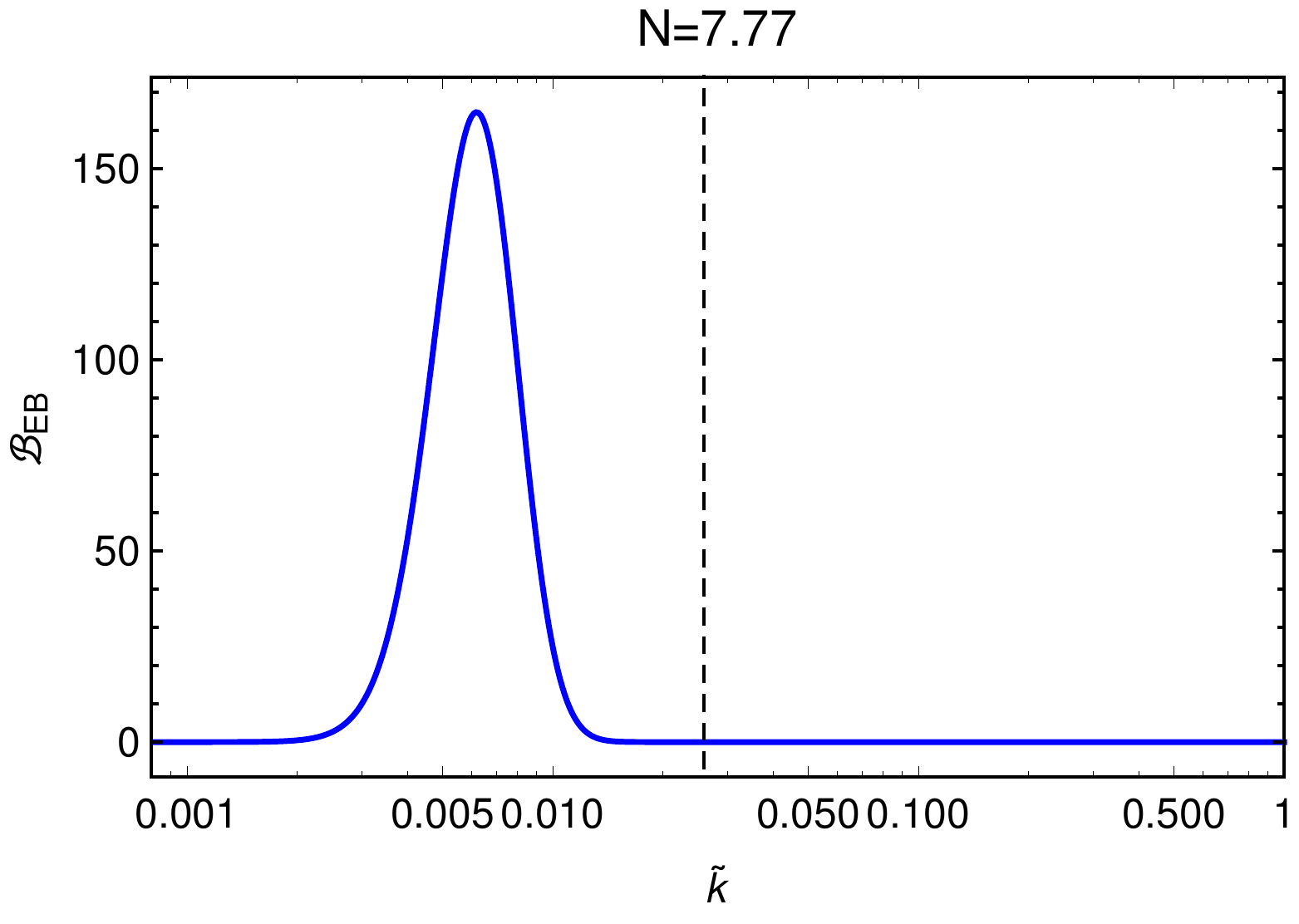}
	\includegraphics[width=0.33\textwidth,angle=0]{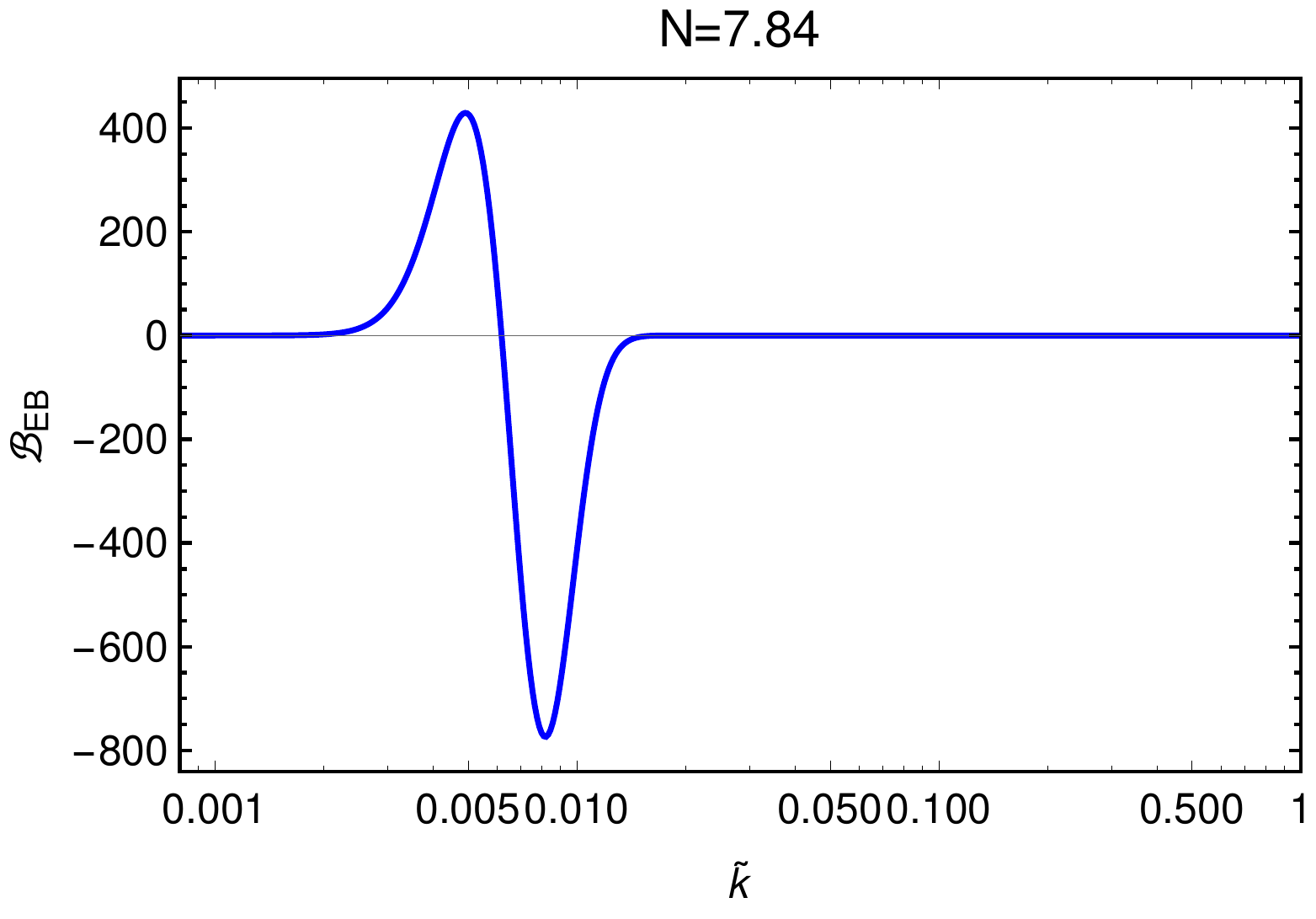}
	\includegraphics[width=0.33\textwidth,angle=0]{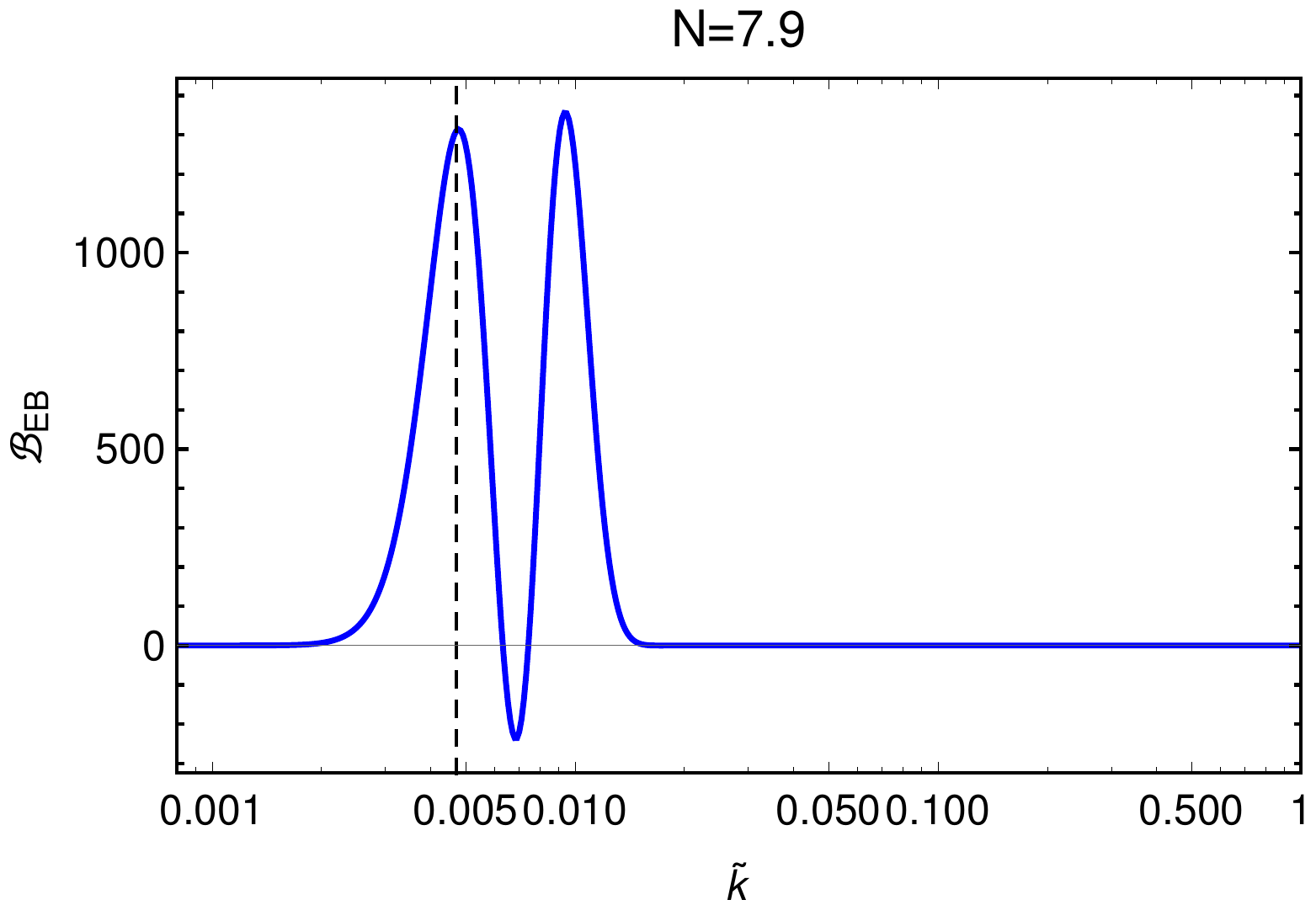}
}
\centerline{
	\includegraphics[width=0.33\textwidth,angle=0]{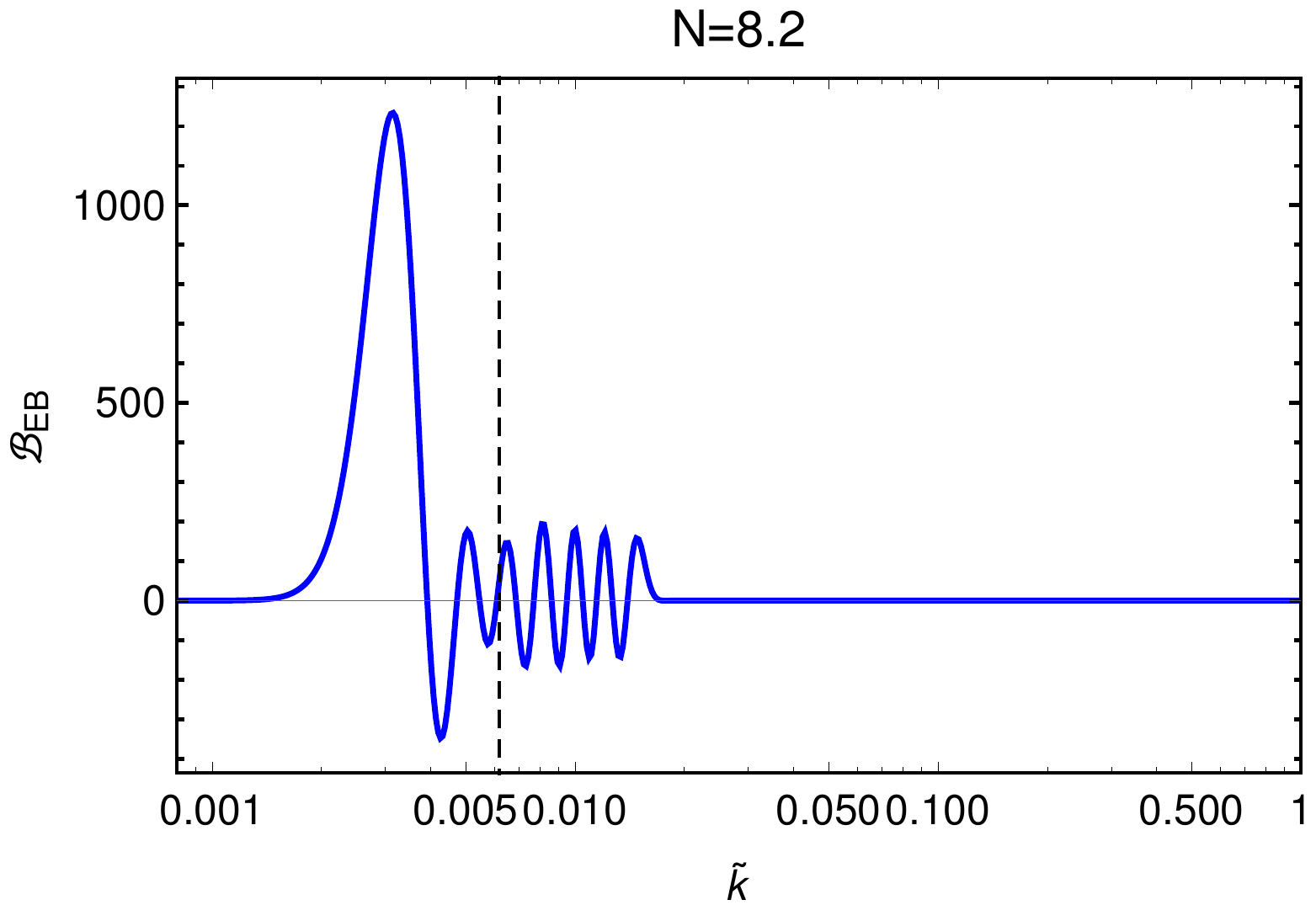}
	\includegraphics[width=0.33\textwidth,angle=0]{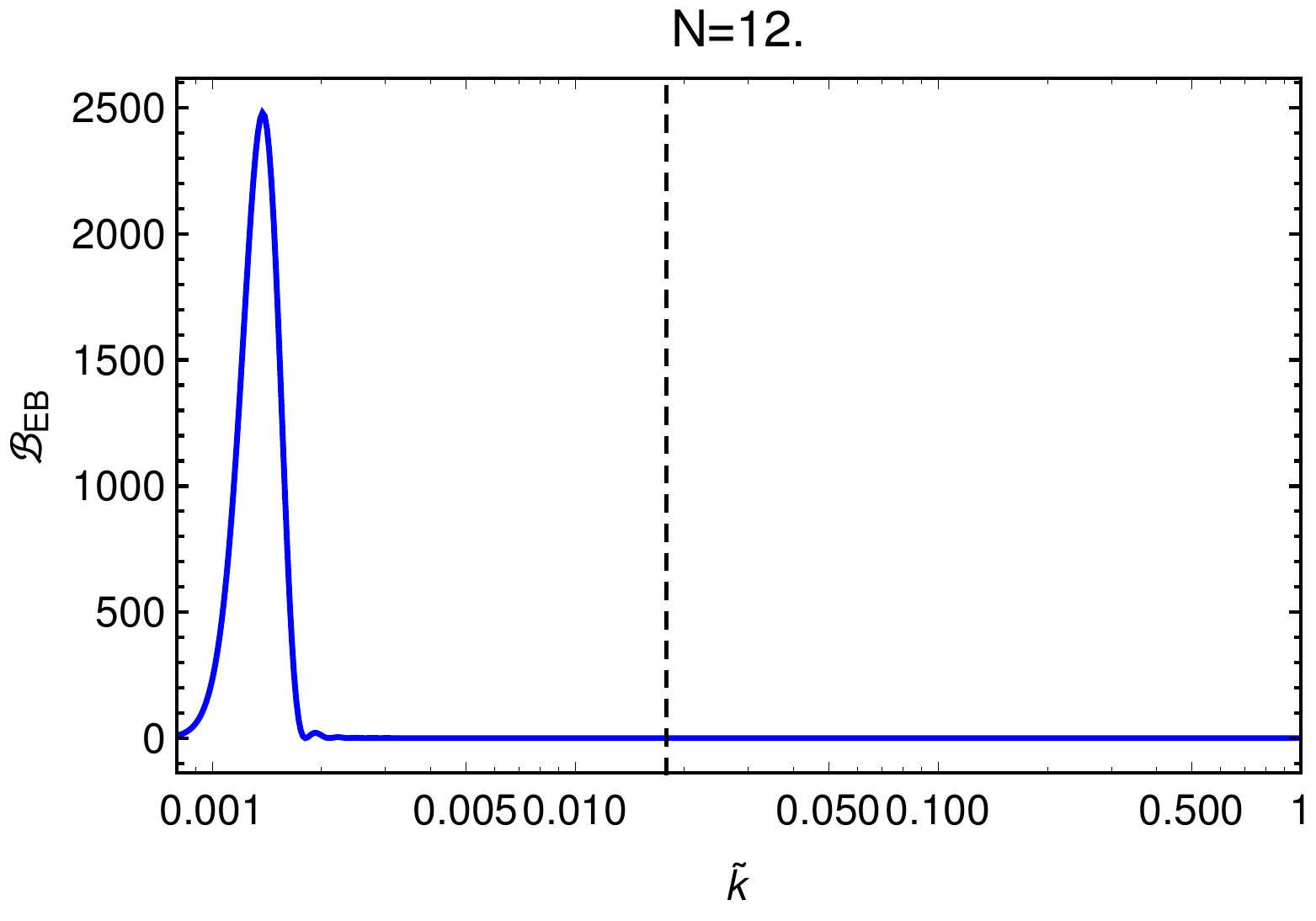}
	\includegraphics[width=0.33\textwidth,angle=0]{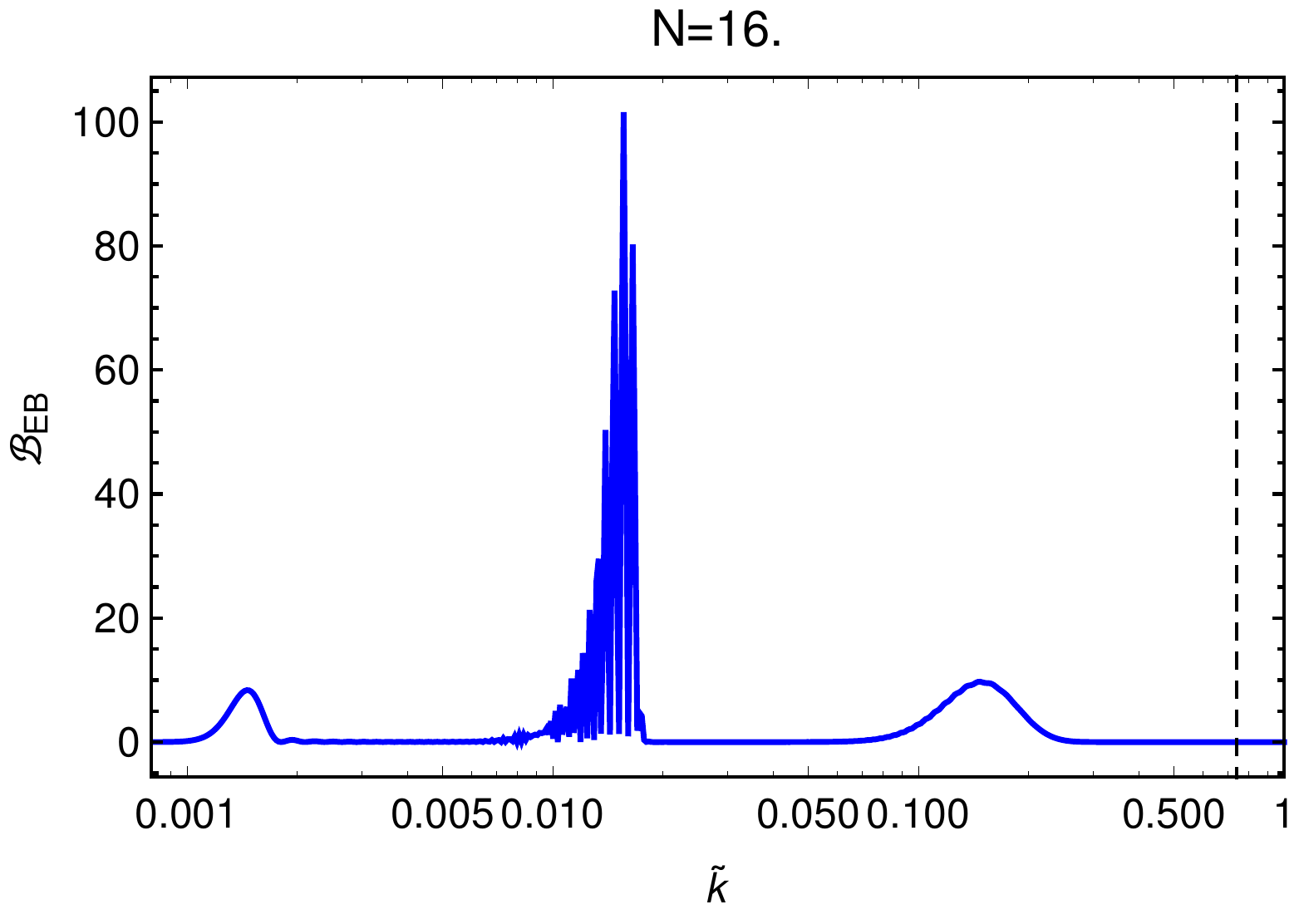}
}
\caption{The normalized integrand defined in (\ref{back}) for various times in the evolution (where time is parametrized by the number of e-folds $N$).
These panels correspond to the same run shown in figure \ref{fig:exponential-AS}.
The vertical dashed lines denote the exact threshold between the stable and unstable modes at the time shown (in the top-middle panel the threshold is at a momentum smaller than the range shown, as explained in the text).}
\label{fig:exponential-integ}
\end{figure}

We discuss here the results obtained for the run discussed in the previous subsection, for the same potential (\ref{potential}) that we have considered so far in this work (choosing $\lambda =1$, and $V_0= 10^{-20} M_p^4$, as described above).
For completeness we performed the analysis assuming an equation of state of matter ($w=0$) as well as radiation ($w=\frac{1}{3}$) for the initially dominating fluid.
The two cases provide qualitatively similar results, with the main (trivial) difference that, starting from the same ratio between the energy density of the initial fluid and that of the scalar field, the transition from fluid domination to inflation takes place sooner in the case of initial radiation domination.
We show in figures \ref{fig:exponential-AS} and \ref{fig:exponential-integ} some of the key quantities obtained in the evolution for the first case (initially, $w_{\rm tot} =0$).

The first run evolved from $N=0$ until $N=16$ e-folds, including gauge field modes with momenta in the range $8 \cdot 10^{-4} \le \tilde{k} \le 1$.
The second run was performed from $N=16$ until $N=29$ with momenta $8\cdot 10^{-4} \le \tilde{k} \le 6\cdot 10^6$.

An interesting feature that characterizes the later inflationary stage, is the appearance of different stages in which the scalar field evolves more slowly, separated by quicker stages of faster evolution.
This is visible in the top-left panel of figure \ref{fig:exponential-AS}, which shows the evolution of the scalar field, as well as in the bottom-right panel, which shows the evolution of the gauge production parameter $\xi \equiv \frac{\partial_\tau \phi}{2 a f H}$.
The evolution shows signs of self-similarity, and the different spikes in $\xi \left( N \right)$ (each spike corresponding to a stage of enhanced $\dot{\phi}$) have a nearly identical amplitude.
The dashed line in the bottom right panels of both figures denotes the value for the particle production parameter obtained from the Anber--Sorbo analytical computation reproduced in our eq.~(\ref{AS-attractor}).
Qualitatively speaking, we see that the Anber--Sorbo predicted value is achieved in an ``averaged'' way, but that the evolution does not reach a stage in which eq.~(\ref{AS-attractor}) is valid at all times.
We have also experimented with various values of parameters in the code, and, particularly, with the amplitude of the potential varying between $V_0=10^{-20} M_p^4$ and $V_0=10^{-120} M_p^4$.
This patterns repeated in all cases that we studied, with a frequency that decreases with decreasing value of $V_0$.

The Anber--Sorbo solution denotes the case in which the effect of gauge particle production and the backreaction to the equation of motion of the scalar field are in perfect equilibrium with each other.
In our case, the transition to inflation is achieved far away from that equilibrium, as testified by the presence of the spikes in the gauge production parameter $\xi $.
 Another way to describe the observed effect is the following.
If, at some point of the evolution of the scalar field the gauge field production parameter is below the Anber--Sorbo solution, that implies that there is less friction in the equation of motion of the scalar field and therefore the scalar field accelerates.
Since the speed of the field increases, the gauge field production also increases.
The production keeps growing until it overshoots the equilibrium level defined by the AS solution.
This leads to a quick increase of the backreaction, that then significantly slows down the scalar field and the gauge production.
This leads to a quick decrease of the gauge field amplification.
The gauge modes produced up to that moment are then subject to redshift, which decreases the backreaction, and it allows the scalar field to speed up again.
This process repeats itself and it does not show any sign of relaxation to an equilibrium (constant or nearly constant $\dot{\phi}$ state) during the evolution that we could cover in our numerical integration.

The six panels of figure \ref{fig:exponential-integ} demonstrate how the integrand defined in (\ref{back}) varies with time.
The first four panels are chosen to be around the transition time $N\simeq 7.8$ whereas the last two show later stages of the evolution.
The vertical dashed line is plotted at the exact threshold between the stable and unstable modes at the time shown.
In the top left panel the backreaction is still subdominant, and one observes a similar shape for the particle production as displayed in figure \ref{fig:integrand}.
In the top middle panel the backreaction has just become dominant and the speed of the inflaton has just decreased sharply.
As a consequence the threshold between stability and instability decreased as well to a value on the left of the range shown in the panel\footnote{As we discussed, the threshold between stable and unstable modes grows adiabatically while the scalar field moves slowly due to Hubble friction, and backreaction effects are subdominant.
Due to this one might be tempted to includee in the numerical evolution only the gauge modes that are unstable at any one given time.
The sharp decrease of the threshold that takes place between the first two panels shows that this would be a mistake.
The third panel shows modes that were previously amplified, and that are now from some time in the stability region.
Their contribution to the backreaction is highly significant at this stage.}. 
The threshold between stability and instability then proceeds to fluctuate wildly for the next $\Delta N\sim 1.5$ e-folds\footnote{This is a consequence of the rapid oscillations of $\dot{\phi}$.
The same qualitative behavior can be observed in the last panel of Figure \ref{fig:samplerun}).} with a positive average value that is slightly greater than $0.002$.
For this reason, the contribution to the backreaction from the modes at $\tilde{k}\simeq 0.002$ continues to grow with respect to what shown in the first three panels.
This growth is visible in the fourth and fifth panel.
On the contrary, the modes corresponding to the second positive peak in the third panel are stable during the evolution between the fourth and the fifth panel, and their contribution to the backreaction decreases (due to cosmological redshift).
In fact, we observe that all the contributions with momentum greater than $\tilde{k}\simeq 0.002$ have been redshifted away as $a^{-4}$ and they are too small to see in a linear scale\footnote{Modes with $\tilde{k} \ll 0.002$ are formally unstable; namely, they have a tachyonic effective frequency in eq.~(\ref{eq-A}).
However, due to the smallness of $k$, this instability is extremely mild, and it is dominated by the dilution due to the redshift.}.

Finally, we can make a number of observations on the last panel shown in the figure: firstly, the bump around $\tilde{k}\simeq 0.002$ has decreased due to cosmological redshift.
Secondly, we notice an enhancement present at around $\tilde{k}\simeq 0.010$.
These are the highest modes that were enhanced immediately before the backreaction became important.
These modes then became stable (and therefore their amplitude oscillated) while simultaneously redshifting away.
Due to the redshift, the contribution of these modes does not appear to differ from zero in the fifth panel.
Nonetheless, their amplitude is greater than that of the surrounding modes, and it acts as a seed for the amplification that leads to the highest bump seen in the last panel.
Thirdly, we observe a new emerging bump at around $\tilde{k}\simeq 0.2$.
This enhancement corresponds to newly unstable modes that are enhanced from their vacuum configuration only in the latest stages of the evolution shown.
This is the bump that grows the fastest at the times immediately after those shown in the figure, and it will eventually become the single dominant bump, much like as in the first panel of the figure.
This generates a new cycle in the evolution, in which the features shown in these panels repeat themselves for the higher, ${\tilde k} = {\rm O } \left( 10^{1} \right)$, comoving momenta.

\subsection{Numerical results for a sinusoidal potential}

In this subsection we want to examine whether the (approximate) self-similar inflationary evolution shown in Figure \ref{fig:exponential-AS} is a consequence of the assumed exponential potential.
We therefore replace the potential with a sinusoidal one, 
\begin{equation}
V\left(\phi\right)=V_0 \left[\cos \left(\frac{\phi}{f}\right)+1\right] \;, 
\end{equation}
which is more typical for axion fields, and that was also chosen by Anber ad Sorbo \cite{Anber:2009ua}.
Since $f$ now denotes the scale of the potential, we follow \cite{Anber:2009ua} and we parametrize the coupling in the topological term with an additional positive parameter $\alpha$, 
\begin{equation}
{\cal L} \supset -\frac{\alpha}{4 f}\,\phi\,F^{\mu\nu} \tilde{F}_{\mu\nu} \;. 
\label{AS-coupling} 
\end{equation} 
To study the proposed Anber--Sorbo solution, we assume the presence of ${\cal N}$ Abelian gauge fields, all coupled to $\phi$ with an identical coupling (\ref{AS-coupling}).
Quite nontrivially, the presence of ${\cal N}$ gauge fields provides an increased friction term also in the evolution equation for the perturbation of $\phi$, leading to a nearly scale invariant (as long as the Anber--Sorbo solution is achieved) spectrum of the primordial density perturbations with amplitude $P_\zeta \simeq \frac{0.05}{{\cal N} \xi^2} $ \cite{Anber:2009ua}.
If this mechanism is realized in nature, the combination ${\cal N} \xi^2$ therefore needs to be fixed so to produce the observed value $P_\zeta \simeq 2.1 \cdot 10^{-9}$ \cite{Akrami:2018odb}.
In the presence of ${\cal N}$ gauge fields, equation (\ref{AS-attractor}) should also present the rescaling $\alpha \rightarrow {\cal N} \, \alpha$, so that we need to enforce 
\begin{equation}
\ln \left[\frac{9}{{\cal I} \,\alpha\, {\cal N}}\frac{M_p^4\, f\, |V'\left(\phi\right)|}{V^2\left(\phi\right)}\right] \simeq \frac{3 \cdot 10^4}{{\cal N}^{1/2}} \;.
\label{AS-Nfields}
\end{equation} 
We fix $\xi \simeq 20$ as the benchmark value chosen in the discussion by Anber and Sorbo.
This then requires ${\cal N}= 6\cdot 10^4$ identically coupled gauge fields.
Eq.~(\ref{AS-Nfields}) can be then turned into a relation for the scale of the potential: 
\begin{equation}
V_0 \simeq \frac{1.68 \cdot 10^{-55} \, \sin\left(\frac{\varphi}{\tilde{f}}\right)}{\alpha \left[1+\cos\left(\frac{\varphi}{\tilde{f}}\right)\right]\,} M_p^4 \;, 
\end{equation}
where $\varphi$ needs to be evaluated when the CMB modes left the horizon.
We choose, $\varphi \simeq \tilde{f}$, and, after parametrically identifying the value of the potential with the energy scale of inflation, we find that this corresponds to ``low-scale'' inflationary scale $V_0 = {\cal O } \left( 10 \, {\rm TeV} \right)$.

\begin{figure}[ht!]
\centerline{
	\includegraphics[width=0.5\textwidth,angle=0]{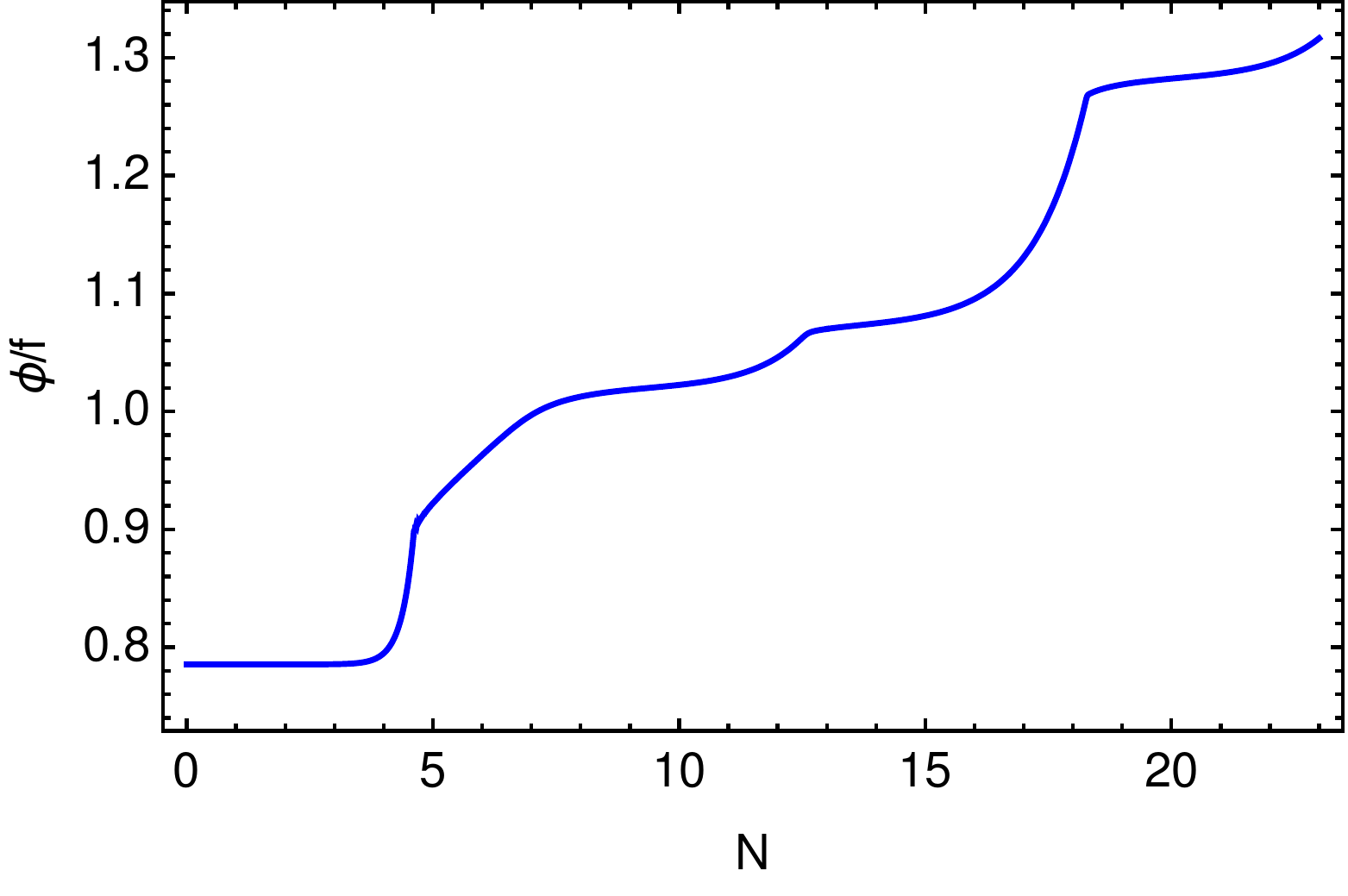}
	\includegraphics[width=0.5\textwidth,angle=0]{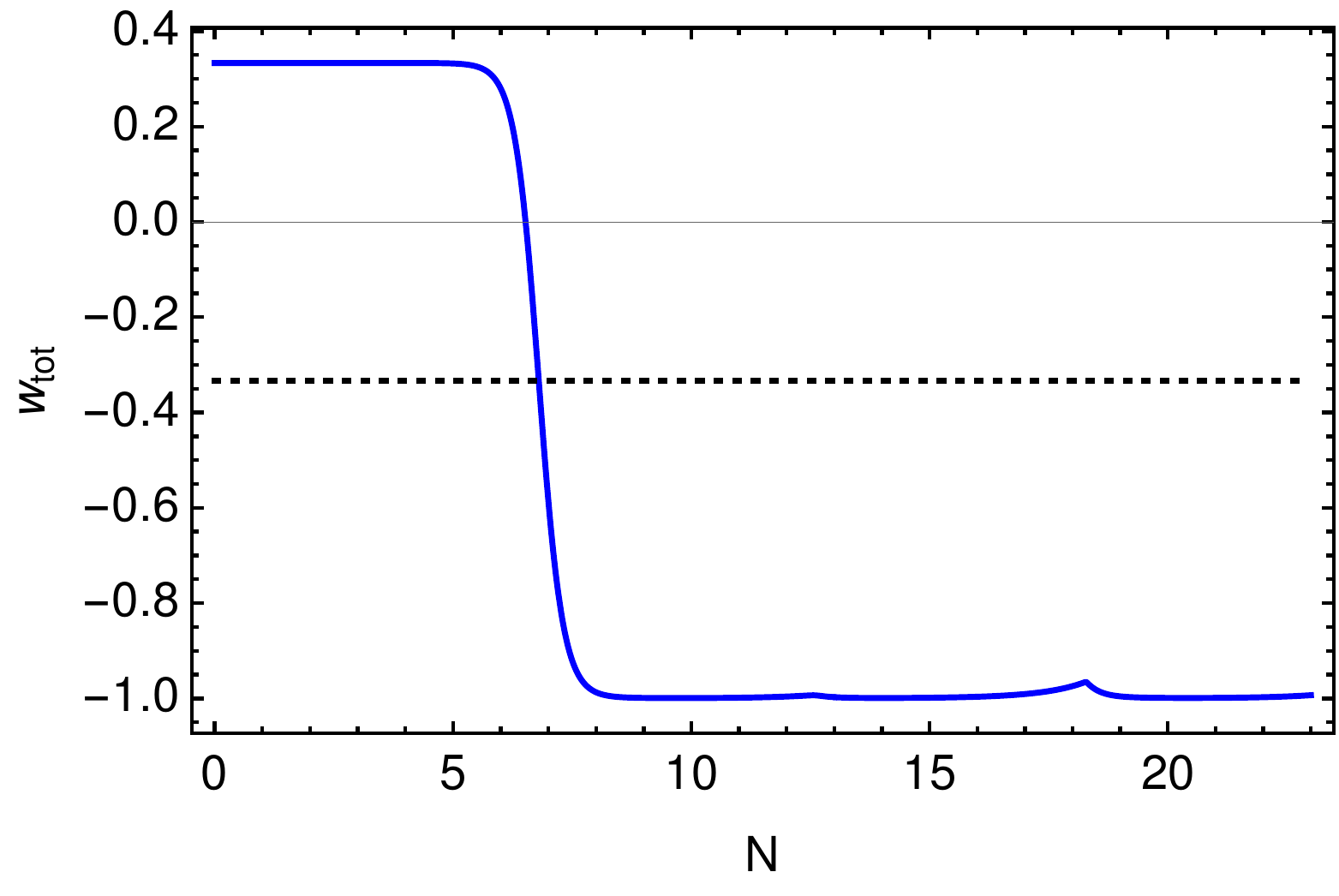}
}
\centerline{
	\includegraphics[width=0.5\textwidth,angle=0]{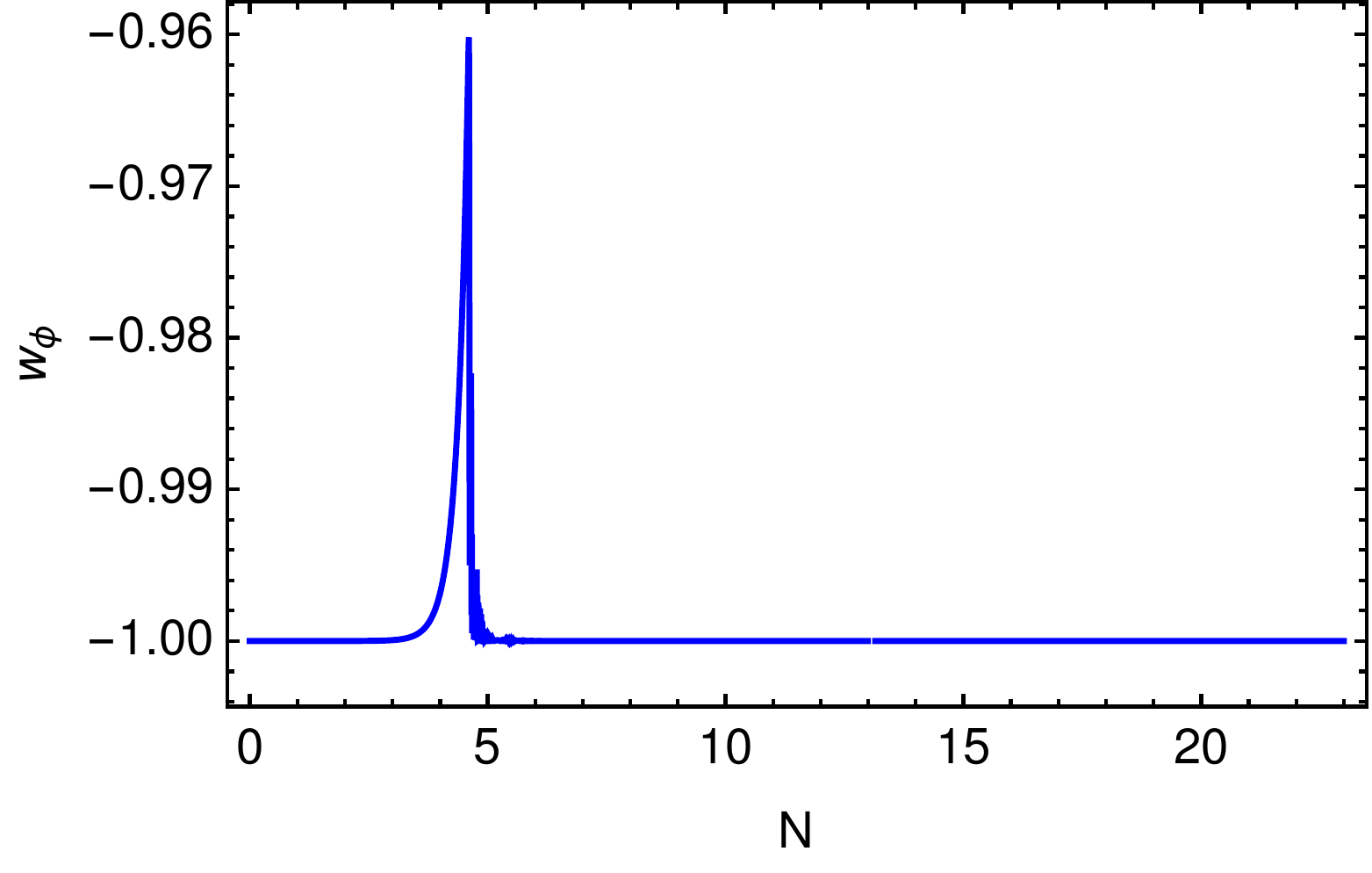}
	\includegraphics[width=0.5\textwidth,angle=0]{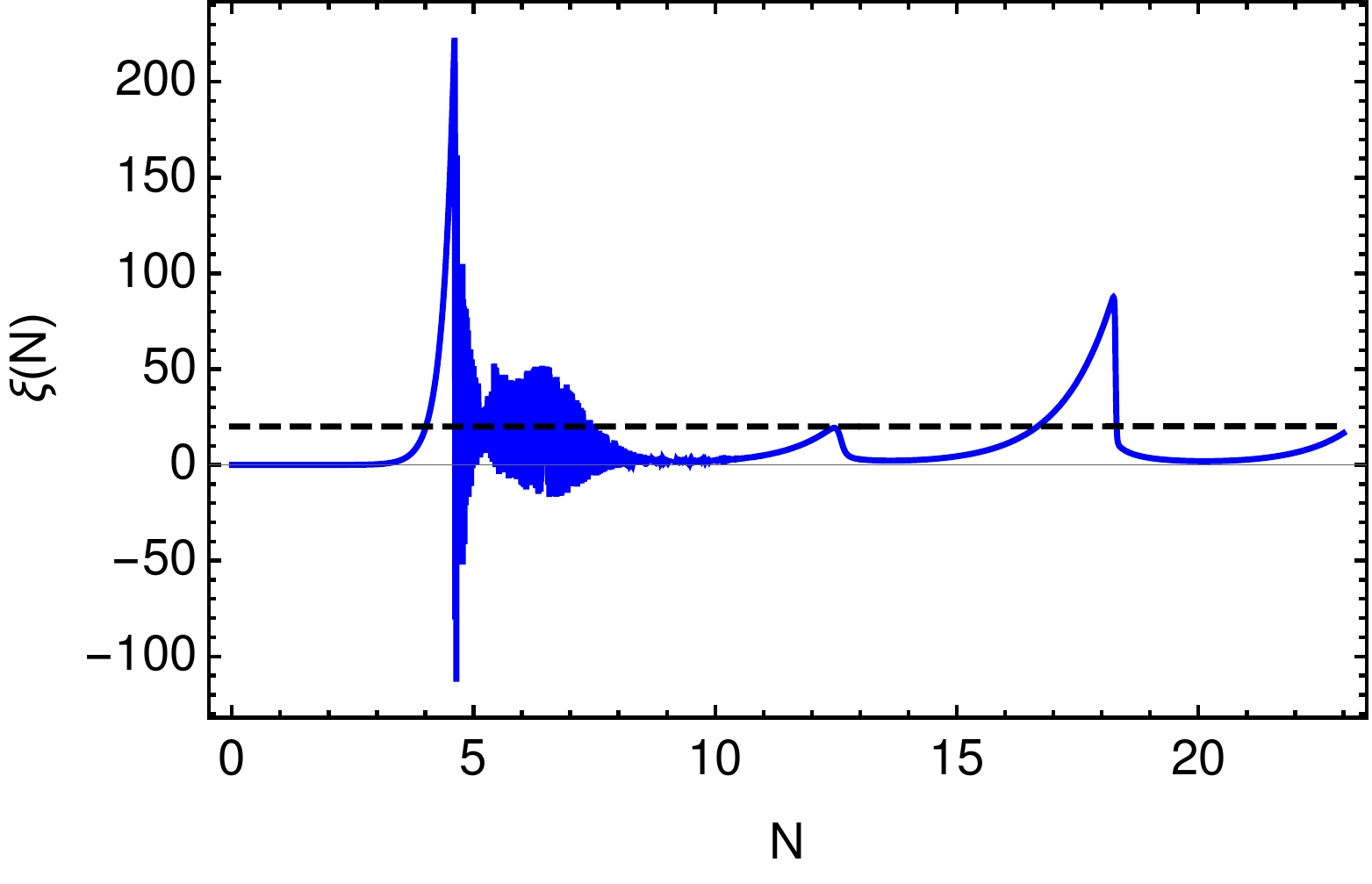}
}
\caption{
The parameters used for this run are $V_0=6\cdot 10^{-59} M_p^4$, $\bar{\rho}_m=10^{12}$, $\tilde{f}=10^{-2}$, $w=1/3$ and $\alpha=10^3$.
\textit{Top left panel}: We observe similar steps in the evolution of the field as in the case of the exponential potential.
\textit{Top right panel}: The equation of state of the Universe evolves from radiation domination to inflation.
\textit{Bottom left panel}: Equation of state of the scalar field.
After the backreaction becomes important the scalar field behaves like an effective cosmological constant.
\textit{Bottom right panel}: The particle production parameter displays similar features to the exponential potential.
The transient effects are different but the "spikes" of particle production are still present.
Again we can see that the AS-Solution (dashed line) is qualitatively achieved in an average way.}
\label{fig:sin-ASrad}
\end{figure}

We modified our numerical code to the choices of parameters just described, and performed a numerical evolution, starting for definiteness from $\phi_{\text{in}}=\frac{\pi}{4}f$.
In Figure \ref{fig:sin-ASrad} we display the results for the case of transition from radiation domination to inflation.
The code was separated in two runs like in the case of the exponential potential.

The first run was performed from $N=0$ until $N=13$ e-folds and in a range of momenta between $2 \cdot 10^{-6} \le \tilde{k} \le 1\cdot10^{-3}$.
The second run lasted from $N=13$ until $N=25$ and the range of momenta was $ 2 \cdot 10^{-6} \le \tilde{k} \le 50$.

Unfortunately, running the code deeper into inflation would require including higher value of the momenta, which slows the numerical evolution to a prohibitive level.
However, the amount of inflation that we could cover shows that the bursts of particle production are also present for the case of the sinusoidal potential, and that the AS inflationary solution is reached only in a (qualitatively) average fashion.
The spikes in this case appear with a slower frequency than in the case of the exponential potential.
This is a consequence of the smallness of the potential rather than the shape of the potential.
We also experimented with various values of $V_0$ between $\sim 10^{-20} M_p^4$ and $\sim 10^{-120} M_p^4$.
These increased values are inconsistent with the CMB normalization, but they allow to obtain more bursts of gauge field amplification in the numerical integration.
We could verify that also in this case sufficiently large values of $V_0$ lead to a qualitatively self-similar spikes that are analogous to the ones that appear in the exponential potential example (although clearly, in this case, the self similarity will disappear when $\phi$ approaches the minimum of the potential).

\section{Conclusions} 
\label{sec:conclusions} 

Cosmological data suggest that the Universe underwent two periods of accelerated expansion: one responsible for setting the initial conditions (background and perturbations) of the ``standard'' big-bang era, and the current one.

The observed cosmological perturbations require that the primordial expansion is not an exact de Sitter one.
This is successfully explained by models of slow-roll scalar field inflation.
On the contrary, the cosmological data are consistent with the assumption that the present Universe is dominated by a vacuum energy density, and strongly constrain any departure from this simplest possibility \cite{Akrami:2018odb}.
Despite of this, our inability to naturally obtain such a small vacuum energy as the one required by the Universe has motivated the construction of quintessence models where the vacuum energy is set to zero (typically, by some unspecified symmetry), and the current expansion is due to some additional source, denoted by ``dark energy''.
Borrowing from the success of scalar field inflation, the simplest models of dark energy make use of scalar fields rolling in sufficiently flat potentials.
Of particular interest are models of tracking quintessence, in which the evolution of the scalar field tracks that of the dominant background source (matter and radiation) for most of the cosmic history, and emerges only at late times \cite{Steinhardt:1999nw}.
The tracking mechanism can allow for an explanation of the coincidence problem, namely why the current value of the dark energy is comparable to that of dark matter.

Also in these interesting models, however, the mass of the quintessence field (namely, the curvature of the potential) needs to be set to an extremely small value (as compared to typical scales that arise in particle physics) set by the current Hubble parameter $H_0 = {\rm O } \left( 10^{-42} \right) M_p$.
In addition, as already discussed in the introduction, several theoretical considerations appeared in the recent literature pointing to the requirement that the slope of the scalar potential might be too steep to support accelerated expansion.
This observation provides one of the main motivations for the present analysis.
Also in this case we borrow an idea that, despite not being as popular as standard slow-roll inflation, is well known in the inflationary community, namely that of warm inflation \cite{Berera:1995ie}.
In this framework, the scalar field potential is too steep to lead to accelerated expansion, if one accounts only for the Hubble friction.
This problem is overcome by the fact that the scalar field is coupled to some additional field.
The coupling, together with the motion of the scalar field, leads to the production of this additional field, at the expenses of the kinetic energy of the rolling scalar.
This can significantly slow down the rolling scalar, allowing for an effective equation of state of the scalar that can provide accelerated expansion.
The more recent literature has implemented this idea in several relatively simple QFT constructions, as for instance the Anber--Sorbo mechanism \cite{Anber:2009ua}, trapped infaltion \cite{Green:2009ds} (see also \cite{Pearce:2016qtn} for a reanalysis of cosmological perturbations in this model) and chromo-natural inflation \cite{Adshead:2012kp}.
Our main goal is to implement the mechanism of \cite{Anber:2009ua} in the context of the present expansion.

In this mechanism, the rolling field is a pseudo-scalar $\phi$, coupled to a U(1) field via the topological term $\phi F {\tilde F}$.
In the first part of this paper we present a novel implementation of this mechanism in the context of extended supergravity models.
Most phenomenologically viable models coming from String Theory assume a compactification to 4 dimensions by means of a Calabi--Yau (CY) manifold as internal space, supplemented by the presence of branes and orientifolds.
The final low-energy effective theory is given by N=1 supergravity, but of a restricted type.
In fact the closed string sector modes experience the underlying CY geometry and their interactions are therefore better described by an N=2 model, possibly further restricted by the geometric details of the chosen CY.
We show that warm dark energy scenarios can be constructed in this N=2 setup by choosing a CY with specific intersection numbers that allow for the appropriate couplings between the vector field and the quintessence scalar (see eqs.~(\ref{cubicprep}) and (\ref{ddd})).
By introducing an orientifold projection we can also consistently embed in this scenario almost any scalar potential via an appropriate choice of superpotential, though some tuning may be required.
While this does not provide a fully consistent uplift of warm dark energy models within string theory, we have clearly identified most of its necessary ingredients.

We account for the backreaction of the produced gauge field by solving the system numerically.
As we disregard the perturbations of  the scalar field in this paper, we can write linear exact equations (in the limit of homogeneous $\phi$) for the U(1) vector field mode functions.
We solve these equations for a dense grid of modes in momentum space.
Focusing for definiteness on an exponential potential, we verified that a sufficiently strong $\phi F {\tilde F}$ interaction 
can indeed successfully account for the current acceleration of the Universe.

The current accelerated stage has been taking place for only about one e-fold of expansion.
Therefore, to perform this study it is enough to reach only the beginning of the accelerated stage.
In our numerical simulations the expansion is initially dominated by the energy density of matter.
This strongly suppresses the motion of the scalar field (just due to Hubble friction) and the amplification of the gauge fields. \footnote{In fact, due to the dominating matter field, the gauge modes are initially stable at sufficiently high momenta and this ensures that the problem is well regulated in the UV.
Moreover, the expansion of the Universe dilutes the energy density of the produced gauge field when their momentum redshifts to small values, so that this mechanism is also regulated in the IR.} The scalar field comes eventually to dominate the energy density of the Universe, as the energy of matter decreases due to the expansion.
Compatibility with data in a steep potential requires that the backreaction from the produced gauge fields becomes important before the transition from matter to scalar field domination, allowing for at least an e-fold of sufficiently accelerated expansion.

While not necessarily for this quintessence application, we nonetheless exploited our numerical code to also study the evolution well inside the accelerated regime.
As it is well known, simulations of inhomogeneous fields during inflation on a fixed grid are extremely expensive since, due to the rapid growth of the scale factor, they must cover a progressively larger and larger dynamical range as the inflationary expansion goes on.
Therefore we could only simulate a limited number of e-folds of accelerated expansion.
The simplest expectation from this study is that, in the strong backreaction regime, the background evolution should approach that of the Anber--Sorbo solution \cite{Anber:2009ua} (in a very different energy regime, in the quintessence application, or in  the original regime, if we change the parameters of the code to simulate the primordial inflationary stage).
The Anber--Sorbo solution is characterized by an adiabatically evolving inflaton speed, and by a nearly identical evolution of the different gauge modes.
Specifically the evolution of a gauge mode of comoving momentum $k_1$ at the time $t_1$ is nearly identical to that of a mode of comoving momentum $k_2$ at the time $t_2$ when we plot it in terms of the physical momentum of the two modes, namely when $\frac{k_1}{a \left( t_1 \right)} = \frac{k_2}{a \left( t_2 \right)}$).
Therefore, in this solution, the system evolves in a nearly steady state fashion, which is the typical slow-roll inflationary expectation, characterized by only a very small breaking of time invariance (this is what generates nearly scale invariant signals).
 
On the contrary, the inflationary stages that emerged from our simulations are characterized by a highly nontrivial evolution of $\phi \left( t \right)$.
As we approach the transition between matter and scalar field domination, the scalar field starts accelerating.
The backreaction from the produced gauge fields, while still subdominant, grows extremely rapidly with time, and it becomes dominant all of a sudden, giving rise to a large burst of gauge field amplification.
This suddenly slows down the gauge field, and actually makes it oscillate back and forth for some time.
In this phase, the energy density of the produced gauge field redshifts away due to the expansion of the Universe, and no new fields are efficiently produced, due to the decreased speed of $\phi$ (the gauge field amplification is exponentially sensitive to the speed of $\phi$, so even a small decrease of the speed can result in a very strong decrease of the gauge field production).
Due to this, the backreaction becomes rapidly subdominant again, and the scalar field starts accelerating once more.
The cycle we just described repeats itself over and over in our simulations, with no sign of convergence toward the Anber--Sorbo steady state solution.
A useful diagnostic for this study is the evolution of the parameter $\xi \left( t \right) \equiv \frac{\dot{\phi} \left( t \right)}{2 f \, H \left( t \right)}$, which controls the gauge field amplification.
While this quantity is nearly constant in the Anber--Sorbo solution, in our numerical evolutions $\xi \left( t \right)$ ``oscillates'' about the Anber--Sorbo value, without showing signs of convergence.

This evolution has some similarities with the transition between overshooting and undershooting evolution in tracking quintessence \cite{Steinhardt:1999nw}, with the big difference that in this second case the scalar field eventually reaches the steady state tracking solution, in which the Hubble friction balances against the slope of the potential.
As a consequence, the tracking solution is an attractor for these models.
The same is true for the inflationary solution in standard slow roll inflation.
Our numerical evolutions do show sign of this convergence for the Anber--Sorbo solutions.
Analogous ``oscillations'' between fast and slow evolution were also observed in refs.~\cite{Cheng:2015oqa}, that studied the transition during inflation between Hubble dominated friction and gauge field production dominated friction, and in refs.~\cite{Notari:2016npn}, that assumed that the scalar inflaton field had zero initial speed.
These studies were limited to the inflationary case.
One could have hoped that in our study, the extra friction due to the initially dominating matter field could have ``gently'' accompanied the system toward the steady state Anber--Sorbo evolution.
However, we could not see this, within the limits of what we could numerically simulate.
We believe that this is worth further study.

\vskip.25cm
\section*{Acknowledgements} 

We thank Massimo Pietroni and especially Lorenzo Sorbo for useful discussions.
The work of GD is supported in part by MIUR-PRIN contract 2017CC72MK003.

\vskip.25cm

\appendix

\section{Gauge field solutions, and early-time growth of the backreaction term}
\label{app:Asol-early}

In this appendix we derive the results for the backreaction of the vector field on the scalar field dynamics that we have presented in Subsection \ref{subsubsec:backgrowth}.
The starting point is eq.~(\ref{A-eq-early}) which assumes early time matter domination, and that the backreaction is negligible.
We rewrite this equation as 
\begin{equation}
\frac{d^2 {\hat A}}{d y^2} + \left( 1 - \epsilon \, y^5 \right) {\hat A} = 0 \;, \qquad y \equiv \frac{4 \, \lambda \, {\tilde k} \, {\tilde \tau}}{3 \, {\tilde f}} \;\;,\;\; \epsilon \equiv \left( \frac{3 \, {\tilde f}}{4 \, \lambda \, {\tilde k}} \right)^5 \, \frac{1}{\tilde k} \;\;,\;\; {\hat A} \equiv \sqrt{2 k} \, A \;.
\label{eomA-y}
\end{equation} 
where we have normalized the gauge filed mode function so to have magnitude $1$ at early (rescaled) times, $y \to 0$.
This is the equation of an oscillator with the time dependent frequency $\omega \left( y \right) \equiv \sqrt{1-\epsilon \, y^5}$.
At the initial time, the frequency varies adiabatically, $\frac{d \omega}{d y} \ll \omega^2$, and the early time evolution of ${\hat A}$ is well described by the adiabatic solution 
\begin{equation}
{\hat A}_{\rm adiabatic} = {\rm e}^{-i \int^y d y' \omega} = {\rm exp} \left[ - i \left( \frac{2 y \sqrt{1-\epsilon y^5}}{7} + \frac{5}{7} \, y \; _2F_1 \left[ \frac{1}{5} ,\, \frac{1}{2} ,\, \frac{6}{5} ,\, \epsilon y^5 \right] \right) \right] \;, 
\label{A-adiabatic}
\end{equation} 
where $_2F_1$ is an ordinary hypergeometric function (according to the usual vacuum, prescription, we keep only the positive frequency term).
The adiabaticity condition is violated at $y \simeq \epsilon^{-1/5}$, but it is then valid again at later times.
At these late times we therefore expect the solution to be a linear combination of the growing mode $ {\rm e}^{+ \int^y d y' \sqrt{-\omega^2}}$ and of the decreasing mode  $ {\rm e}^{- \int^y d y' \sqrt{-\omega^2}}$.
To obtain an accurate late time solution one should find the two coefficients of this linear combination, by studying the solution in the non adiabatic stage.

Our present goal is to obtain a rough estimate of the backreaction term, to estimate the range of parameters needed for the full numerical evolution (which, eventually, is the only computation that we want to perform accurately).
Therefore, we simply use the growing term as an estimate, 
\begin{equation}
	\begin{split}
{\hat A}_{\rm adiabatic ,estimate} &= {\rm e}^{\int_{\epsilon^{{-1/5}}}^y d y' \sqrt{\epsilon \, y^{'5} -1}} \\[2mm]
&= {\rm exp} \left[ \frac{2 y \sqrt{\epsilon y^5-1}}{7} - i \frac{5}{7} \, y \; _2F_1 \left[ \frac{1}{5} ,\, \frac{1}{2} ,\, \frac{6}{5} ,\, \epsilon y^5 \right] + \frac{5 i \sqrt{\pi} \, \Gamma \left( \frac{6}{5} \right)}{7 \, \Gamma \left( \frac{7}{10} \right) \, \epsilon^{1/5}} \right] \;, 		
	\end{split}
\label{adiab-est}
\end{equation} 
which is real (as it can be inspected directly).
The integrand in the $\vec{E} \cdot \vec{B}$ backreaction term is proportional to the time derivative of the magnitude square of the mode function, namely to $\frac{d}{dy} \vert {\hat A} \vert^2$, up to constant rescalings.

We computed this quantity for the (\ref{adiab-est}) estimate, and we compared it with the one obtained from a numerical integration of eq.~(\ref{eomA-y}).
We found that a much better agreement is obtained when $\frac{d}{dy} \vert {\hat A} \vert^2$ is divided by $\sqrt{\epsilon \, y^5 -1}$.
This leads to the approximation 
\begin{equation} 
\begin{split}
	\left( \frac{d}{dy} \left\vert A \right\vert^2 \right)_{\rm approx.} &\equiv \frac{\partial_y \left\vert A_{\rm adiabatic ,estimate}  \right\vert^2}{\sqrt{\epsilon y^5-1}} \\[2mm]
	&= 2 \,  {\rm exp} \left[ \frac{4 y \sqrt{\epsilon y^5-1}}{7} - i \frac{10}{7} \, y \; _2F_1 \left[ \frac{1}{5} ,\, \frac{1}{2} ,\, \frac{6}{5} ,\, \epsilon y^5 \right] + \frac{10 i \sqrt{\pi} \, \Gamma \left( \frac{6}{5} \right)}{7 \, \Gamma \left( \frac{7}{10} \right) \, \epsilon^{1/5}} \right] \;.
\end{split}
\label{A2p-approx} 
\end{equation}

\begin{figure}[ht!]
\centerline{
\includegraphics[width=0.5\textwidth,angle=0]{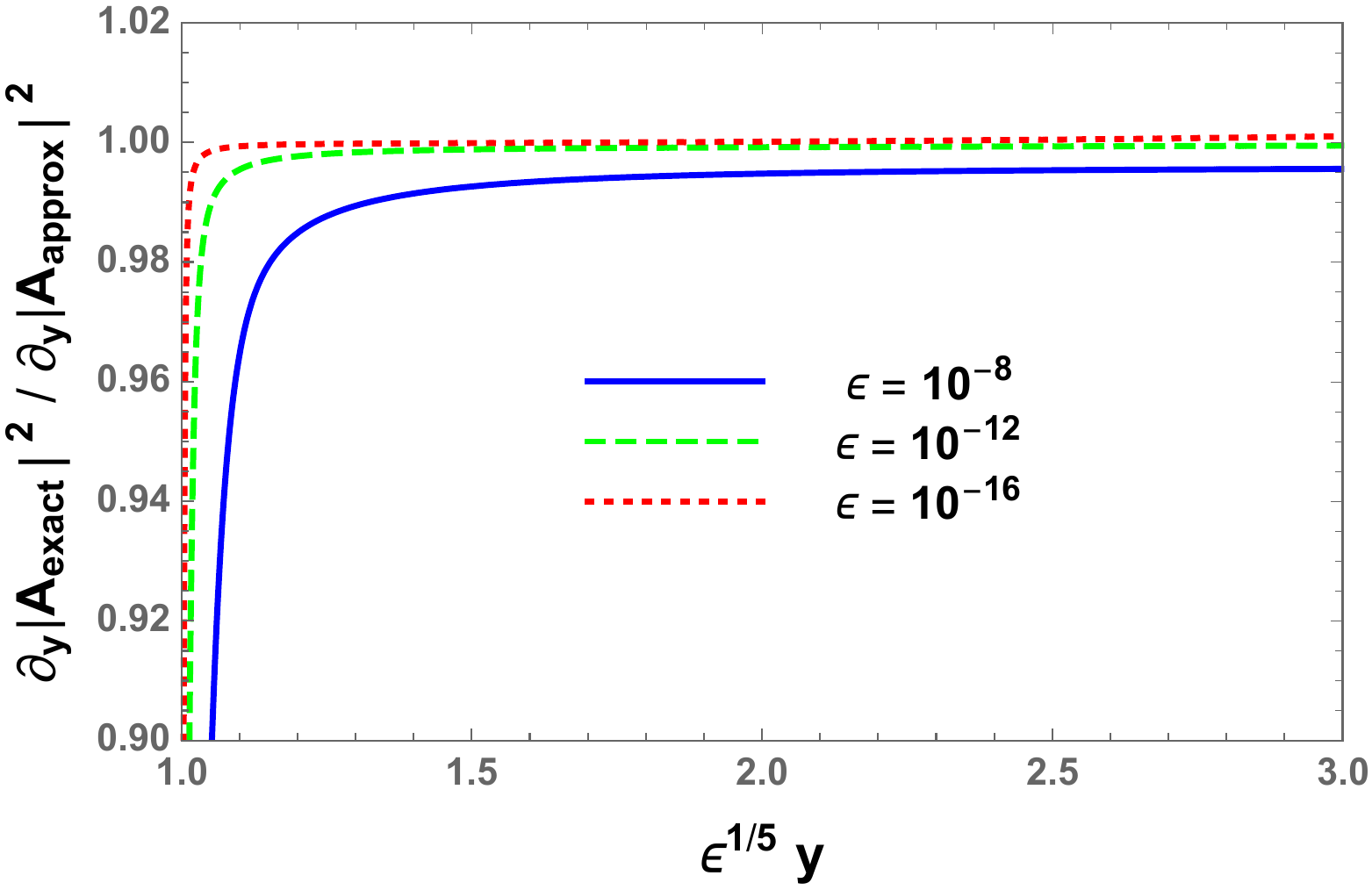}
}
\caption{Comparison between the analytic approximation (\ref{A2p-approx}) and the corresponding quantity obtained from the numerical integration of eq.~(\ref{eomA-y}) for three different values of the parameter $\epsilon$ and as a function of rescaled time $y$.
}
\label{fig:dy-AAp}
\end{figure}

In Figure \ref{fig:dy-AAp} we compare the approximation (\ref{A2p-approx}) against the numerical integration of eq.~(\ref{eomA-y}).
As we are interested in $\lambda ,\, {\tilde k} = {\rm O} \left( 1 \right)$, the parameter $\epsilon$ is of ${\rm O} \left( \frac{f^5}{M_p^5} \right)$, which, as we should in the main text, could be as small as $10^{-20}$ in the cases that we consider.
As ${\tilde \tau}$ ranges from $0$ to a few, the combination $\epsilon y^5$ also ranges in this interval.
We see that the approximation (\ref{A2p-approx}) is very accurate, and clearly adequate for our estimates.

In particular, we use this result to estimate the backreaction term in the evolution equation (\ref{eom-phi}) of the scalar field, that in rescaled variables, reads 
\begin{equation}
\frac{d^2 \varphi}{d {\tilde \tau}^2} + \frac{2}{a} \, \frac{d a}{d {\tilde \tau}} \, \frac{d \varphi}{d {\tilde \tau}} - \frac{12}{{\bar \rho}_m^{2/3}} \, \lambda \, a^2 \, {\rm e}^{-\lambda \varphi} = - \frac{2 \, \lambda^3}{81 \, \pi^2 \, a^2} \, \frac{{\bar \rho}_m^{2/3}}{{\tilde f}^4} \, \frac{V_0}{M_p^4} \, \int d {\tilde k} \, {\tilde k}^2 \, \frac{d}{d {\tilde \tau}} \, \left\vert {\hat A} \right\vert^2 \;.
\end{equation} 

We define as measure of backreaction the ratio between the fourth and third term in this equation, that we write as in eq.~(\ref{back}) as the integral $\int d {\tilde k} \; {\cal B}_{EB}$, with 
\begin{equation}
{\cal B}_{EB} =  \frac{2 \, \lambda^3 \, {\rm e}^{\lambda \varphi}}{729 \, \pi^2 \, a^4} \, \frac{{\bar \rho}_m^{4/3}}{{\tilde f}^5} \, {\tilde k}^3 \, \frac{V_0}{M_p^4} \, \frac{d}{d y} \, \left\vert {\hat A} \right\vert^2 \;. 
\label{BEB} 
\end{equation} 
We plot this quantity in Figure \ref{fig:integrand} in the main text.

\section{System of equations solved }
\label{app:eqs}

In this Appendix we present the system of equations that we use in our numerical integrations, using the number of e-folds of expansion $N$ as ``time variable'', so that our numerical results can be more easily reproduced.
To derive this system, we started from the equations in physical time, $t$, rescaled to the dimensionless combination 
\begin{equation}
{\tilde t} \equiv \frac{\sqrt{3 V_0}}{2 M_p} \, t \;.
\end{equation} 
so that ${\tilde t} = 1$ corresponds to the estimated transition time (\ref{tau-transition}).~\footnote{The physical time is related to the conformal time by $d t = a \, \tau$; integrating the early time solution (\ref{a-early}) leads to $t = \frac{\tau^3}{3 \, \tau_{\rm in}^2}$, so that the transition value for the scale factor $a_{\rm transition} \simeq {\bar \rho}_m^{1/3} $ is attained at the physical time $t_{\rm transition} = \frac{2 M_p}{\sqrt{3 V_0}}$, corresponding to ${\tilde t}_{\rm transition} = 1$.} This time variable is used to introduce the dimensionless Hubble rate 
\begin{equation}
{\tilde H} \equiv \frac{1}{a} \, \frac{d a}{d {\tilde t}} = \frac{2 M_p}{\sqrt{3 \, V_0}} \, H \;.
\end{equation} 

As in the main text, we introduced the dimensionless scalar field $\varphi \equiv \phi / M_p$, and we rescale the axion decay constant analogously, ${\tilde f} \equiv f / M_p$.
We then rescale the comoving momentum of the gauge modes as 
\begin{equation}
{\tilde k} \equiv \frac{3 \sqrt{3} \, {\tilde f}}{2 \, \lambda \, {\bar \rho}_m^{1/3}} \, \frac{M_p}{\sqrt{V_0}} \, k \;.
\end{equation}
As shown in the previous appendix, ${\tilde k} = 1$ corresponds to our estimate for the greatest momentum that is excited at the estimated transition time ${\tilde t} = 1$.
Finally, we rescale the gauge field mode functions as 
\begin{equation}
{\tilde A} \equiv \sqrt{2 \, k} \, A \, {\rm e}^{i k \, \tau} \, = \sqrt{2 k} \, A \, {\rm e^{\frac{4 i \lambda {\bar \rho}_m^{1/3}}{9 {\tilde f}} \, {\tilde k} \, \int_0^N \frac{d N}{{\rm e}^N \, {\tilde H}}}} \;.
\end{equation} 
Due to the $\sqrt{2 k}$ factor, the rescaled mode function ${\tilde A}$ has initially magnitude one.
Moreover, the last factor rescales away the phase due to the early time solution, so that the mode function is nearly constant in the early time regime (namely, it only varies due to the last term in (\ref{A-eom}), which vanishes at asymptotically early times); this considerably speeds up the numerical evaluation of the gauge modes in the early time / UV regime.

Using the definition (\ref{efolds}) it is immediate to see that a derivative wrt (rescaled) physical time can be related to a derivative wrt the number of e-folds via $\frac{d}{d {\tilde t}} = {\tilde H} \, \frac{d}{d N}$.
We can then write the evolution equations for the scalar field, a second order equation for the scale factor (obtained from a combination of the $00$ and of the spatial diagonal part of the background Einstein equations~\footnote{We choose a linear combination that eliminates the $E^2+B^2$ term from the equation.}), and the equation for the gauge field modes, as differential equations in terms of $N$.
Some lengthy but straightforward algebra gives 
\begin{eqnarray} 
&& \!\!\!\!\!\!\!\! \!\!\!\!\!\!\!\! \!\!\!\!\!\!\!\! 
\frac{d^2 \varphi}{d N^2}  + \left[ 1 
- \frac{1}{6} \left( \frac{\partial \varphi}{\partial N} \right)^2 + \frac{2}{9} \frac{ {\bar \rho}_m}{{\rm e}^{3 N} \, {\tilde H}^2 }  + \frac{8}{9 \, {\tilde H}^2} {\rm e}^{-\lambda \varphi} \right] \frac{d \varphi}{d N}  - \frac{4}{3} \, 
\frac{\lambda \, {\rm e}^{-\lambda \varphi} }{ {\tilde H}^2 } 
= \frac{4}{3 \, {\tilde H}^2} \, \frac{\left\langle \vec{E} \cdot \vec{B} \right\rangle}{ V_0  \, {\tilde f} } \;, \nonumber\\ 
&& \!\!\!\!\!\!\!\! \!\!\!\!\!\!\!\! \!\!\!\!\!\!\!\! 
\frac{d {\tilde H}}{d N} = -{\tilde H} \left[ 2 + 
\frac{1}{6} \left( \frac{\partial \varphi}{\partial N} \right)^2 - \frac{2}{9} \, \frac{{\bar \rho}_m}{{\rm e}^{3 N} \, {\tilde H}^2 }  - \frac{8}{9 {\tilde H}^2} \, {\rm e}^{-\lambda \varphi} \right] \;, \nonumber\\ 
&& \!\!\!\!\!\!\!\! \!\!\!\!\!\!\!\! \!\!\!\!\!\!\!\! 
\frac{d^2 {\tilde A}}{d N^2} 
+ \left[ - 1 - 
\frac{1}{6} \left( \frac{\partial \varphi}{\partial N} \right)^2 + \frac{2}{9} \, \frac{{\bar \rho}_m}{{\rm e}^{3 N} \, {\tilde H}^2 }  + \frac{8}{9 {\tilde H}^2} \, {\rm e}^{-\lambda \varphi} 
 -i \frac{8 \, \lambda \, {\bar \rho}_m^{1/3}}{9 \, {\tilde f}} \, \frac{\tilde k}{{\rm e}^N \, {\tilde H}} \right] \frac{d {\tilde A}}{d N} \nonumber\\
&& \quad\quad\quad\quad \quad\quad\quad\quad \quad\quad\quad\quad \quad\quad \quad\quad\quad\quad 
- \frac{4 \, \lambda \, {\bar \rho}_m^{1/3}}{9 \, {\tilde f}^2} \, \frac{\tilde k}{{\rm e}^N \, {\tilde H}} \, \frac{d \varphi}{d N} {\tilde A} = 0 \;, 
\end{eqnarray} 
where
\begin{equation}
\frac{ \left\langle \vec{E} \cdot \vec{B} \right\rangle}{V_0 \, {\tilde f}} = -
\frac{\lambda^3 \, {\bar \rho}_m}{162 \pi^2 {\tilde f}^4} \, \frac{V_0}{M_p^4} \, {\tilde H} \, {\rm e}^{-3 N} 
\int d {\tilde k} \, {\tilde k}^2 \, \frac{\partial}{\partial N} \left\vert {\tilde A} \right\vert^2 \;.
\end{equation} 

Since we are evolving a second order differential equation for the scale factor, the first order Friedman eqaution (\ref{eom-00}) has to be imposed as an initial condition.
This, together with the other initial conditions in the early time regime that we have set in the main text, reads 
\begin{eqnarray} 
&& {\tilde H}_{\rm in}  = {\tilde H}_{\rm in} = \frac{2}{3} \sqrt{ 1 + {\bar \rho}_m + \frac{\left\langle E^2 + B^2 \right\rangle_{\rm in}}{2 V_0} + \frac{2 \, \lambda^2}{81 \, {\bar \rho}_m} } \;, 
\nonumber\\ 
&& \varphi_{\rm in} = 0 \;, \qquad \frac{d \varphi}{d N} \vert_{\rm in} = \frac{4 }{9} \, \frac{\lambda}{\sqrt{{\bar \rho}_m}} \, \frac{1}{{\tilde H}_{\rm in}} \;, \\ 
&& {\tilde A}_{{\rm in}} = 1 \;, \qquad \frac{d {\tilde A}}{d N} \vert_{\rm in} = 0 \;, \nonumber
\end{eqnarray}
where, in the UV early time regime, 
\begin{equation}
 \frac{\left\langle E^2 + B^2 \right\rangle_{\rm in}}{2 V_0} = \frac{4}{729 \, \pi^2} \, \frac{\lambda^2 \, {\bar \rho}_m^{4/3}}{{\tilde f}^4} \, \frac{V_0}{M_p^4} \, \int d {\tilde k} \, {\tilde k}^3 \;.
\label{EB-in}
\end{equation} 

This whole set of equations is written in terms of dimensionless quantities, and it is ready to be integrated numerically (starting from the initial moment $N=0$).
We note that the two backreaction terms introduce the parameter $V_0 / M_p^4$.
The vector field energy density enters in this system only through the initial condition.
This quantity is divergent, being the vacuum energy density of the gauge fields before they undergo any amplification.
In principle, one should subtract it via some regularization scheme \cite{Ballardini:2019rqh}.
However, we only include modes in our numerical evolution for which this term is completely negligible, so, for our practical purposes, this term can simply be disregarded.

\end{document}